\documentclass[prd,aps,showpacs,preprintnumbers,amssymb]{revtex4}
\usepackage{graphicx}
\usepackage{epsf}
\usepackage{amsmath}
\usepackage{epstopdf}
\usepackage{multirow}
\usepackage{bm}
\usepackage{dcolumn}

\def\e3p{$\eta \rightarrow 3 \pi$}

\newcolumntype{M}[1]{>{\centering\arraybackslash}m{#1}}
\newcolumntype{N}{@{}m{0pt}@{}}

\begin{document}

\title{%
\hfill{\normalsize\vbox{%
\hbox{}
 }}\\

{Quark and glue spectroscopy of scalars and pseudoscalars in SU(3) flavor limit}}

\author{Amir H. Fariborz
$^{\it \bf a}$~\footnote[1]{Email:
 fariboa@sunyit.edu}}

\author{Renata Jora
$^{\it \bf b}$~\footnote[2]{Email:
 rjora@theory.nipne.ro}}

\author{Maria Lyukova
	$^{\it \bf a}$~\footnote[2]{Email:
		lyukovm@sunyit.edu}}

\affiliation{$^{\bf \it a}$ Department of Matemathics/Physics, SUNY Polytechnic Institute, Utica, NY 13502, USA}
\affiliation{$^{\bf \it b}$ National Institute of Physics and Nuclear Engineering PO Box MG-6, Bucharest-Magurele, Romania}

\date{\today}

\begin{abstract}
	
Within the framework of the generalized linear sigma model with glueballs recently proposed \cite{Jora25,Jora26}, we study the schematic spectroscopy of scalar and pseudoscalar mesons in the SU(3) flavor limit and explore their quark and glue contents.  	In this framework, for both scalars and pseudoscalars,   the two octet physical states are admixtures of quark-antiquark and four-quark components,  and the three singlet states contain quark-antiquark, four-quark and glue components.  We identify the two scalar octets with $a_0(980)$ and $a_0(1450)$ and the two pseudoscalar octets with $\pi(137)$ and $\pi(1300)$.
We show that, as expected,  the light pseudoscalar  octet is made dominantly of  quark-antiquarks whereas the light scalar octet has a reversed substructure with a dominant four-quark component.   The case of singlets is more complex due to surplus of states up to around 2 GeV.  We consider all 35 permutations for identifying the three pseudoscalar singlets of our model with three of the seven experimental candidates.   Our numerical simulation unambiguously identifies the lightest and the heaviest pseudoscalar singlets with $\eta(547)$ and $\eta(2225)$, respectively, and  favors the identification of the middle singlet with either $\eta(1295)$ or $\eta(1405)$ [or, to a lesser extent, with $\eta(1475)$] and thereby allows a probe of their substructures.   
We then estimate the quark and glue components and find that the three pseudoscalar singlets (from lightest to heaviest) are  mainly of quark-antiquark, four-quark and glue substructure, while the corresponding three scalar singlets (from lightest to heaviest) are of four-quark, quark antiquark and glue contents.  The masses of pure pseudoscalar and scalar glueballs are estimated around 2.0 and 1.6 GeV, respectively.

\end{abstract}
\maketitle

\section{Introduction}

Despite the extensive depth and breath of investigations in low-energy QCD, a complete theoretical framework that can answer everything we like to know about strong interaction in non-perturbative region is yet to emerge \cite{PDG}.   Nevertheless,  a great deal of progress has been made which  has advanced our knowledge of the field, or at least certain aspects of it.   These include
chiral perturbation theory \cite{ChPT1} and its extensions such as chiral unitary approach \cite{ChUA1}-\cite{ChUA8} and inverse amplitude method \cite{IAM1}-\cite{IAM3};   lattice QCD  \cite{LQCD1}-\cite{McNeile:2000xx}; QCD sum-rules \cite{SVZ}-\cite{Huang}; linear sigma models \cite{Schechter1}-\cite{Giacosa22}; as well as many other inspiring  non-perturbative techniques \cite{jaffe}-\cite{Amsler:1995tu}.

At low energies the main degrees of freedom are mesons and baryons, bound states of two, three or more quarks,  or glueballs which are gauge invariant bound states of gluons with different possible quantum numbers \cite{Fritsch}-\cite{Mathieu}.
Among many challenges in low-energy QCD,  understanding the spectroscopy and dynamics of some of the scalar  and pseudoscalar mesons using various tools is known to be an exceptionally formidable task that has been undertaken both theoretically and experimentally in works spanned over several decades (see \cite{PDG} and \cite{07_KZ,Pelaez:2015qba} for comprehensive reviews).  Particularly, there have been intensified efforts over the past couple of decades on the spectroscopy of light scalar mesons, and while the exact status of some of these states is not yet pinned down, many illuminating trends about their characteristics have come to light. 
Their deviations from conventional quark-antiquark states raises the possibility  of these states having  more complex substructures such as four-quark composites \cite{jaffe} or molecular structures  \cite{Weinstein:1990gu}
or a mixed combination of the two.   Furthermore,  some of these states [those above 1 GeV such as $f_0(1500)$ and $f_0(1710)$] may contain significant glue components and that significantly adds to the already long list of challenges in understanding these states.   On the pseudoscalar sector too there are open questions.  For example,  even though  the light pseudoscalar mesons below 1 GeV are  well understood, the situation above 1 GeV remains unclear with  an overpopulation of isosinglet pseudoscalars that include $\eta(1295)$, 
$\eta(1405)$, $\eta(1475)$, $\eta(1760)$ and $\eta(2225)$ which are all listed in PDG \cite{PDG} (there are also disucssions in literature \cite{07_KZ} about $\eta(1440)$ which can be the radial excitation of $\eta$ and that it may be the state that appears as $\eta(1405)$ and $\eta(1475)$).     Some of these states are considered to be
possible non quark-antiquark candidates, such as,  for example,  the interpretation of $\eta(1295)$ as an exotic particle (multiquark, glueball or hybrid) is not excluded in \cite{07_KZ}, or the possibility of $\eta(1405)$ and $\eta(1475)$ being dynamically generated in $\eta f_0(980)$ and $\pi a_0(980)$ channels  investigated in \cite{ChUA8}.

Theoretical frameworks that aim to address the properties of scalar and pseudoscalar mesons up to around 2 GeV  should be mindful of the glue contents in addition to quark-antiquarks and four-quark components.   A generalized linear sigma model for understanding the properties of scalars and pseudoscalars,  particularly their quark-antiquark and four-quark admixtures,  was proposed in \cite{Jora1} and further developed in \cite{Jora2}-\cite{Jora7}. This model is formulated in terms of two chiral nonets (a quark-antiquark and a four-quark) and exhibits chiral symmetry and its breakdown,  both spontaneously and explicitly through quark mass terms.    The model also incorporates an effective instanton term that exactly mocks up the U(1)$_{\rm A}$ anomaly.    If the  potential of the model is not specified (i.e. the least model-independent approach), certain general results can be inferred solely  based on the underlying symmetry \cite{Jora1}, however,  in order to make complete predictions for the scalar and pseudoscalar masses and mixings, the potential has to be modeled.  Relying exclusively on chiral symmetry, its breakdown,  and anomalies,  is not sufficient to define a practical framework,  for in principle there are infinite number of terms that contribute to the potential.  To make the framework practical,  in \cite{Jora5} a systematic procedure was defined in which the contributing terms with fewer quark-antiquark lines were considered favored.    In this scheme, the leading potential contains terms with eight or fewer quark lines.  Various low-energy processes were studied in this framework in its leading order and it was found that while light pseudoscalar mesons below 1 GeV are dominantly quark-antiquark states, the  light scalars have reversed substructure, typically with  dominant four-quark components.

The cases of isosinglet scalar and pseudoscalar states above 1 GeV are overshadowed by mixing with glueballs,  and as a result,   their investigation within the generalized linear sigma model \cite{Jora5} (which did  not contain glueballs) is incomplete.      In a recent work \cite{Jora25},  the generalized linear sigma model was extended to include two glueballs (a scalar and a pseudoscalar glueball). In \cite{Jora25},  the most general Lagrangian was formally developed that embodies, in addition to  chiral symmatry (and its spontaneous and explicit breakdown), terms that represent interaction of glueballs with quarkonia in a manner that exactly realizes the axial and trace anomalies of QCD.    This most general Lagrangian contains (even in the leading order) a large set of free parameters that need to be determined by fitting the theoretical predictions of the model to available experimental data. Since various theoretical predictions are  nonlinear functions of the model parameters, a fit of these predictions to available data by a brute force numerical method is not a reliable starting point due to the fact that the minimization of functions that depend on many parameters are often a jagged function with many local minima.   Therefore, in such minimizations it is possible to end up at a non-physical minimum in the multidimensional parameter space of the model.   Our general strategy is to first push the model to  exactly solvable limits by imposing physically meaningful conditions, and once the parameters are determined in these solvable limits, gradually relax the conditions and study the evolution of the parameter space.      In \cite{Jora25}, we considered a decoupling limit, in which the two glueballs (the scalar and the pseudoscalar glueballs), which exist in the Lagrangian, do not interact with quark composite fields.    In this exactly solvable limit, scalar glueball still plays an important role in stabilizing the QCD vacuum but refrains from mixing with quarkonia,  and hence,  becomes a pure glueball.  Moreover, we showed in \cite{Jora26} that the model predicts  the pure scalar glueball mass $m_h = 2 h_0$ where $h_0$ is the glueball condensate,  which upon comparison with QCD sum-rules determination of dimension four gluon condensate, we found that $h_0$ is in the range of  0.80 to 1.0 which results in an estimate of the pure scalar glueball mass in the range of 1.6-2.0 GeV.     We also determined,  in the decoupling limit,  the initial model predictions for masses as well as for quark and glue contents.   In the decoupling limit, the model is pushed to SU(3)$_{\rm V}$ subgroup with additional non-interaction restrictions imposed on the glueballs.  In the present work, we take the first step away from the decoupling limit and while we still work in the flavor SU(3) limit, we relax the additional conditions that prevented the glueballs from interacting with quark-antiquark and four-quark nonets.    Since glueballs are flavor singlets,  SU(3) symmetry is considered to be a reasonable starting point for understanding their interactions.      Of course,  for a complete spectroscopy of scalars and pseudoscalars, the SU(3) breaking effects should be included.

In Sec. II, we give a brief overview of the theoretical framework developed in \cite{Jora25} as well as our notation and 
setup in the SU(3) flavor limit.   We then give  the details of our exact parameter determination  in Sec. III, which in turn leads to the results and predictions in Sec. IV (including the details of the numerical analysis as well as the relaxation of inputs using parallel and Monte Carlo computations).  We conclude with a summary and a brief outline of  future works in Sec. V.

\section{Brief Review of the Theoretical framework}

In this section we give a brief review of the theoretical model developed in \cite{Jora1}-\cite{Jora7} and recently extended to include scalar and pseudoscalar glueballs in \cite{Jora25}.   The model is constructed in terms of 3$\times$3 matrix
chiral nonet fields:
\begin{equation}
M = S +i\phi, \hskip 2cm
M^\prime = S^\prime +i\phi^\prime,
\label{sandphi1}
\end{equation}
which are in turn defined in terms of ``bare'' scalar meson nonets $S$ (a quark-antiquark scalar nonet) and $S'$ (a four-quark scalar nonet), as well as ``bare'' pseudoscalar meson nonets $\phi$ (a quark-antiquark pseudoscalar nonet) and $\phi'$ (a four-quark pseudoscalar nonet).
Chiral fields $M$ and $M'$ transform in the same way under
chiral SU(3) transformations
\begin{eqnarray}
M &\rightarrow& U_L\, M \, U_R^\dagger,\nonumber\\
M' &\rightarrow& U_L\, M' \, U_R^\dagger,
\end{eqnarray}
but transform differently under U(1)$_A$
transformation properties
\begin{eqnarray}
M &\rightarrow& e^{2i\nu}\, M,  \nonumber\\
M' &\rightarrow& e^{-4i\nu}\, M'.
\label{U1A}
\end{eqnarray}
There are several possible four-quark substructures for $M'$ (such as diquark-antidiquark types  or molecular type), however, the model does not distinguish these different types of four-quark substructures and can only probe the percentages of quark-antiquark versus four-quark components through the U(1)$_A$ transformation according to (\ref{U1A}).   

The effective Lagrangian is constructed in terms $M$ and $M'$ and should display chiral symmetry (and its beakdown), explicit symmetry breaking due to quark masses as well as the axial and trace anomalies of QCD according to:
\begin{eqnarray}
&&\partial^{\mu}J^5_{\mu}=\frac{g^2}{16\pi^2}N_F\tilde{F}F=G,
\nonumber\\
&&\theta^{\mu}_{\mu}=\partial^{\mu}D_{\mu} = -\frac{\beta(g^2)}{2g}FF=H,
\label{intr64553}
\end{eqnarray}
where $F$ is the SU(3)$_{\rm C}$ field tensor, $\tilde{F}$ is its dual, $N_F$ is the number of flavors, $\beta(g^2)$ is the beta function for the coupling constant, $J^5_{\mu}$ is the axial current and $D_{\mu}$ is the dilatation current.

With identifications $H=h^4$, $G=g h^3$, where $h$ and $g$ respectively represent the scalar and pseudoscalar glueball fields, the generalized linear sigma model Lagrangian augmented to include scalar and pseudoscalar glueballs has the general structure \cite{Jora25}
\begin{eqnarray}
{\cal L}&=&-\frac{1}{2}{\rm Tr}(\partial^{\mu}M\partial_{\mu}M^{\dagger})-\frac{1}{2}{\rm Tr}(\partial^{\mu}M'\partial_{\mu}{M'}^{\dagger})
-{1\over 2} (\partial_{\mu} h)(\partial_{\mu} h) - {1\over 2} (\partial_{\mu} g) (\partial_{\mu} g)
-V, \nonumber \\
- V&=& f + f_{\rm A} + f_{\rm S} + f_{\rm SB}.
\label{inlgr567}
\end{eqnarray}
where $f(M, M', g, h)$ is invariant under chiral, axial and scale transformations, and in leading order is
\begin{eqnarray}
f &=&
- \left(
u_1 h^2 {\rm Tr}[MM^{\dagger}]
+ u_2{\rm Tr}[MM^{\dagger}MM^{\dagger}]+
u_3 h^2 {\rm Tr}[M^{\prime}M^{\prime \dagger}]+ u_4 h (\epsilon_{abc}\epsilon^{def}M^a_dM^b_eM^{\prime c}_f+h.c.)+
\right.
\nonumber \\
&&
\left.\hskip .5cm  u_5 h^4 + u_6  h^2 g^2  + \cdots\right),
\label{pot201867} 
\end{eqnarray}
where $u_1\cdots u_6$ are the unknown constants that need to be determined from experiment.    The instanton term $f_{\rm A}$ breaks axial symmetry and is given by
\begin{eqnarray}
f_{\rm A} &=&
i{G\over 12} \left[ \gamma_1\ln \left(\frac{\det M}{\det M^{\dagger}}\right)+\gamma_2\ln\left(\frac{{\rm Tr}(MM^{\prime\dagger})}{{\rm Tr}(M^{\prime}M^{\dagger})}\right)\right],
\label{axial2018967}
\end{eqnarray}
where $\gamma_1$ and $\gamma_2$  are arbitrary parameters that must satisfy the constraint: $\gamma_1+\gamma_2=1$ \cite{Jora1}-\cite{Jora6}.   The next term $f_{\rm S}$ breaks scale symmetry
\begin{eqnarray}
f_{\rm S} &=&
-H \left\{
\lambda_1 \ln\left(\frac{H}{\Lambda^4}\right)
+\lambda_2 \left[\ln\left(\frac{\det M}{\Lambda^3}\right)+\ln\left(\frac{\det M^{\dagger}}{\Lambda^3}\right)\right]
\right.
\nonumber \\
&&
\left.
\hskip .8cm
+ \lambda_3\left[ \ln\left(\frac{{\rm Tr} MM^{\prime\dagger}}{\Lambda^2}\right)+\ln\left(\frac{{\rm Tr}M' M^{\dagger}}{\Lambda^2}\right)
\right]\right\}.
\label{scale7756}
\end{eqnarray}
where $\lambda_1$, $\lambda_2$ and $\lambda_3$ are also arbitrary parameters that must fulfill the condition: $4\lambda_1+6\lambda_2+4\lambda_3=1$ \cite{Jora7}.
The potential is invariant under ${\rm U(3)}_{\rm L} \times {\rm  U(3)}_{\rm R}$ with the exception of the $f_A$ term which breaks ${\rm U(1)}_{\rm A}$.
The leading explicit symmetry breaking term has the form of quark mass term:
\begin{eqnarray}
f_{SB}=2 {\rm Tr}[AS]
\label{sym3528637}
\end{eqnarray}
where $A={\rm diag}(A_1,A_2,A_3)$ is a  matrix proportional to the three light quark masses.  We are interested in the SU(3) limit where
\begin{eqnarray}
&&\alpha_3=\alpha_1=\alpha
\nonumber\\
&&\beta_3=\beta_1=\beta
\label{u3_sol}
\end{eqnarray}
In this limit the minimum equations are:
\begin{eqnarray}
\left\langle {{\partial V} \over {\partial S_1^1}} \right\rangle_0  &=&
\left\langle {{\partial V} \over {\partial S_3^3}} \right\rangle_0 =
8\, u_4\, h_0\,\alpha\,\beta + 2\, u_1 \, { h_0}^{2}\alpha
+ 4\, u_2 \, {\alpha}^{3} +
{2\over 3}\,{\frac {{h_0}^{4} \left( 3\,\lambda_2+\lambda_3 \right) }{
		\alpha}}
-2\,A = 0,
\label{E_Vmin_qq}
\\
\left\langle {{\partial V} \over {\partial {S'}_1^1}} \right\rangle_0  &=&
\left\langle {{\partial V} \over {\partial {S'}_3^3}} \right\rangle_0  =
{{2h_0}\over{3\beta}}
\left(
6\,\alpha^{2}\beta\, u_4 + 3\, u_3\,h_0\, {\beta}^{2}+{
		 h_0}^{3}\lambda_3
\right) = 0,
\label{E_Vmin_4q}
\\
\left\langle {{\partial V} \over {\partial h}} \right\rangle_0  &=&
8\,\ln  \left( {\frac {3\alpha\,\beta}{{\Lambda}^{2}}} \right) 
{ h_0}^{3}\lambda_3 + 8\,\ln  \left( {\frac {{\alpha}^{3}}{{\Lambda}^{3
		}}} \right) { h_0}^{3}\lambda_2 + 4\,\ln  \left( {\frac {{ h_0}^{
		4}}{{\Lambda}^{4}}} \right) { h_0}^{3}\lambda_1\nonumber \\
&&
+ 4\, \left( \lambda_1 +  u_5 \right) { h_0}^{3}
+ 6\, \left(  u_1 \,{\alpha}^{2} + u_3\,{\beta}^{2} \right)  h_0
+12\,{\alpha}^{2}\beta\, u_4  = 0.
\label{E_Vmin_h}
\end{eqnarray}
The first two equations represent the stability of vacuum with respect to variation of quark-antiquark  field $S_1^1$ (or $S_3^3$), and four-quark field ${S'}_1^1$ (or ${S'}_3^3$), respectively.  The last equation determines  the stability of vacuum with respect to the scalar glueball field $h$.

The octet ``8'' and singlet ``0'' mass matrices in the SU(3) limit are (where $Y$ refers to the scalars and $N$ to the pseudoscalars):
\begin{eqnarray}
\left(Y^2_8\right)_{11} &=&
{1\over \alpha^2}
\left(12\,u_2\,{\alpha}^{4}-4\, u_4 \,  h_0\,\beta\,{
	\alpha}^{2}+2\, u_1\,{ h_0}^{2}{\alpha}^{2}-2\,{ h_0}^{4}
\lambda_2\right)
\nonumber\\
\left(Y^2_8\right)_{12} &=& 		
{{2h_0}\over {3\alpha\beta}}
\left( -6\,{\alpha}^{2}\beta\,  u_4 +{h_0}^{3}\lambda_3 \right)
\nonumber\\
\left(Y^2_8\right)_{22} &=& 2 u_3 h_0^2		
\end{eqnarray}
\begin{eqnarray}
\left(Y^2_0\right)_{11} &=&
{1 \over {3\alpha^2}}\left[
\left( -6\,\lambda_2-2\,\lambda_3 \right) h_0^{4}+6
\, u_1\, h_0^2{\alpha}^{2}+24\, u_4 \, h_0\,\beta\,
{\alpha}^{2}+36\, u_2\,{\alpha}^{4}
\right]
\nonumber\\
\left(Y^2_0\right)_{12} &=&
8\,u_4\, h0\,\alpha
\nonumber\\
\left(Y^2_0\right)_{13} &=&
{4\over {\sqrt{3}\alpha}}\,
\left( 6\,{\alpha}^{2}\beta\,{\it u_4}+3\,{\it
	u_1}\,{\it h_0}\,{\alpha}^{2}+6\,{h_0}^{3}\lambda_2+2\,{h_0}^{3
}\lambda_3 \right)
\nonumber\\
\left(Y^2_0\right)_{22} &=&
-{{2h_0^2}\over {3 \beta^2}} \,
\left( -3\, u_3\, \beta^{2}+ h_0^2\lambda_3 \right)
\nonumber\\		
\left(Y^2_0\right)_{23}		&=&	
{{4\sqrt{3}}\over {3\beta}}
\left( 3\,{\alpha}^{2}\beta\,{\it u_4}+3\, u_3\, h_0\,{\beta}^{2}+2\, h_0^{3}\lambda_3 \right)
\nonumber\\
\left(Y^2_0\right)_{33}		&=&	
24\,\ln  \left( {\frac {{\alpha}^{3}}{{\Lambda}^{3}}} \right) h_0^2 \lambda_2
+12\,\ln  \left( {\frac {h_0^4}{{\Lambda}^{4}}}
\right) {h_0}^{2}\lambda_1+24\,\ln  \left( {\frac {\alpha\,\beta
	}{{\Lambda}^{2}}} \right) {h_0}^{2}\lambda_3+24\,\ln  \left( 3
\right) { h_0}^{2}\lambda_3 + \left( 28\,\lambda_1 + 12\,{\it u_5}
\right) {h_0}^{2}
\nonumber\\
&&
+6\, u_1 \,{\alpha}^{2}+6\, u_3 \,{\beta}
^{2}
\end{eqnarray}
\begin{eqnarray}
\left(N^2_8\right)_{11} &=&
{1\over {\alpha^2}}
\left(
4\, u_2 \, \alpha^4 + 4\,  u_4 \, h_0 \, \beta \, \alpha^2 + 2\,  u_1 \, h_0^2 \alpha^2 + 2\, h_0^4
\lambda_2
\right)
\nonumber\\
\left(N^2_8\right)_{12} &=&
{{2 h_0}\over {3 \alpha \beta}}
\left( 6\, \alpha^2 \beta\,  u_4 +  h_0^3 \lambda_3 \right)
\nonumber\\
\left(N^2_8\right)_{22} &=&
2\, u_3\,  h_0^2
\end{eqnarray}
\begin{eqnarray}
\left(N^2_0\right)_{11} &=&
{1\over {3\alpha^2}}
\left[
\left(
6\,\lambda_2 + 2\,\lambda_3 \right) h_0^4 + 6 \,  u_1 \, h_0^2 \alpha^2 - 24\, u_4 \,  h_0 \, \beta\,
\alpha^2 + 12\, u_2 \, \alpha^4
\right]
\nonumber\\
\left(N^2_0\right)_{12} &=& -8\, u_4 \, h_0\, \alpha
\nonumber\\
\left(N^2_0\right)_{13} &=&
{{\sqrt{3}h_0^3}\over {18\alpha}}
\left( 2\,\gamma_1 + 1 \right)
\nonumber\\
\left(N^2_0\right)_{22} &=&
{{2h_0^2}\over {3\beta^2}}\,
\left( 3\,u_3\,\beta^2 + h_0^2\lambda_3 \right)
\nonumber\\
\left(N^2_0\right)_{23} &=&
{{\sqrt {3} h_0^3}\over {18\beta}}\,  \left( -1+\gamma_1 \right)
\nonumber\\
\left(N^2_0\right)_{33} &=& 2\, u_6\, h_0^2.
\nonumber\\
\end{eqnarray}

The octet physical states
\begin{equation}
\Psi_{8^+}  =
\left[
\begin{array}{cc}
\psi_{8^+}^{(1)}\\
\psi_{8^+}^{(2)}
\end{array}
\right],
\hskip .75cm
\Psi_{8^-}  =
\left[
\begin{array}{cc}
\psi_{8^-}^{(1)}\\
\psi_{8^-}^{(2)}
\end{array}
\right],
\end{equation}
diagonalize $\left[Y_8^2\right]$ and $\left[N_8^2\right]$ respectively and are related to the octet ``bare'' states
\begin{equation}
B_{8^+}=
\left[
\begin{array}{c}
f_8\\
f'_8
\end{array}
\right],
\hskip .75cm
B_{8^-}=
\left[
\begin{array}{c}
\eta_8\\
\eta'_8
\end{array}
\right],
\label{F_eta8_OS}
\end{equation}
by
\begin{eqnarray}
\Psi_{8^+} &=&
\left[{K_{8^+}}\right]^{-1}
B_{8^+},
\nonumber\\
\Psi_{8^-} &=&
\left[{K_{8^-}}\right]^{-1}
B_{8^-},
\end{eqnarray}
therefore
\begin{eqnarray}
{\widetilde \Psi}_{8^+}
 \left[ Y_8^2\right]_{\rm diag}  \Psi_{8^+} & = &
{\widetilde  B_{8^+}}  \left[Y_8^2\right] B_{8^+}, \nonumber \\
{\widetilde \Psi}_{8^-}
 \left[ N_8^2\right]_{\rm diag}  \Psi_{8^-} & = &
 {\widetilde  B_{8^-}}  \left[N_8^2\right] B_{8^-},
\end{eqnarray}
where $\left[ Y_8^2\right]_{\rm diag} = {\rm diag}\left(m_{8^+}, {m'}_{8^+}\right)$ and
$\left[ N_8^2\right]_{\rm diag} = {\rm diag}\left(m_{8^-}, {m'}_{8^-}\right)$ are diagonalized mass matrices that contain the physical octet masses.

Similarly, the singlet physical states
\begin{equation}
\Psi_{0^+}  =
\left[
\begin{array}{cc}
\psi_{0^+}^{(1)}\\
\psi_{0^+}^{(2)}\\
\psi_{0^+}^{(3)}
\end{array}
\right],
\hskip .75cm
\Psi_{0^-}  =
\left[
\begin{array}{cc}
\psi_{0^-}^{(1)}\\
\psi_{0^-}^{(2)}\\
\psi_{0^-}^{(3)}
\end{array}
\right],
\end{equation}
diagonalize $\left[Y_0^2\right]$ and $\left[N_0^2\right]$ respectively and are related to the singlet ``bare'' states
\begin{equation}
B_{0^+}=
\left[
\begin{array}{c}
f_0\\
f'_0\\
h
\end{array}
\right],
\hskip .75cm
B_{0^-}=
\left[
\begin{array}{c}
\eta_0\\
\eta'_0\\
g
\end{array}
\right],
\label{F_eta_OS}
\end{equation}
by
\begin{eqnarray}
\Psi_{0^+} &=&
\left[{K_{0^+}}\right]^{-1}
B_{0^+},
\nonumber\\
\Psi_{0^-} &=&
\left[{K_{0^-}}\right]^{-1}
B_{0^-},
\end{eqnarray}
therefore
\begin{eqnarray}
{\widetilde \Psi}_{0^+}
 \left[ Y_0^2\right]_{\rm diag}  \Psi_{0^+} & = &
{\widetilde  B_{0^+}}  \left[Y_0^2\right] B_{0^+}, \nonumber \\
{\widetilde \Psi}_{0^-}
 \left[ N_0^2\right]_{\rm diag}  \Psi_{0^-} & = &
 {\widetilde  B_{0^-}}  \left[N_0^2\right] B_{0^-}.
\end{eqnarray}
where $\left[ Y_0^2\right]_{\rm diag} = {\rm diag}\left(m_{0^+}, {m'}_{0^+}, {m''}_{0^+}\right)$ and
$\left[ N_0^2\right]_{\rm diag} = {\rm diag}\left(m_{0^-}, {m'}_{0^-}, {m''}_{0^-}\right)$ are diagonalized mass matrices that contain the physical singlet masses.

\section{Parameter determination}
In this section we give our general strategy for determining the unknown model parameters. The minimum equation 
(\ref{E_Vmin_qq}) can be used to determine $A$. The minimum equation  (\ref{E_Vmin_4q}) establishes the relationship:
\begin{eqnarray}
	2\alpha^2u_4+\beta h_0u_3+\frac{\alpha h_0^3\lambda_3}{3\alpha\beta}=0,
	\label{firdtmin76775}
\end{eqnarray}
to write
\begin{eqnarray}
	[N_8^2]_{12}=-2h_0^2\frac{\beta }{\alpha}u_3.
	\label{E_N8_12_initial}
\end{eqnarray}
At the same time,  from the parameterization of rotation matrices we can write:
\begin{eqnarray}
	\left[N_8^2\right]_{12}&=& \sin \theta_{8^-} \cos \theta_{8^-} \left( {m'}_{8^-}^2 - m_{8^-}^2\right)
	=-2h_0^2\frac{\beta }{\alpha}u_3,
	\label{E_N8_12}\\
	\left[N_8^2\right]_{22}	&=&  \cos^2 \theta_{8^-} \, {m'}_{8^-}^2 +
	                                              \sin^2 \theta_{8^-} \, m_{8-}^2 = 2\, h_0^2\, u_3,
\label{E_N8_22}
\end{eqnarray}
and equations for decay constants give:
\begin{eqnarray}
	&&f_{8^-}=2\alpha\cos \theta_{8^-}-2\beta\sin\theta_{8^-},
	\nonumber\\
	&&{f'}_{8^-}=2\alpha\sin \theta_{8^-} + 2\beta\cos \theta_{8^-}.
	\label{E_fpi_fpip}
\end{eqnarray}
For an input of $h_0$, ${m}_{8^-}$, ${m'}_{8^-}$, ${f}_{8^-}$  and ${f'}_{8^-}$,  equations  (\ref{E_N8_12}), (\ref{E_N8_22}) and (\ref{E_fpi_fpip}) can be solved for four unknowns $\alpha$, $\beta$, $u_3$ and $\theta_{8^-}$.   For example, from (\ref{E_fpi_fpip})
\begin{eqnarray}
	&&\frac{\beta}{\alpha}=\frac{-\sin\theta_{8^-}\,
		f_{8^-} + \cos\theta_{8^-}\, f'_{8^-}}{\cos\theta_{8^-}\, f_{8^-} + \sin\theta_{8^-}f'_{8^-}},
	\label{E_beta_over_alpha}
\end{eqnarray}
which, upon substitution of this result, together with the solution of $u_3$ from Eq. (\ref{E_N8_22}), into  Eq. (\ref{E_N8_12}) we can solve for $\cos \theta_{8^-}$:
\begin{eqnarray}
	\cos\theta_{8^-}=f_{8^-}m^2_{8^-}\Bigg[f_{8^-}^2\, m_{8^-}^4 + {f'}_{8^-}^2 \, {m'}_{8^-}^4\Bigg]^{-1/2}.
	\label{E_cos_theta_8m}
\end{eqnarray}
This result, when substituted into (\ref{E_fpi_fpip}), leads to determination of $\alpha$ and $\beta$ and subsequently $u_3$ from (\ref{E_N8_22}), all  in terms of $h_0$, $f_{8^-}$, ${f'}_{8^-}$, $m_{8-}$ and $m_{8-}'$.

\begin{figure}[!htb]
	\centering
	\includegraphics[width=3.1in]{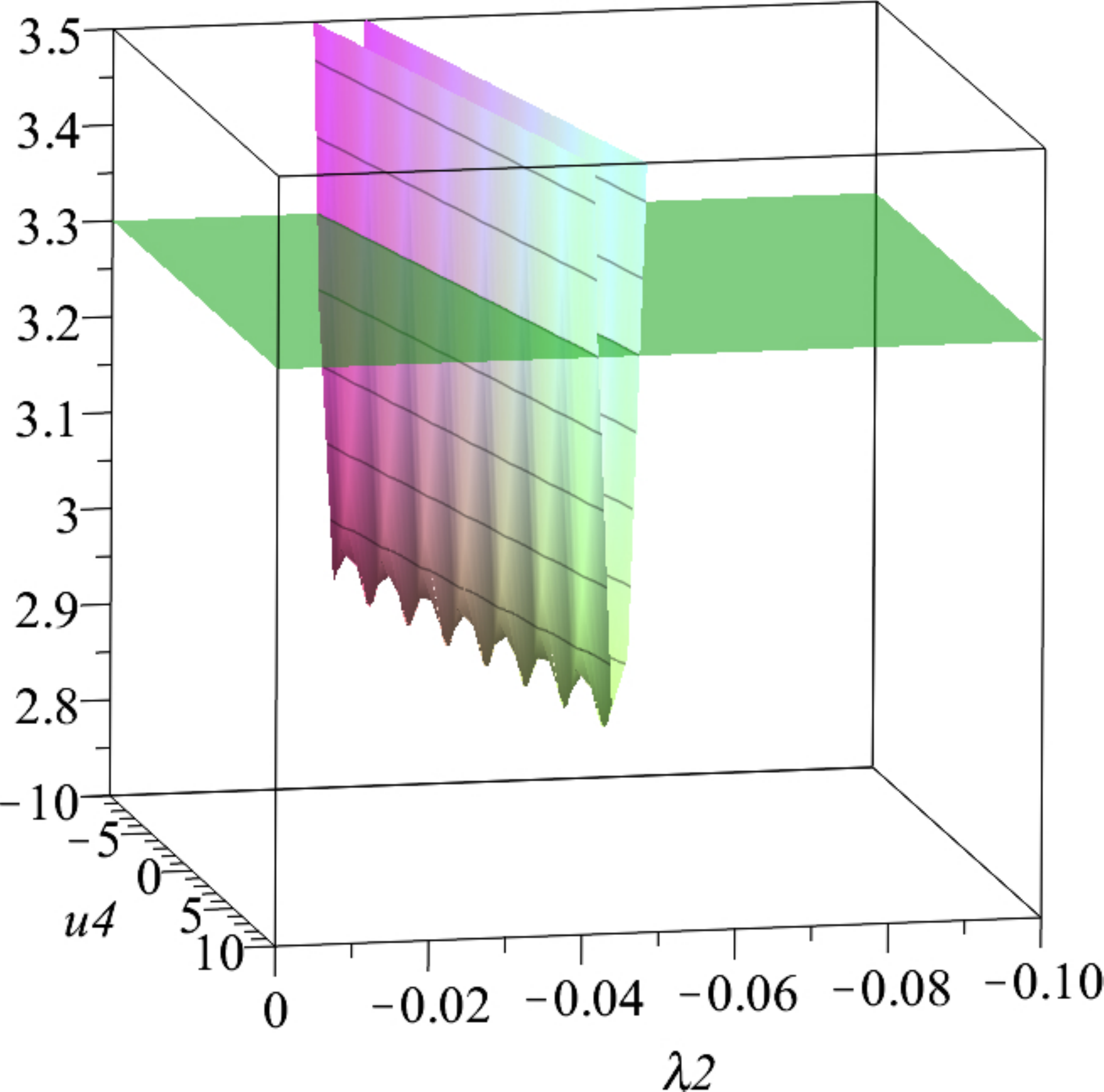}
	\includegraphics[width=3.1in]{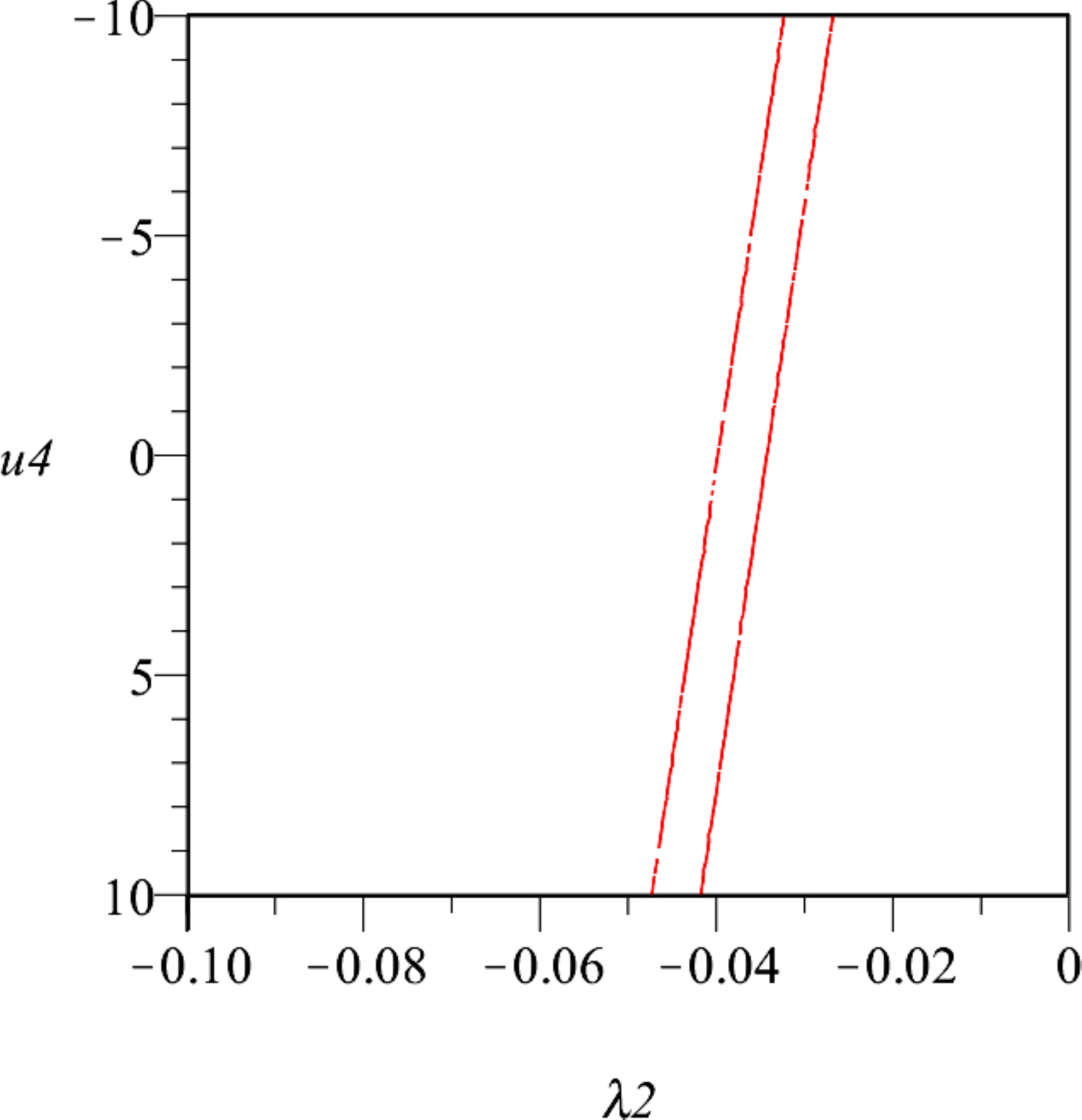}

	\caption{Trace of $Y_8^2$ versus $u_4$ and $\lambda_2$  (left) which also includes the physical flat plane
${\rm Tr}\left( Y_8^2\right) = m_{8^+}^2 + {m'}_{8^+}^2$.   The intersection of the surface with the plane is shown on the right.}
	\label{F_Tr_Y8}
\end{figure}

Having determined $\alpha$, $\beta$, $u_3$ and mixing angle $\theta_{8^-}$, we can now determine $u_1$, $u_2$, $u_4$ and $\lambda_2$ from the trace and determinant of the octets.   Note that with $\lambda_1=\frac{11}{36}$  (from a first order trace anomaly result), together with  the condition $4\lambda_1+6\lambda_2+4\lambda_3=1$,  we can then calculate $\lambda_3$.   We first apply the determinant equations
\begin{eqnarray}
\det \left(N_8^2\right) &=&  m^2_{8^-} {m'}^2_{8^-},
\label{E_det_N8}\\
\det\left( Y_8^2\right)	&=&  m^2_{8^+} {m'}^2_{8^+},
\label{E_det_Y8}
\end{eqnarray}
both linear in $u_1$ and $u_2$,  to solve for these two variables:
\begin{eqnarray}
u_1 &=& 
\left( 
144\,\alpha^4 \beta^2  h_0^2 u_4^2 
+288 \, \alpha^2 \beta^3 h_0^3 u_3 \, u_4 
-48 \, \alpha^2 \beta\,h_0^5 \lambda_3\, u_4 
-108 \,\beta^4 h_0^4 u_3^2 
+144 \, \beta^2 h_0^6 \lambda_2 \, u_3 
+4 \, h_0^8 \lambda_3^2 \right. \nonumber \\
&& \left. 
-27 \, m_{8^-}^2  m_{8^-}'^2 \alpha^2 \beta^2 
+9 \, m_{8^+}^2 m_{8^+}'^2 \alpha^2 \beta^2 \right) / 
\left( 72\,\alpha^2 \beta^2 h_0^4 u_3 \right) ,
\nonumber \\
u_2	&=&  \left( 
144 \, \alpha^4 \beta^2 h_0^2 u_4^2
+144 \, \alpha^2 \beta^3 h_0^3 u_3\, u_4
-48 \, \alpha^2 \beta \, h_0^5 \lambda_3 \, u_4
-36 \, \beta^4 h_0^4 u_3^2 
+72 \, \beta^2 h_0^6 \lambda_2 \, u_3 \right. \nonumber \\
&& \left. 
+4 \, h_0^8 \lambda_3^2 
-9\, m_{8^-}^2 m_{8^-}'^2\alpha^2 \beta^2
+9\, m_{8^+}^2 m_{8^+}'^2\alpha^2 \beta^2 \right) /
\left( 144\,\alpha^4\beta^2 h_0^2 u_3 \right),
\label{E_u1_u2}
\end{eqnarray}
These solutions identically satisfy the trace of pseudoscalar octet (since in dtermination of $\alpha$, $\beta$, $u_3$ and $\theta_{8^-}$ this trace is implicitly implemented).  However, substitution of $u_1$ and $u_2$ from (\ref{E_u1_u2}) into trace of scalar octet leads to a quadratic equation in $u_4$ and $\lambda_2$, therefore we cannot uniquely solve for these two unknowns and need to impose physical constrains to limit their ranges of variation. For our numerical analysis we use inputs
\begin{eqnarray}
m_{8^-} &=& 137 \,\,{\rm MeV}
\nonumber\\
{m'}_{8^-} &=& 1300 \pm 100 \,\,{\rm MeV}
\nonumber \\
f_{8^-} &=& 131 \,\,{\rm MeV}
\nonumber\\
{f'}_{8^-} &=&  -0.6939 \pm 0.06939 \,\,{\rm MeV}
\nonumber\\
m_{8^+} &=& 980 \,\,{\rm MeV}
\nonumber\\
{m'}_{8^+} &=& 1474\,\,{\rm MeV}
\nonumber \\
h_0 &=& 0.80 \rightarrow 0.81 \nonumber \\
\lambda_2 &=& -0.04 \rightarrow -0.03
\label{inputs}
\end{eqnarray}
These inputs generate a set of numerical values of parameters $\alpha$, $\beta$, and $u_3$:
\begin{equation}
	S_\mathrm{I} = \left\{
	\alpha, \beta, u_3 \,\Bigg| \, {\rm Generated \,\, from \,\, inputs \,\, of \,\,} (\ref{inputs})
	\right\}
	\label{S_I}
\end{equation}
Since $u_4$ and $\lambda_2$ (and hence $u_1$ and $u_2$) are not uniquely determined,  we seek to generate a set of acceptable values for these parameters that satisfy the basic physical constrains.   We have plotted ${\rm Tr} \left( Y_8^2\right)$ versus $u_4$ and $\lambda_2$ in Fig. \ref{F_Tr_Y8}.  The intersections of this surface with physical plane  ${\rm Tr} \left( Y_8^2\right) = m_{8^+}^2 + {m'}_{8^+}^2$ are in the form of two parallel lines in $u_4 \lambda_2$ plane that can be used to generate a set of  values of $u_4$ and $\lambda_2$.   To further limit this range, we filter out points on these intersections that do not result in real eigenvalues for the octet and singlet scalar systems  as well as impose  additional constrains related to the expected values for the masses of the singlet scalar system.  Since there are several $f_0$ states below 2 GeV we start our numerical search with points at which  ${\rm Tr} \left( Y_8^2\right) =  m_{8^+}^2 + m_{8^+}^2$ (as the two octet scalar masses are well known), along with ${\rm Tr} \left( Y_0^2\right) \le 12$ ,which intends to impose the condition that the scalar singlet masses be less than 2 GeV (since there are several experimental candidates in this range).  These conditions generate an initial set of points for $u_4$ and $\lambda_2$ [as well as $u_1$ and $u_2$ using (\ref{E_u1_u2})]:
\begin{equation}
	S_\mathrm{II} = \left\{
	u_1, u_2, u_4, \lambda_2 \,\Bigg| \, m^2_{8^+} > 0, {m'}^2_{8^+}> 0,
 m^2_{0^+} > 0, {m'}^2_{0^+}> 0, {m''}^2_{0^+}> 0,	
	{\rm Tr}\left(Y_0^2\right)\le 12 \,\,\mathrm{GeV}^2 \right\}.
	\label{S_II}
\end{equation}

This set gets further refined when constrains from the pseudoscalar singlet system which introduces two additional parameters $\gamma_1$ and $u_6$ are considered.
We examine whether there are values of $\gamma_1$ and $u_6$ (together with choices of $\alpha$, $\beta$ and $u_3$ from set $S_\mathrm{I}$ and choices of  $u_1$, $u_2$, $u_4$, $\lambda_2$ from set $S_\mathrm{II}$) such that the three pseudoscalar singlets of the model approach three of the experimentally known $\eta$ states. There are several $\eta$ states listed in PDG \cite{PDG}:
\begin{eqnarray}
m^{\rm exp.}[\eta (547)] &=& 547.862 \pm
0.017\, {\rm
	MeV},\nonumber \\
m^{\rm exp.}[\eta' (958)] &=& 957.78 \pm 0.06
\, {\rm
	MeV}.
\nonumber\\
m^{\rm exp.}[\eta (1295)] &=& 1294 \pm 4\, {\rm
	MeV},\nonumber \\
m^{\rm exp.}[\eta (1405)] &=& 1408.8 \pm 1.8 \,
{\rm
	MeV},
\nonumber \\
m^{\rm exp.}[\eta (1475)] &=& 1476 \pm 4\, {\rm
	MeV},\nonumber \\
m^{\rm exp.}[\eta (1760)] &=& 1751 \pm 15 \,
{\rm
	MeV},
\nonumber \\
m^{\rm exp.}[\eta (2225)] &=& 2221^{+13}_{-10} \,
{\rm 	MeV}.
\label{E_eta_exp}
\end{eqnarray}
We will refer to these $\eta$ states by $\eta_i$ with $i=1\cdots 7$. There are 35 scenarios for selecting three of the $\eta$ states from  this list.  We will see in next section that inclusion of  $\eta_1=\eta(547)$ and $\eta_7=\eta(2225)$ is necessary in order for the three pseudoscalar singlets predicted in our model to approach three of the above known states (i.e. in any acceptable  scenario,  the lightest and the heaviest predicted pseudoscalar singlets will have to be identified with $\eta_1$ and  $\eta_7$, respectively).

\begin{figure}[!htb]
  \centering
   	\includegraphics[width=0.4\textwidth]{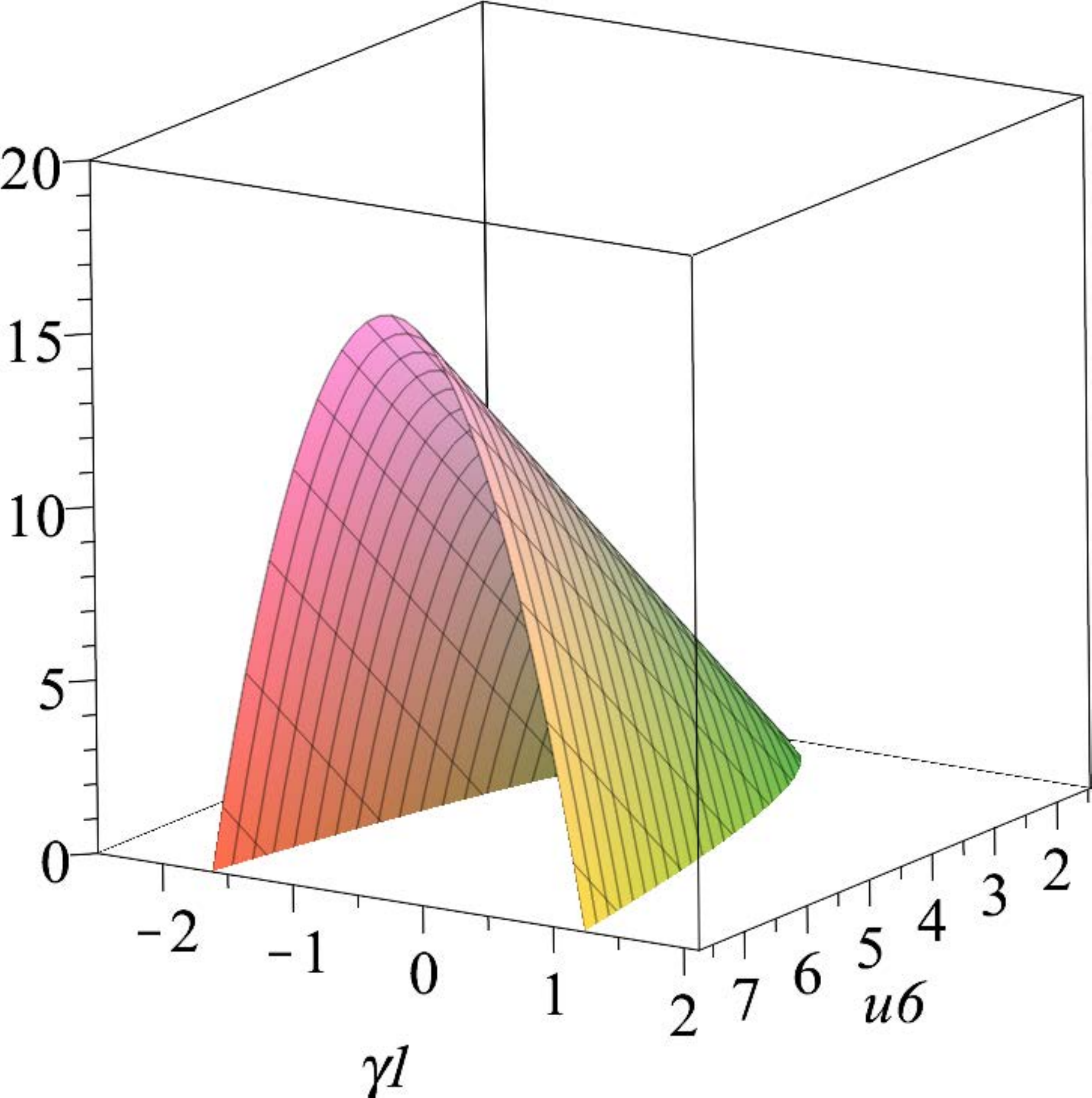} \hskip 2cm
   	\includegraphics[width=0.4\textwidth]{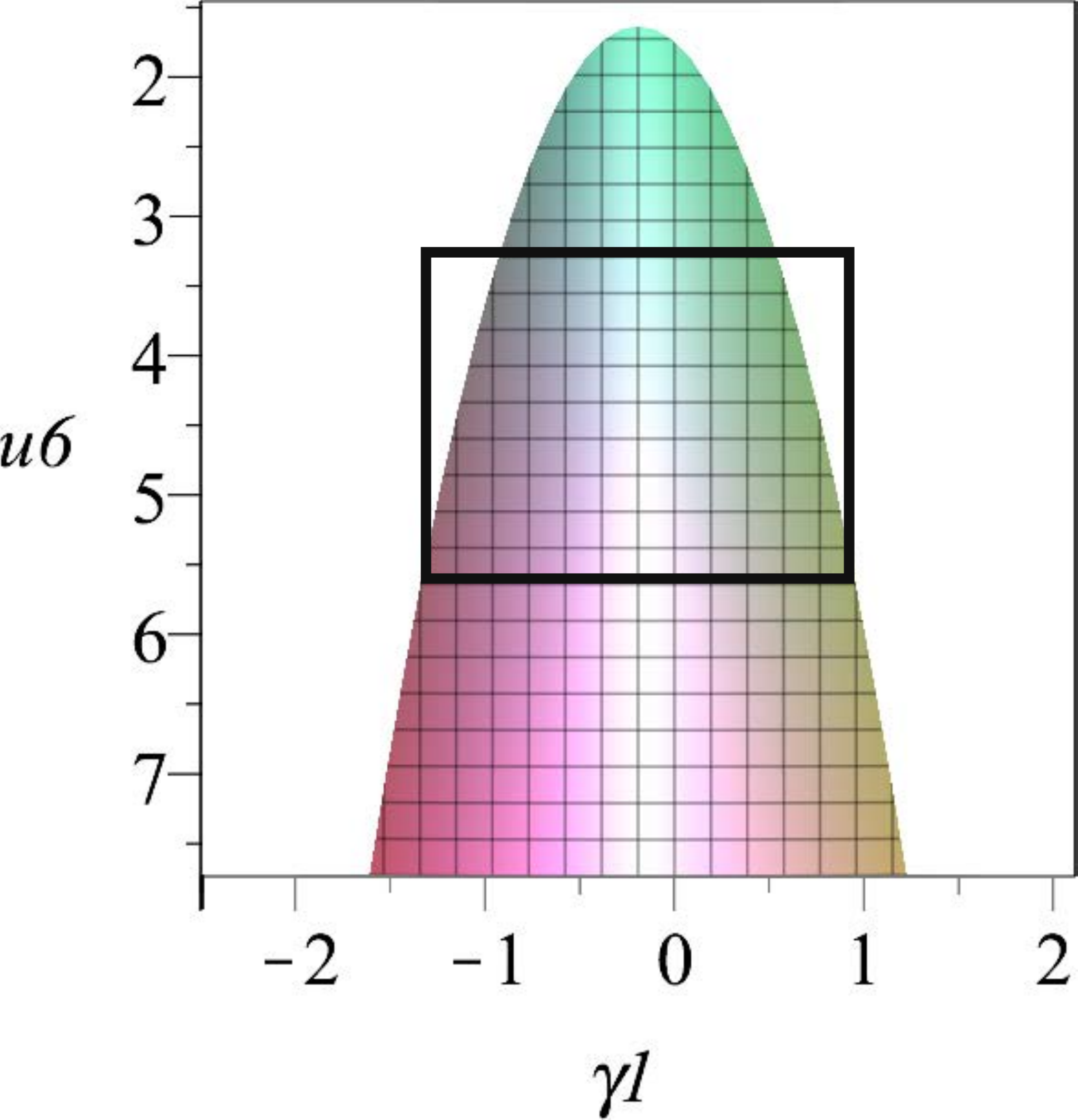}

  \caption{Determinant of $N_0^2$ versus $u_6$ and $\gamma_1$ (left).   The projection of the surface (when the determinant is positive)  onto the flat (zero-determinant) plane is also shown (right). }
  \label{F_det_N0}
\end{figure}

For a choice of target experimental states $\eta_i$, $\eta_j$ and $\eta_k$ from the list (\ref{E_eta_exp}) with $i<j<k$,
the trace and determinant of $N_0^2$ are forced (by numerical search of the model parameters) to approach
\begin{eqnarray}
{\rm Tr} \left( N_0^2\right)	&=&
m^2_{0^-} +  {m'}^2_{0^-} + {m''}^2_{0^-} \rightarrow m^2_{\eta_i} + m^2_{\eta_j} + m^2_{\eta_k}
\label{E_Tr_N0_initial}\\
\det \left(N_0^2\right) &=&  m^2_{0^-} {m'}^2_{0^-}{m''}^2_{0^-} \rightarrow m^2_{\eta_i} m^2_{\eta_j} m^2_{\eta_k},
\label{E_det_N0}
\end{eqnarray}
We consider
\begin{equation}
0 \le {\rm Tr} \left(N_0^2\right) \le 12
\label{E_Tr_N0}
\end{equation}
which can accommodate any of the 35 possible scenarios for targeting three of the experimental $\eta$ states. Since this trace  linearly depends on $u_6$, this imposes an initial limit on $u_6$.
Figure \ref{F_det_N0}\, (left) shows determinant of $N_0^2$ as a function of $u_6$ and $\gamma_1$ and its projection onto $u_6 \gamma_1$ plane is also shown (right).  The two horizontal sides of the rectangular box show the limits on $u_6$ obtained from imposing  condition (\ref{E_Tr_N0}) and the two vertical sides of the same rectangle show the preliminary limits on $\gamma_1$ obtained from intersection of the surface with flat plane on which the determinant vanishes.   This process carves out a  broad region for selecting the initial values of $u_6$ and $\gamma_1$ that can cover all 35 target scenarios in our numerical simulation.    However, not every value of $u_6$ and $\gamma_1$ in this rectangle is necessarily acceptable and additional filters need to be imposed which result in a tighter set of values for these two parameters:
\begin{equation}
	S_\mathrm{III} = \left\{
	u_6, \gamma_1 \,\Bigg| \, m^2_{0^-} > 0, {m'}^2_{0^-}> 0, {m''}^2_{0^-}> 0,
	{\rm Tr}\left(N_0^2\right)\le 12 \,\,\mathrm{GeV}^2 \right\}.
	\label{S_III}
\end{equation}
Note that the singlet pseudoscalar masses also depend on other parameters
which are selected from sets $S_\mathrm{I}$ and $S_\mathrm{II}$.

Having generated our initial parameter sets $S_\mathrm{I}$, $S_\mathrm{II}$ and
$S_\mathrm{III}$ we can now proceed to further refine them.   We investigate
which of the 35 scenarios for identifying our model predictions for the three pseudoscalar singlets with three of the $\eta$ states is favored.  In order to do so we use function $\chi$ (first introduced in \cite{Jora5}),  which measures the agreement of our model prediction for masses $m_{0^-}$, ${m'}_{0^-}$ and   ${m''}_{0^-}$ with the central values of the experimental masses, i.e.
\begin{equation}
\chi_{ijk} =
{
{\left| {m}_{0^-} - {\widehat m}^{\rm exp}_{\eta_i} \right|}
\over
{{\widehat m}^{\rm exp}_{\eta_i} }
}
+
{
{\left| {m'}_{0^-} - {\widehat m}^{\rm exp}_{\eta_j} \right|}
\over
{{\widehat m}^{\rm exp}_{\eta_j} }
}
+
{
{\left| {m''}_{0^-} - {\widehat m}^{\rm exp}_{\eta_k} \right|}
\over
{{\widehat m}^{\rm exp}_{\eta_k} }
}
\end{equation}
where $i,j,k = 1 \cdots 7$ with $i<j<k$ and hatted parameters refer to the central values of the experimental masses, i.e. $m^{\rm exp}_{\eta_i} = {\widehat m}^{\rm exp}_{\eta_i} \pm \delta m_{\eta_i}^{\rm exp}$.    We measure the goodness of each $\chi_{ijk}$ by comparing it with the overall percentage of experimental  uncertainty that we define by
\begin{equation}
\chi^{\rm exp}_{ijk} =
{
{\delta m_{\eta_i}^{\rm exp}}
\over
{{\widehat m}^{\rm exp}_{\eta_i} }
}
+
{
{\delta m_{\eta_j}^{\rm exp}}
\over
{{\widehat m}^{\rm exp}_{\eta_j} }
}
+
{
{\delta m_{\eta_k}^{\rm exp}}
\over
{{\widehat m}^{\rm exp}_{\eta_k} }
}.
\label{E_chi_exp_ijk}
\end{equation}
Clearly, a scenario $ijk$ with $\chi_{ijk} \le \chi^{\rm exp}_{ijk}$ is well within the overall experimental uncertainty.

\section{Results}

\begin{figure}[!htb]
	\centering
	\includegraphics[width=7in]{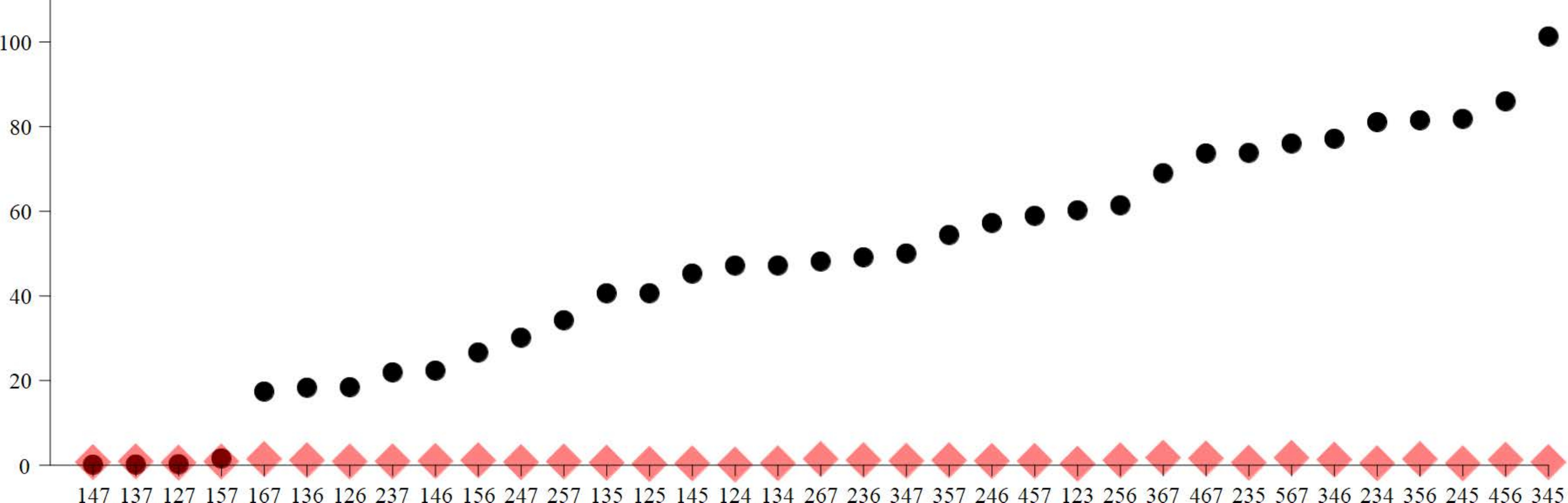} \\ [15pt]
	\includegraphics[width=6in]{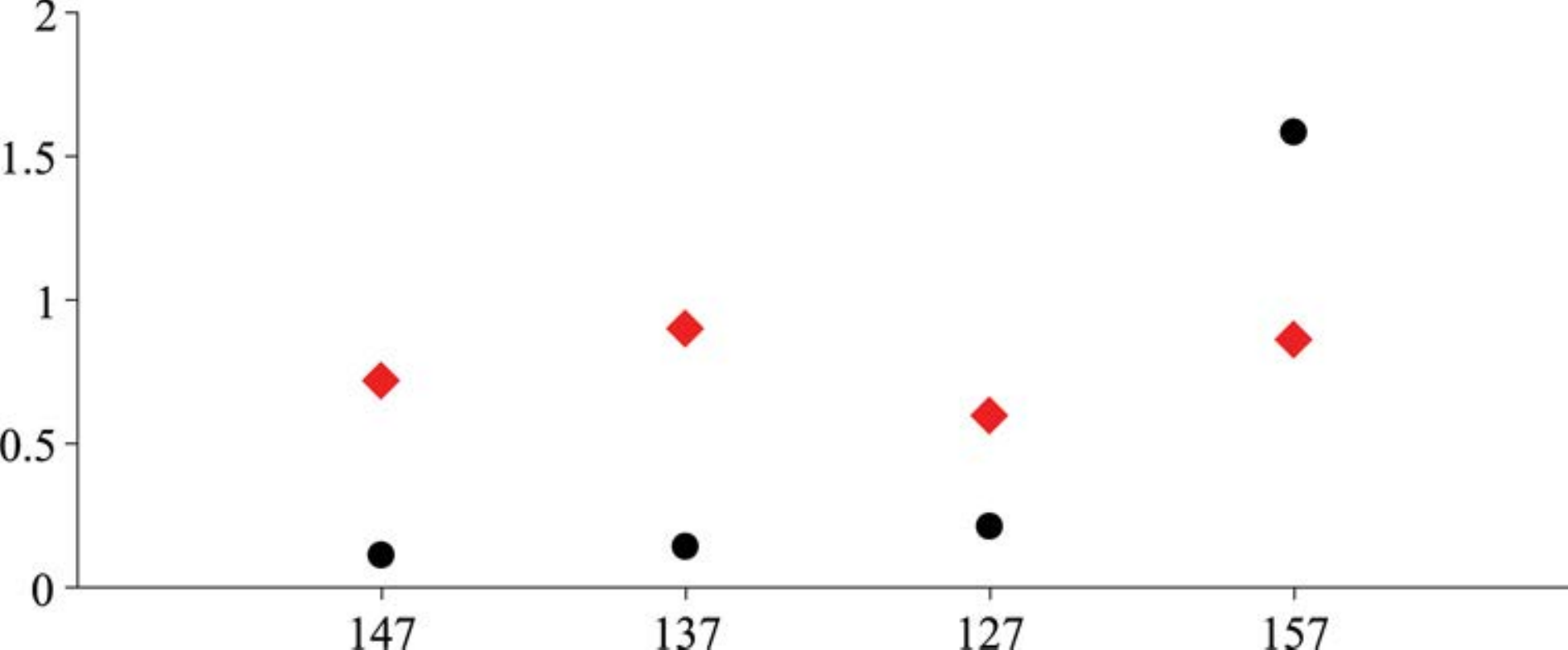}
	
	\caption{The lowest $\chi_{ijk}$ (circles) and the corresponding $\chi^{\rm exp}_{ijk}$ (diamonds) computed from (\ref{E_chi_exp_ijk}) using PDG data \cite{PDG} for all permutations $ijk$. The lower graphs zooms in on permutations that are in agreement with experiment.}
	\label{F_chi_global}
\end{figure}

Our global numerical analysis of all 35 permutations is shown in Fig. \ref{F_chi_global}, with circles representing the lowest $\chi_{ijk}$ for a given permutation $ijk$ and compared with the corresponding $\chi^{\rm exp}_{ijk}$ (diamonds).   The condition $\chi_{ijk} \le \chi^{\rm exp}_{ijk}$ is only met for scenarios 127, 137 and 147 whereas the predictions of the model for other scenarios are clearly far from agreement with experiment (permutation 157 is not too far from experimental range, nevertheless,  is formally outside this range).    We will see later that, based on further analysis of the predictions,  permutation 127 is not favored even though its chi is within the experimental range. For more details,  we have given our simulation data for scenarios 127, 137 and 147 in Appendix A.

The model parameters for these scenarios are given in Table \ref{T_parameters}.  In each case, there is a large subset of $S_\mathrm{I}$, $S_\mathrm{II}$ and $S_\mathrm{III}$ for which  $\chi$ is less than or equal to $\chi^{\rm exp}$ (for permutation 127, the set is considerably smaller than those  for scenarios 137 and 147, therefore, statistically,  the model does not favor this scenario, in addition to more reasons against this scenario that will be discussed below).     For each scenario,   the first column in Table \ref{T_parameters},  corresponds to the lowest $\chi$ whereas the second column represents the averages and standard deviations.   Parameters have different degrees of distribution around the averages, mostly have small ranges of  variation (below 10\%) with the exception of $u_4$,  and to a lesser extent  $\gamma_1$,  which are widely spread.  

\begin{table}[htbp]
	\begin{center}

\caption{Values of model parameters for three scenarios 127 (first three columns), 137 (middle three columns) and 147 (last three columns). For each scenario, the first column corresponds to the values of the  model parameters when $\chi$ is minimum (given in many digits to allow sufficient accuracy for reproduction of the results given in this work), the second column gives the averages and standard deviations  of parameters over all values with $\chi < \chi^{\rm exp}$ and the last column gives the percentages of deviations around the averages.}
\begin{tabular}{ M{40pt} *{3}{|| M{57pt} r @{}M{14pt} @{}l  M{18pt}} N}

\hline
\hline
	
& \multicolumn{5}{c||}{127} & \multicolumn{5}{c||}{137} & \multicolumn{5}{c}{147} &\\ [4pt]

Param &
$\chi_{min}$ & \multicolumn{3}{c}{ave $\pm$ $\sigma$} & \% Dev &
$\chi_{min}$ & \multicolumn{3}{c}{ave $\pm$ $\sigma$} & \% Dev &
$\chi_{min}$ & \multicolumn{3}{c}{ave $\pm$ $\sigma$} & \% Dev &\\
	
\hline
\hline
	
$\chi\%$ & 
 0.21 &  0.463 & $\pm$ &  0.105 & 23 &  
 0.14 &  0.564 & $\pm$ &  0.156 & 28 & 
 0.11 &  0.539 & $\pm$ &  0.125 & 23
&\\ [6pt]
	
$\alpha \times 10^2$ &  
5.904291049 &  5.826 & $\pm$ &  0.080 & 1 & 
5.751680243 &  5.813 & $\pm$ &  0.040 &  0.6 &  
5.661070389 &  5.724 & $\pm$ &  0.038 &  0.6
&\\ [6pt]

 $\beta \times 10^2$ & 
 2.836042326 &  2.987 & $\pm$ &  0.154 & 5 &  
 3.134026160 &  3.016 & $\pm$ &  0.076 & 3 &  
 3.294879326 &  3.182 & $\pm$ &  0.069 & 2
&\\ [6pt]

$u_1 \times 10^{-1}$ & 
1.421012321 &  1.486 & $\pm$ &  0.051 & 3 & 
1.531317349 &  1.497 & $\pm$ &  0.024 & 2 &  
1.604664163 &  1.548 & $\pm$ & 0.030 & 2
&\\ [6pt]
	
$u_2 \times 10^{-2}$ & 
$-$6.164917406 & $-$6.712 & $\pm$ &  0.438 & 7 & 
$-$7.146949710 & $-$6.805 & $\pm$ &  0.223 & 3 & 
$-$7.718529859 & $-$7.233 & $\pm$ &  0.278 & 4
&\\ [6pt]
	
$u_3$ &
1.052435387 &  1.097 & $\pm$ &  0.051 & 5 &  
1.050658660 &  1.122 & $\pm$ &  0.048 & 4 &  
1.089311916 &  1.125 & $\pm$ &  0.025 & 2
&\\ [6pt]
	
$u_4 \times 10^{1}$ & 
1.323861648 & $-$ 0.718 & $\pm$ &  2.304 & 321 & 
$-$6.281688145 & $-$ 1.967 & $\pm$ &  4.559 & 232 &  
1.221714358 & $-$4.452 & $\pm$ &  4.858 & 109
&\\ [6pt]
	
$u_6$ & 
2.306185521 &  2.225& $\pm$ &  0.103 & 5 &  
2.921389969 &  2.779 & $\pm$ &  0.103 & 4 & 
3.010232246 &  3.043 & $\pm$ &  0.064 & 2
&\\ [6pt]
	
$\lambda_2 \times 10^{2}$ &  
$-$3.944010641 & $-$3.940 & $\pm$ &  0.011 & 0.2 & 
$-$3.896911757 & $-$3.934 & $\pm$ &  0.033 &  0.8 & 
$-$3.972089358 & $-$3.926 & $\pm$ &  0.037 &  0.9
&\\ [6pt] 
	
$\lambda_3 \times 10^{3}$ & 
3.604604059 &  3.546 & $\pm$ &  0.174 & 5 & 
2.898120799 &  3.468 & $\pm$ & 0.503 & 15 &  
4.025784814 &  3.348 & $\pm$ &  0.565 & 17
&\\ [6pt] 
	
$\gamma_1 \times 10^{1}$ & 
$-$1.99609884 & $-$3.024 & $\pm$ &  1.136 & 38 & 
$-$0.01232177 &  0.098 & $\pm$ &  0.573 & 584 &  
2.070742550 &  1.402 & $\pm$ &  0.540 & 39
&\\ [6pt]
	
$h_0 \times 10^{1}$ & 
 8.013683748 &  8.048 & $\pm$ &  0.031 &  0.3 &  
 8.083227014 &  8.055 & $\pm$ &  0.026 &  0.3 &  
 8.069665569 &  8.042 & $\pm$ &  0.028 &  0.3
&\\ [6pt]
	
$f_{8_-}' \times 10^{4}$ & 
$-$7.180174999 & $-$7.100 & $\pm$ &  0.386 & 5 & 
$-$7.604198587 & $-$6.981 & $\pm$ &  0.397 & 6 & 
$-$7.609795602 & $-$7.248 & $\pm$ &  0.236 & 3
&\\ [6pt]
	
$m_{8_-}'$ & 
1.291508983 & 1.342 & $\pm$ &  0.038 & 3 &  
1.336510544 &  1.361 & $\pm$ &  0.024& 2 &  
1.380486452 &  1.382 & $\pm$ &  0.009 &  0.6
&\\ [6pt]

\hline
\hline
	
\end{tabular}
\label{T_parameters}
\end{center}

\end{table}

\begin{figure}[!htb]
	\centering
	
	\includegraphics[width=2.1in]{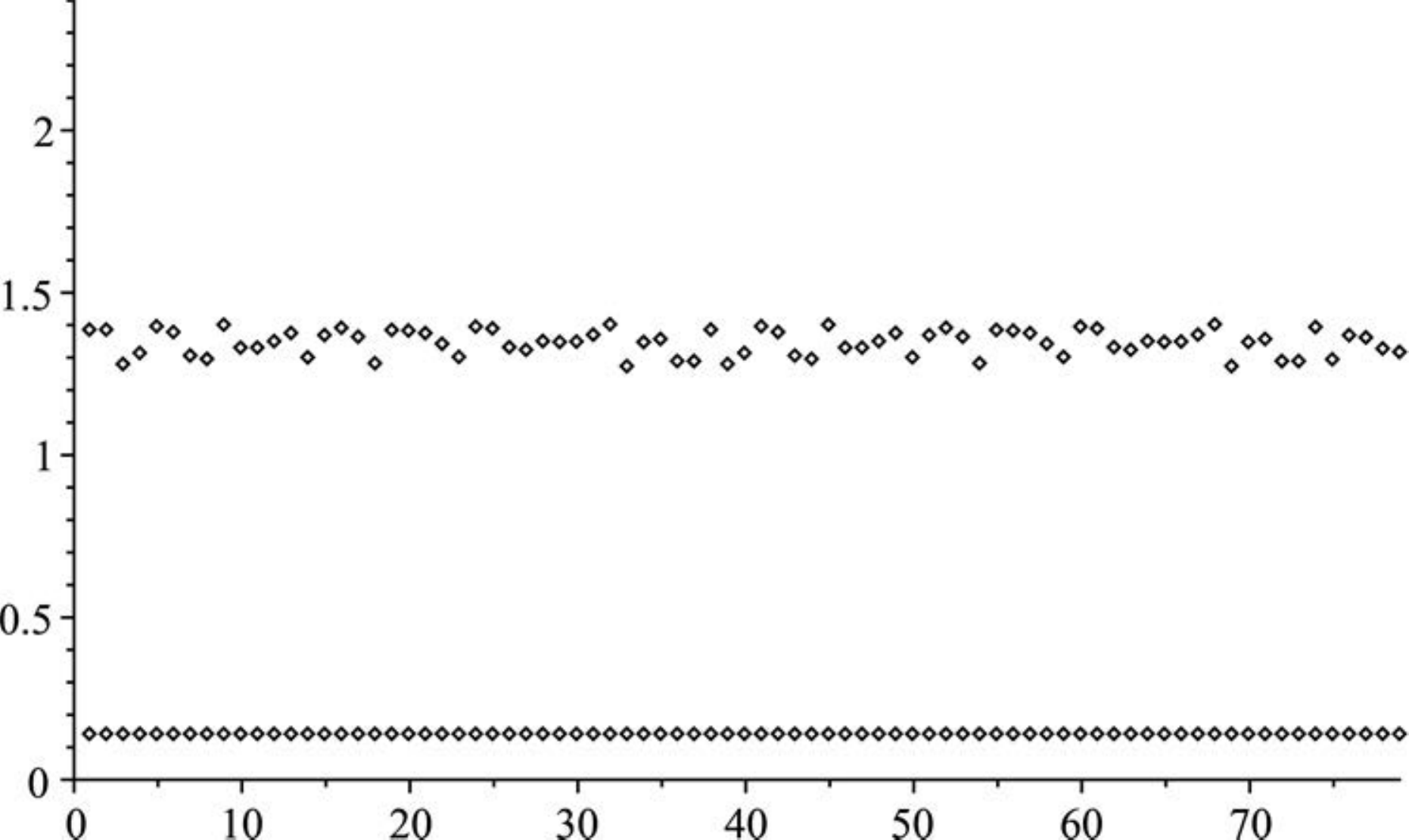}
	\includegraphics[width=2.1in]{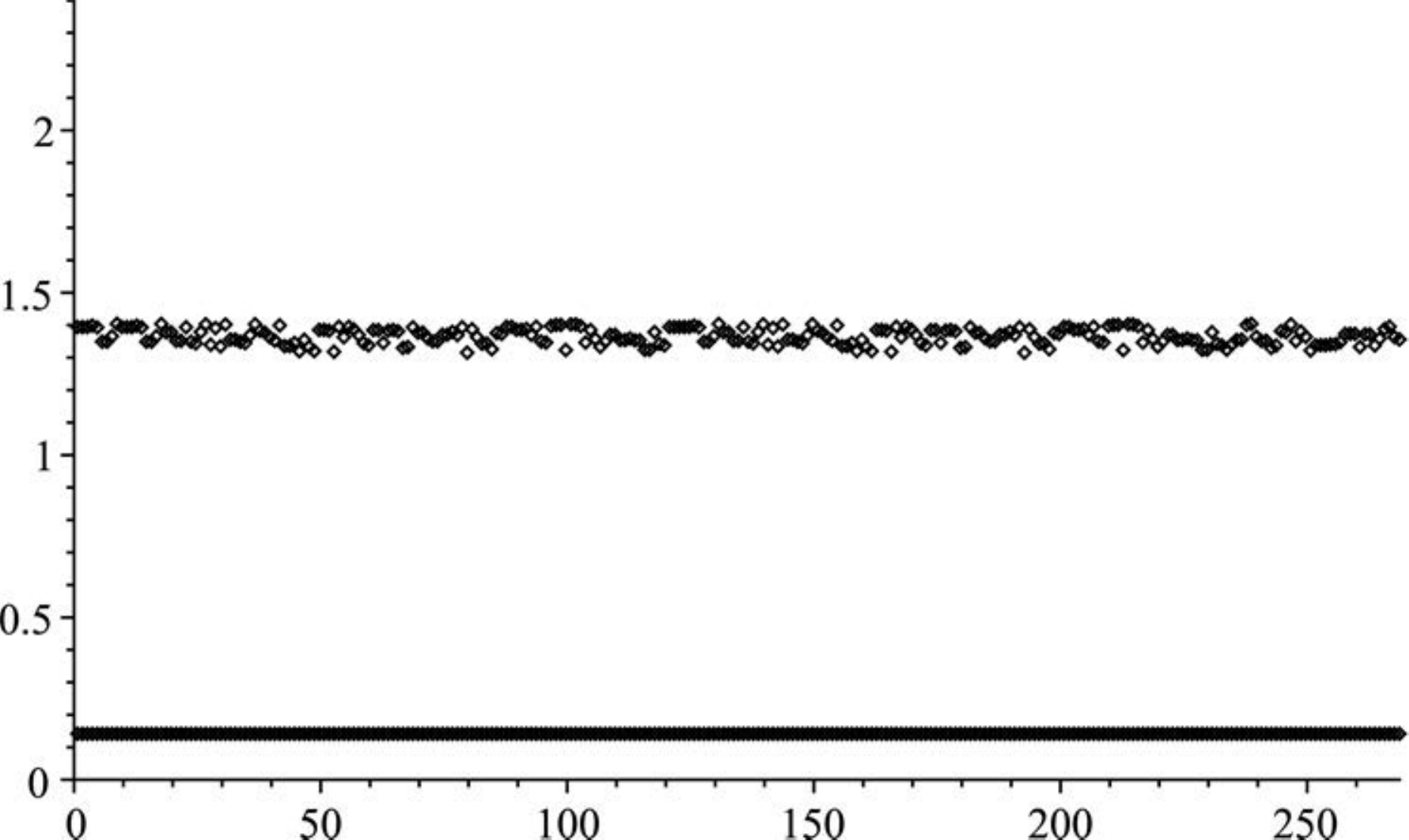}
	\includegraphics[width=2.1in]{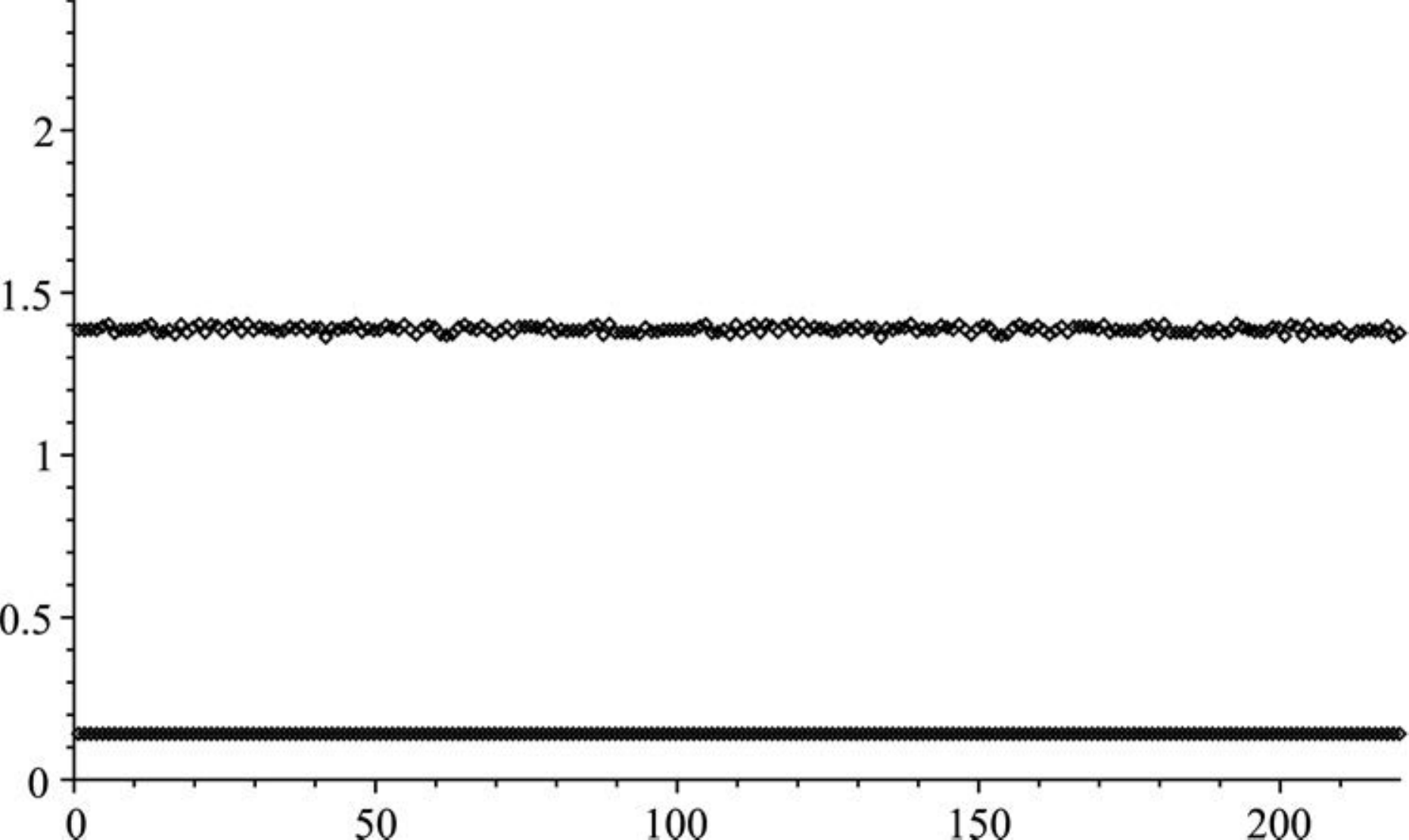}
	
	\includegraphics[width=2.1in]{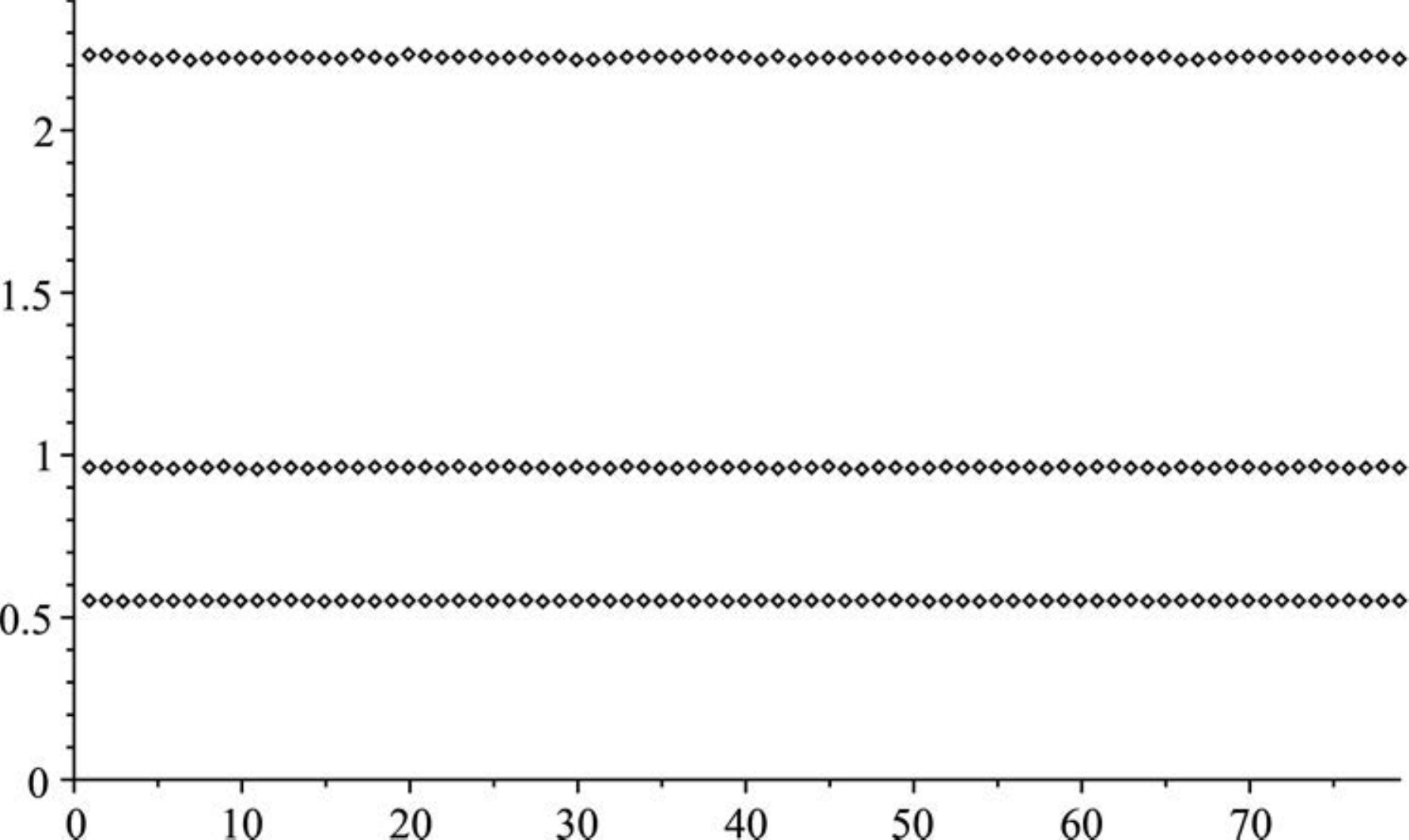}
	\includegraphics[width=2.1in]{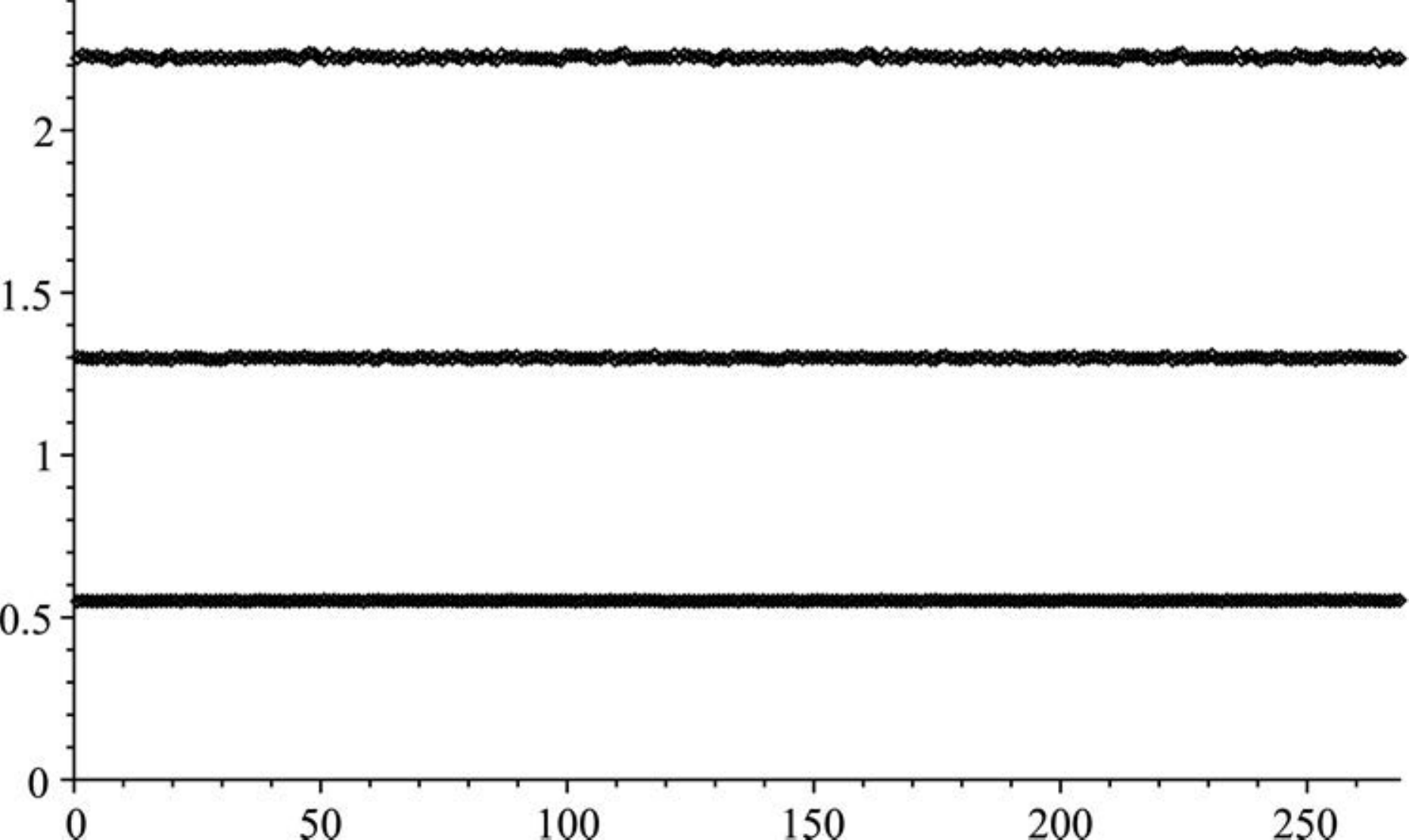}
	\includegraphics[width=2.1in]{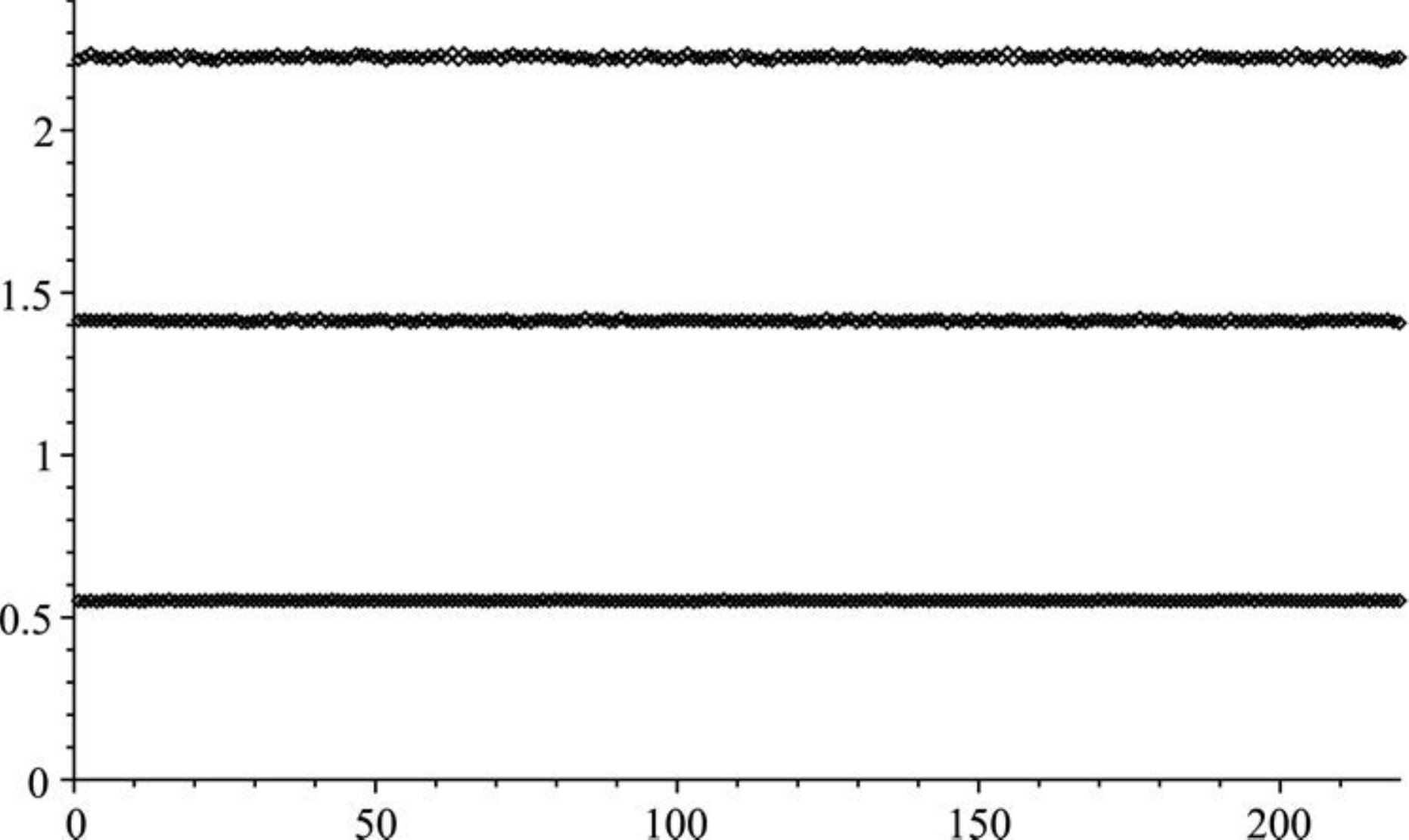}
	
	\includegraphics[width=2.1in]{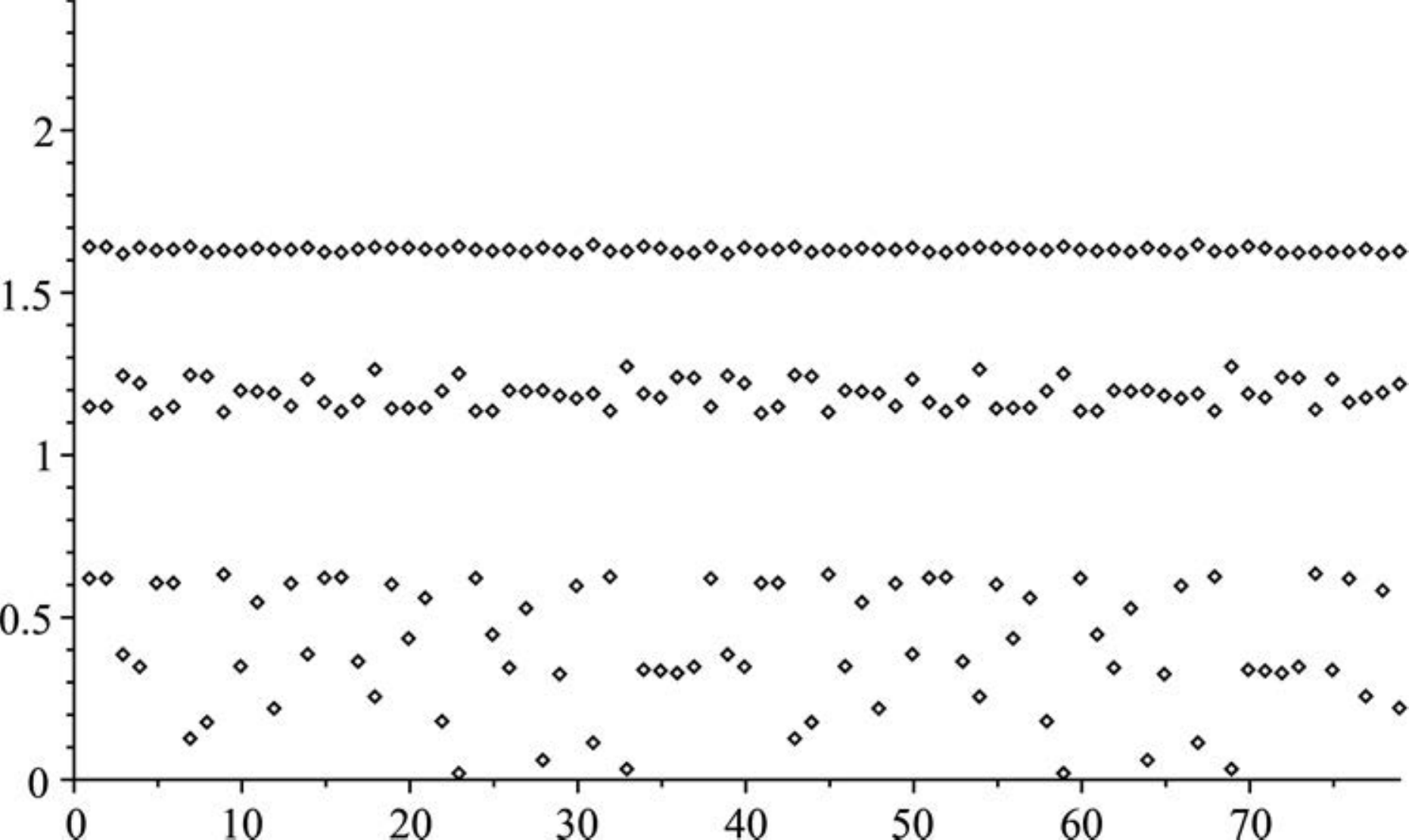}
	\includegraphics[width=2.1in]{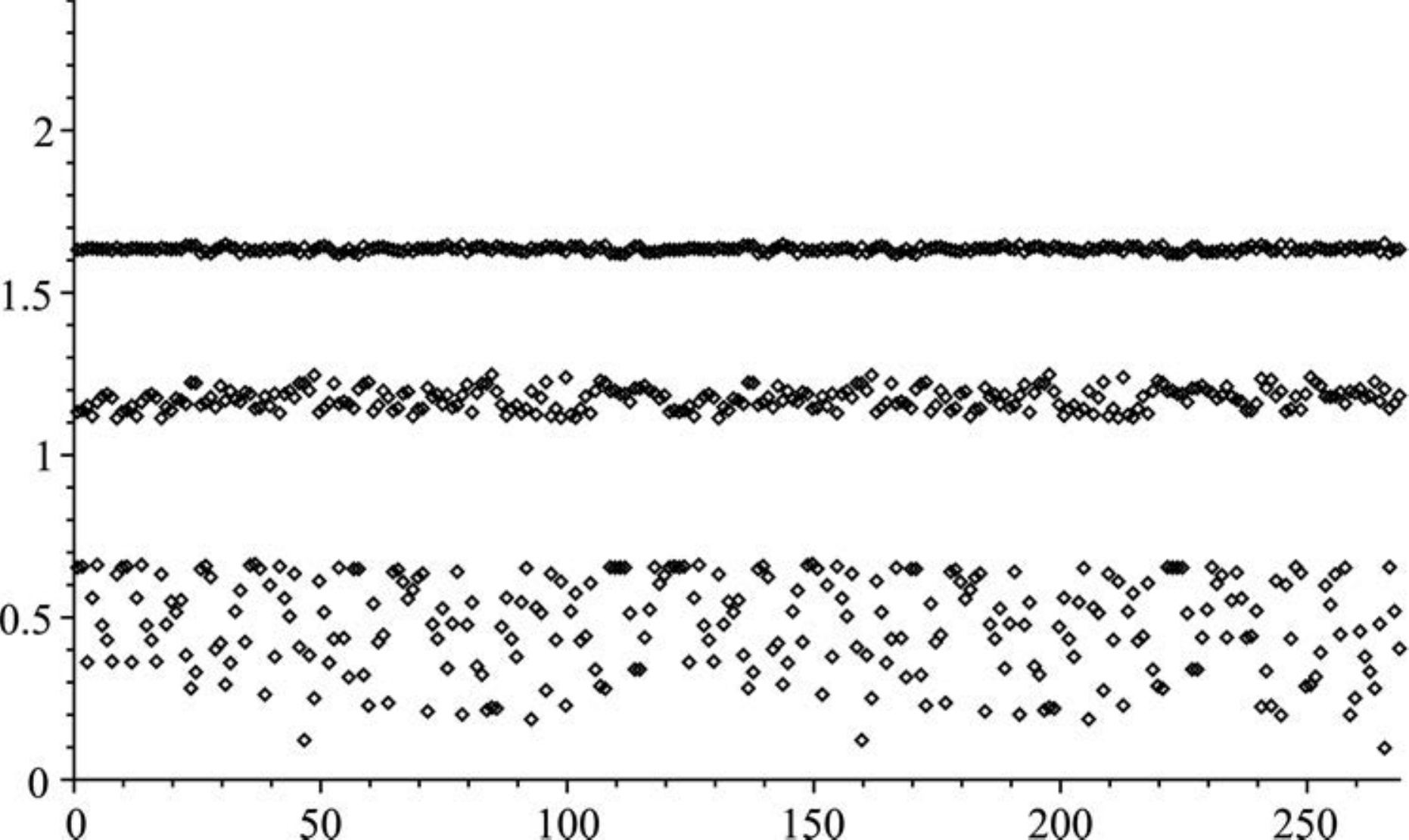}
	\includegraphics[width=2.1in]{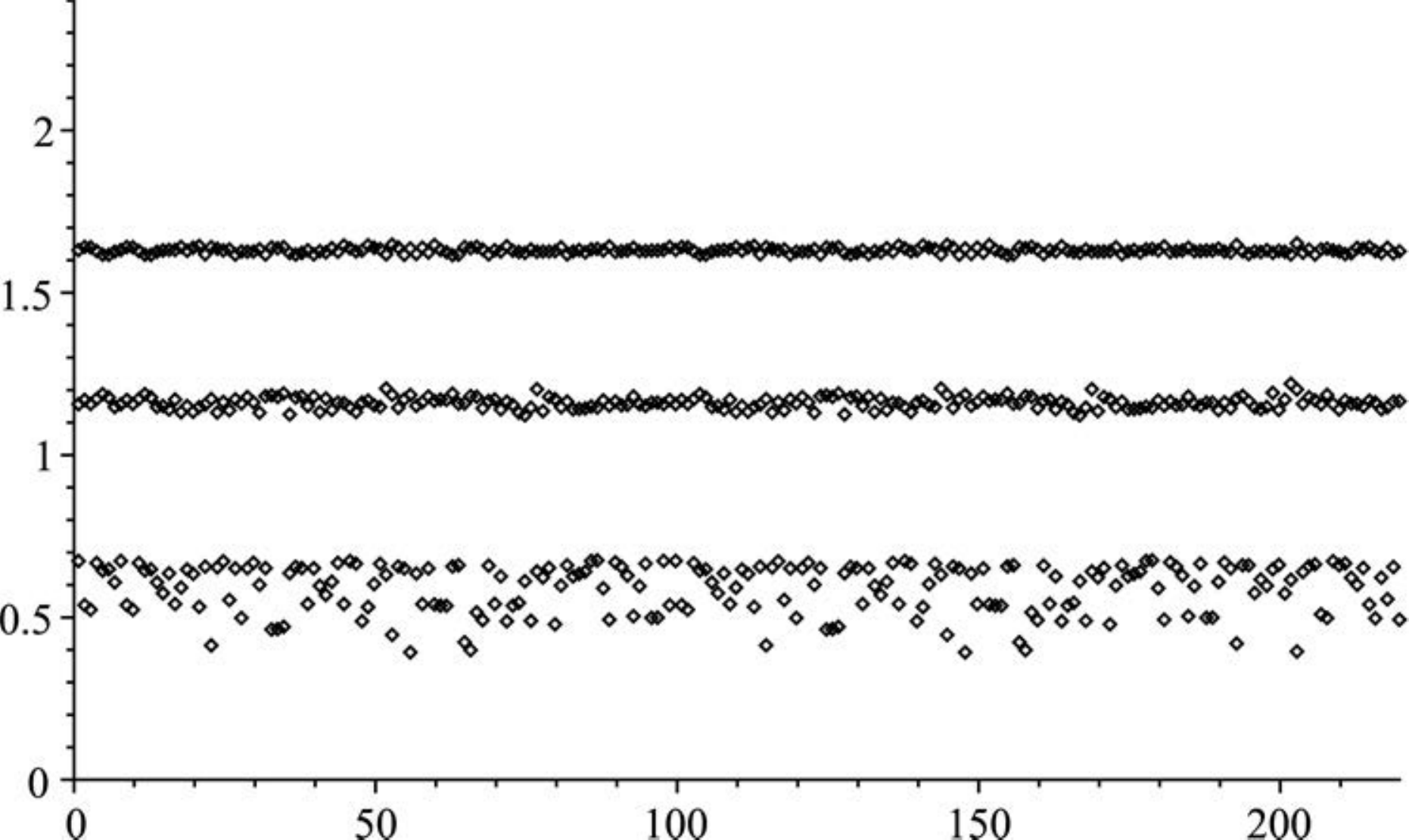}

	\caption{Simulation data for masses (in GeV) of pseudoscalar octets (first row), pseudoscalar singlets (middle row) and scalar singlets (last row)   versus  solution number,  with $\chi< \chi_{\rm exp}$. }
	\label{F_N8N0Y0_masses}
\end{figure}

For the same subsets  $S_\mathrm{I}$, $S_\mathrm{II}$ and $S_\mathrm{III}$ for which  $\chi \le \chi^{\rm exp}$, the computed masses are plotted in Fig. \ref{F_N8N0Y0_masses} showing variations around  their averages that range from 0.001 to 0.191 GeV.   These variations are largest for the lightest scalar singlet around 0.50 GeV which is a clear characteristic of sigma meson [or $f_0(500)$ as listed in PDG \cite{PDG}].    The numerical values of these masses (together with their averages and standard deviations are summarized in Table \ref{T_masses}.    As mentioned before,  for the heavier pseudoscalar octet mass ($m'_{8^-}$) we input the experimental mass of   $\pi(1300)$ which has a large uncertainty in the range of 1.2-1.4 GeV.   In our Monte Carlo simulation,  we have examined this entire experimental range and,  as Fig.  \ref{F_N8N0Y0_masses} (first row) shows,  
 our simulations for all three scenarios favor the upper end of this experimental mass range (i.e. above 1.3 GeV). 
In the three scenarios 127, 137 and 147, the lightest and the heaviest pseudoscalar singlets (Fig.  \ref{F_N8N0Y0_masses},  middle  row) are identified with    $\eta(547)$ and $\eta(2225)$. However, the middle  pseudoscalar singlet differs in these scenarios and may respectively be identified with $\eta'(958)$,
$\eta(1295)$ and $\eta(1405)$.    Therefore, while the model is sensitive to selecting the lightest and heaviest
pseudoscalar singlets, it is not very sensitive in identifying  the middle pseudoscalar singlet.  In each scenario, the  three  scalar singlets are consistent with a very light state, followed by a state with a mass around 1.2 GeV and the heaviest scalar singlet with a mass of about 1.6 GeV, which may be identified with physical states $f_0(500)$, $f_0(1370)$ and $f_0(1500)$ [or $f_0(1710)$], respectively.   Considering the predictions for substructures of these states helps unraveling the proximity of these states to the actual experimental candidates.

In addition to the physical masses, we can also extract the bare (unmixed) masses of distinct components.   These are given in Table \ref{T_bare_masses}.   For both octet and singlet pseudoscalars,  the pure quark-antiquark bare masses are lighter than the pure four-quark bare masses.   Contrarily,  this is reversed for scalar octets and singlets where the pure four-quark masses are in fact lighter than the pure quark-antiquark masses consistent with MIT bag model of Jaffe \cite{jaffe} where the scalars are strongly bound.      Of particular interest are the pure pseudoscalar glueball mass ($m_g$) and pure scalar glueball mass ($m_h$).   For the two favored scenarios 137 and 147, Table \ref{T_bare_masses} gives $m_g\approx 2.0$ GeV and $m_h \approx$ 1.6 GeV.

\begin{table}[htbp]
	\begin{center}

\caption{Predicted masses (in GeV)  for three scenarios 127 (first two columns), 137 (middle two columns) and 147 (last two columns). For each scenario, the first column corresponds to the values of the masses  when $\chi$ is minimum and the second column gives the averages and standard deviations  of masses over subsets of $S_\mathrm{I}$, $S_\mathrm{II}$ and $S_\mathrm{III}$ for which $\chi \le \chi^{\rm exp}$.}
		
\begin{tabular}{ M{40pt} *{3}{|| M{40pt} r @{}M{12pt} @{}l} N }
\hline
\hline

& \multicolumn{4}{c||}{127} & \multicolumn{4}{c||}{137} & \multicolumn{4}{c}{147} &\\ [4pt]

Mass &
$\chi_{min}$ & \multicolumn{3}{c||}{ave $\pm$ $\sigma$} &
$\chi_{min}$ &\multicolumn{3}{c||}{ave $\pm$ $\sigma$} &
$\chi_{min}$ & \multicolumn{3}{c}{ave $\pm$ $\sigma$} &\\
	
\hline
\hline

$\chi\%$ & 
 0.21 &  0.463 & $\pm$ &  0.105 &  
 0.14 &  0.564 & $\pm$ &  0.156 & 
 0.11 &  0.539 & $\pm$ &  0.125 
&\\ [6pt]

$m_{8^-}'$ & 
1.291 &  1.342 & $\pm$ &  0.038 &  
1.336 &  1.361 & $\pm$ &  0.024 &  
1.380 &  1.382 & $\pm$ &  0.009
&\\ [6pt]

$m_{0^-}$ & 
0.547 &  0.546 & $\pm$ &  0.001 &  
0.5470 &  0.546 & $\pm$ &  0.001 &  
0.5470 &  0.547 & $\pm$ &  0.001
&\\ [6pt]

$m_{0^-}'$ &
0.956 &  0.957 & $\pm$ &  0.002 &  
1.295 &  1.294 & $\pm$ &  0.003 &  
1.408 &  1.408 & $\pm$ &  0.003
&\\ [6pt]

$m_{0^-}''$ & 
2.216 &  2.220 & $\pm$ &  0.004 &  
2.219 &  2.220 & $\pm$ &  0.005 &  
2.219 &  2.220 & $\pm$ &  0.005
&\\ [6pt]

$m_{0^+}$ & 
0.172 &  0.399 & $\pm$ &  0.191 &  
0.628 &  0.470 & $\pm$ &  0.147 &  
0.510 &  0.585 & $\pm$ &  0.075
&\\ [6pt]

$m_{0^+}'$ &
1.237 &  1.182 & $\pm$ &  0.041 &  
1.176 &  1.168 & $\pm$ &  0.033 &  
1.174 &  1.156 & $\pm$ &  0.017
&\\ [6pt]

$m_{0^+}''$ & 
1.620 &  1.627 & $\pm$ &  0.006 &  
1.628 &  1.629 & $\pm$ &  0.007 &  
1.636 &  1.625 & $\pm$ &  0.008
&\\ [6pt]
\hline
\hline

\end{tabular}

\label{T_masses}

\end{center}

\end{table}

\begin{table}[htbp]
	\begin{center}

\caption{Predicted bare masses (in GeV) for three scenarios 127, 137, and 147.   For each scenario, the first column corresponds to the values of the masses  when $\chi$ is minimum and the second column gives the averages and standard deviations  of masses over subsets of $S_\mathrm{I}$, $S_\mathrm{II}$ and $S_\mathrm{III}$ for which $\chi \le \chi^{\rm exp}$.}
		
\begin{tabular}{ M{40pt} *{3}{|| M{40pt} r @{}M{12pt} @{}l} N }
\hline
\hline

& \multicolumn{4}{c||}{127} & \multicolumn{4}{c||}{137} & \multicolumn{4}{c}{147} &\\ [4pt]

Bare &
$\chi_{min}$ & \multicolumn{3}{c||}{ave $\pm$ $\sigma$} &
$\chi_{min}$ &\multicolumn{3}{c||}{ave $\pm$ $\sigma$} &
$\chi_{min}$ & \multicolumn{3}{c}{ave $\pm$ $\sigma$} &\\
	
\hline
\hline

$m_{\eta_8}$ &  
0.578 &  0.632 & $\pm$ &  0.238 &  
0.657 &  0.645 & $\pm$ &  0.142 &  
0.711 &  0.689 & $\pm$ &  0.143
&\\ [6pt]

$m_{\eta_8'}$ &  
1.162 &  1.192 & $\pm$ &  0.256 &  
1.171 &  1.207 & $\pm$ &  0.250 &  
1.191 &  1.206 & $\pm$ &  0.171
&\\ [6pt]

$m_{f_8}$ &  
1.334 &  1.308 & $\pm$ &  0.256 & 
1.326 &  1.294 & $\pm$ &  0.250 &  
1.309 &  1.295 & $\pm$ &  0.171
&\\ [6pt]

$m_{f_8'}$ &  
1.162 &  1.192 & $\pm$ &  0.256 &  
1.171 &  1.207 & $\pm$ &  0.250 &  
1.191 &  1.206 & $\pm$ &  0.171
&\\ [6pt]

$m_{\eta_0}$ &  
0.763 &  0.846 & $\pm$ &  0.346 & 
0.933 &  0.873 & $\pm$ &  0.332 &  
0.906 &  0.947 & $\pm$ &  0.333
&\\ [6pt]

$m_{\eta_0'}$ &  
1.607 &  1.595 & $\pm$ &  0.407 & 
1.487 &  1.591 & $\pm$ &  0.446 &  
1.570 &  1.542 & $\pm$ &  0.418
&\\ [6pt]

$m_g$ &  
1.721 &  1.698 & $\pm$ &  0.362 & 
1.953 &  1.899 & $\pm$ &  0.358 &  
1.980 &  1.984 & $\pm$ &  0.275
&\\ [6pt]

$m_{f_0}$ &  
1.238 &  1.181 & $\pm$ &  0.323 &  
1.148 &  1.152 & $\pm$ &  0.342 &  
1.182 &  1.120 & $\pm$ &  0.324
&\\ [6pt]

$m_{f_0'}$ &  
0.345 &  0.546 & $\pm$ &  0.391 &  
0.730 &  0.617 & $\pm$ &  0.421 &  
0.608 &  0.728 & $\pm$ &  0.405
&\\ [6pt]

$m_h$ &  
1.591 &  1.597 & $\pm$ &  0.144 &  
1.605 &  1.598 & $\pm$ &  0.129 &  
1.597 &  1.594 & $\pm$ &  0.132
&\\ [6pt]
\hline
\hline

\end{tabular}

\label{T_bare_masses}

\end{center}

\end{table}

\begin{figure}[!htb]
	\centering
	\includegraphics[width=2.3in]{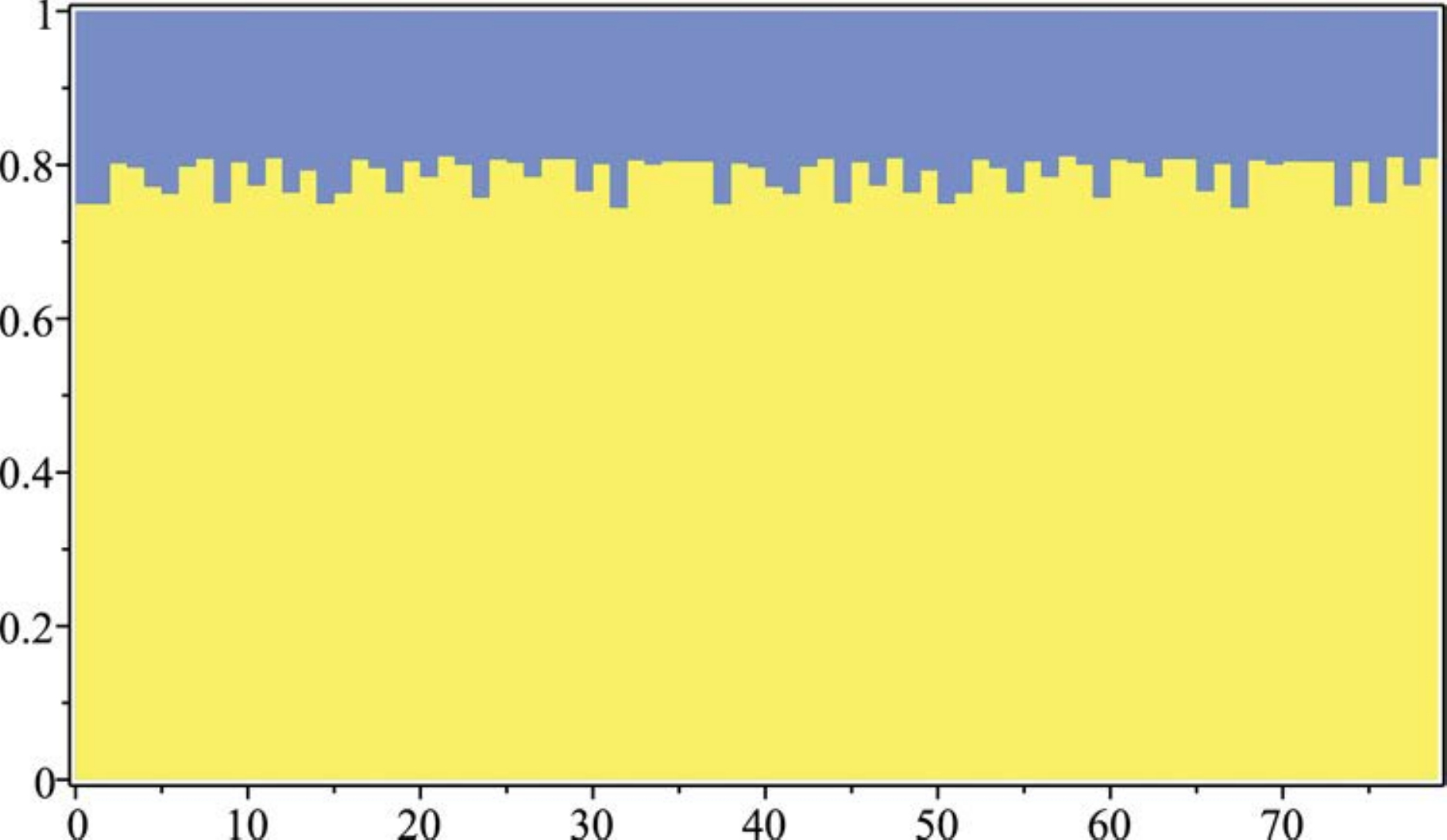}
	\includegraphics[width=2.3in]{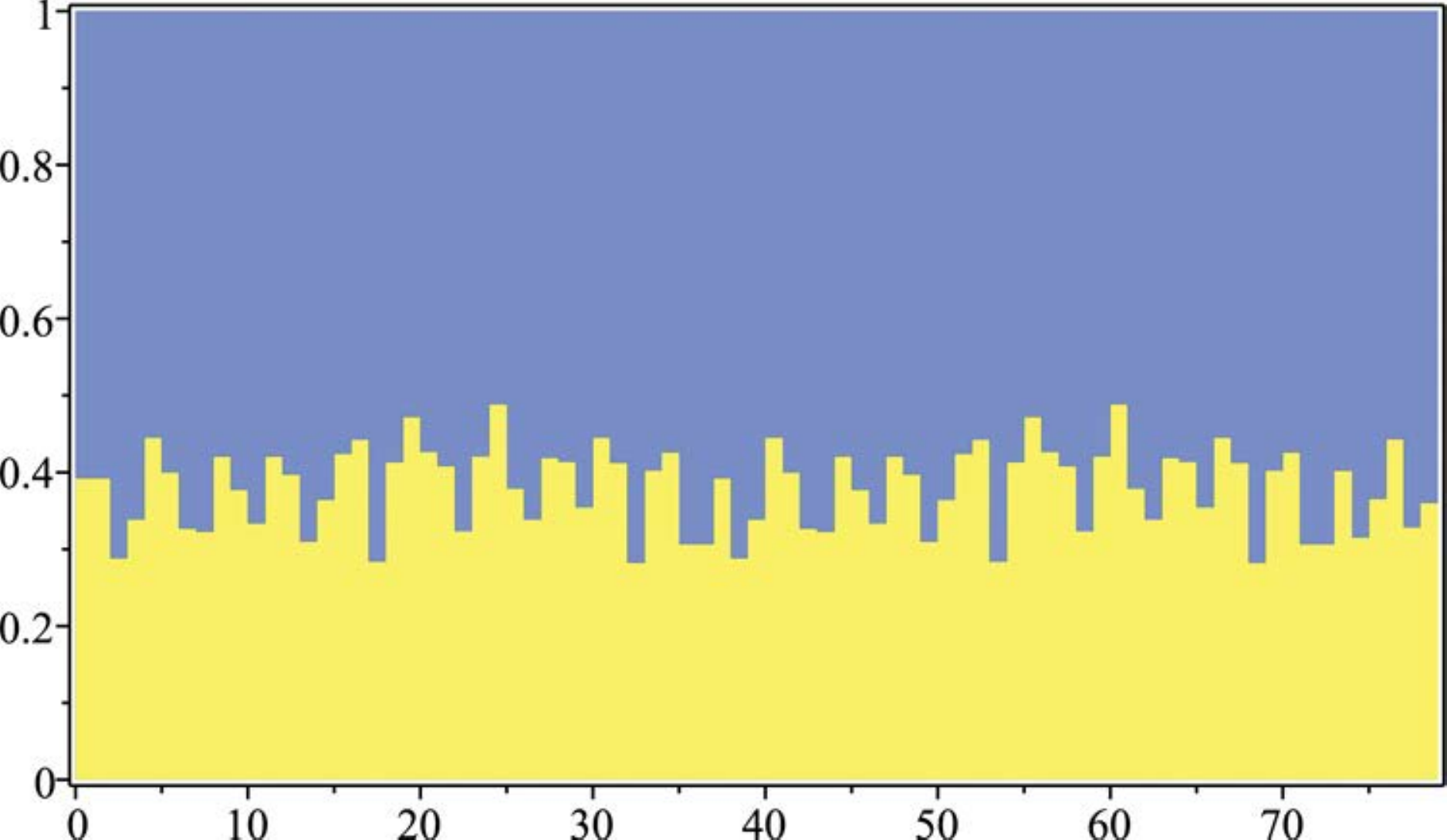}
	
	\includegraphics[width=2.3in]{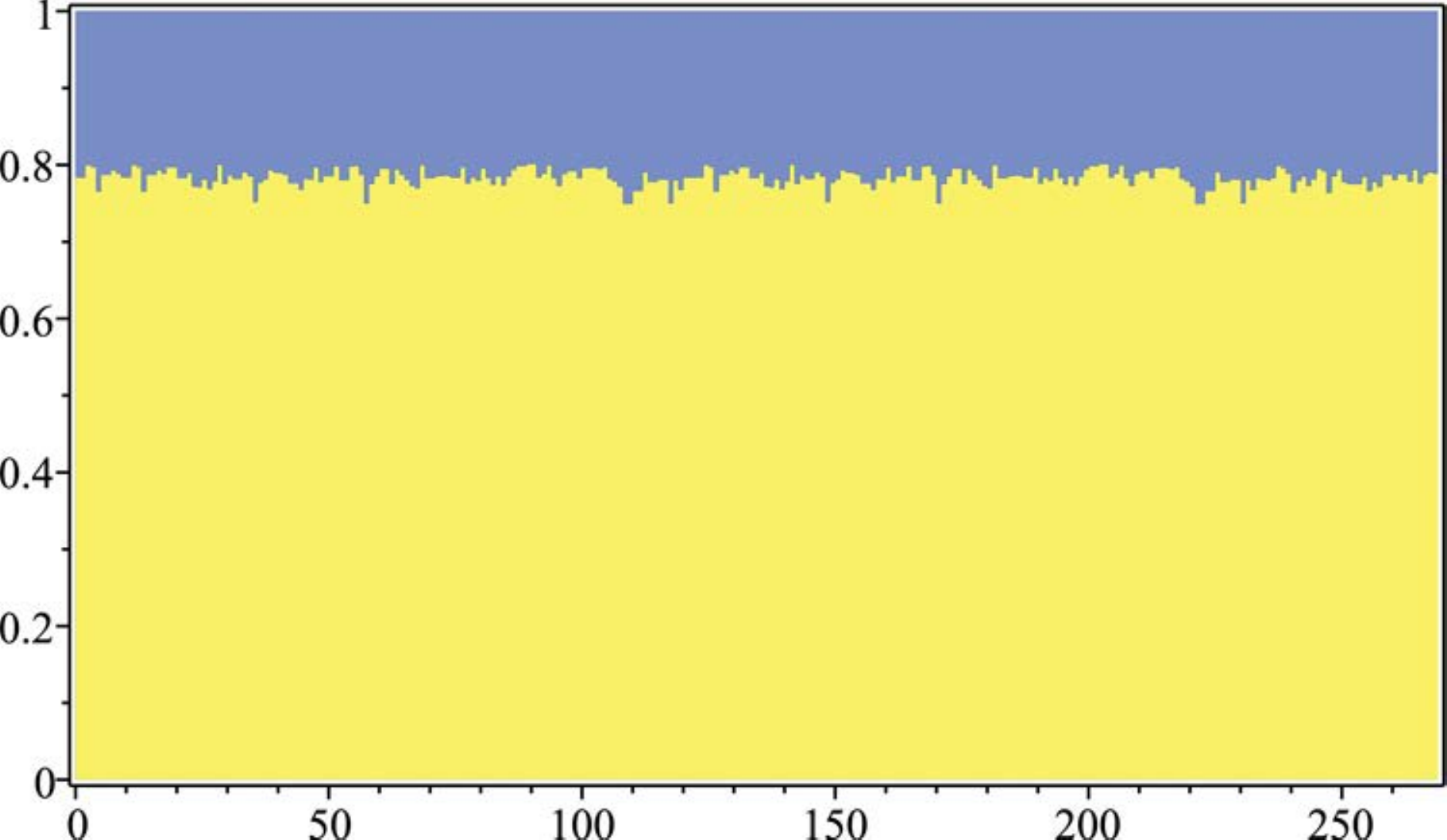}
	\includegraphics[width=2.3in]{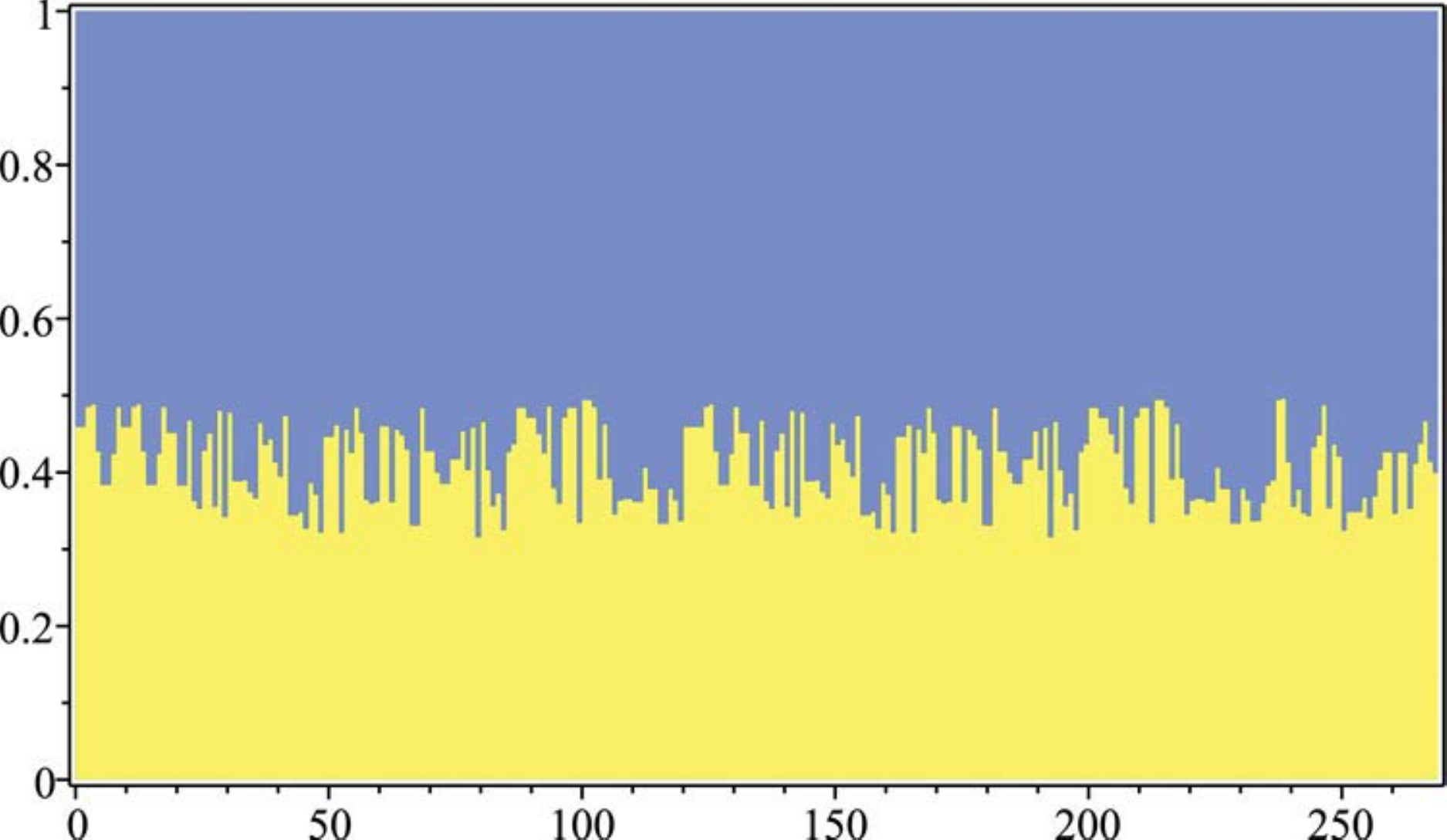}
	
	\includegraphics[width=2.3in]{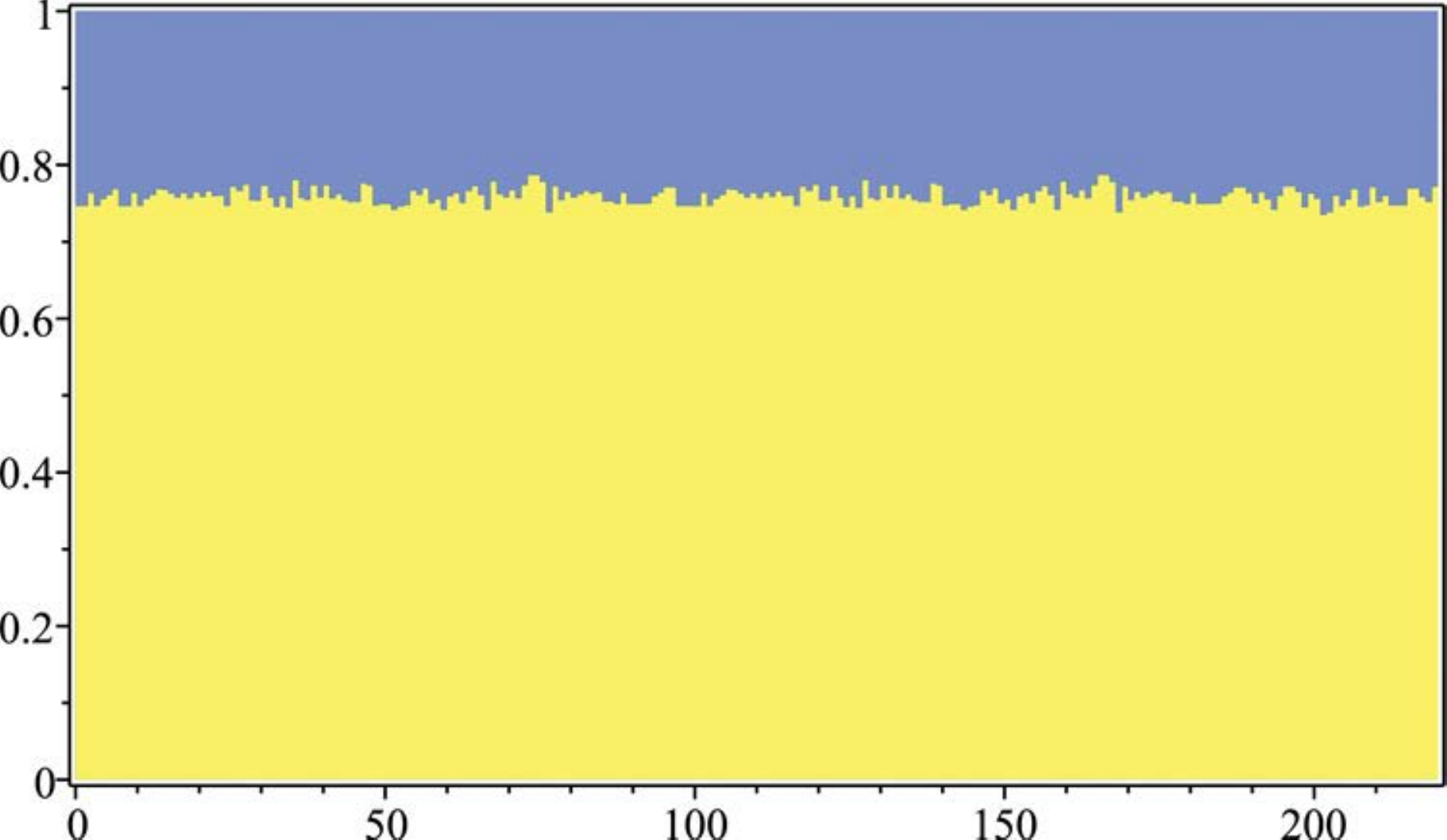}
	\includegraphics[width=2.3in]{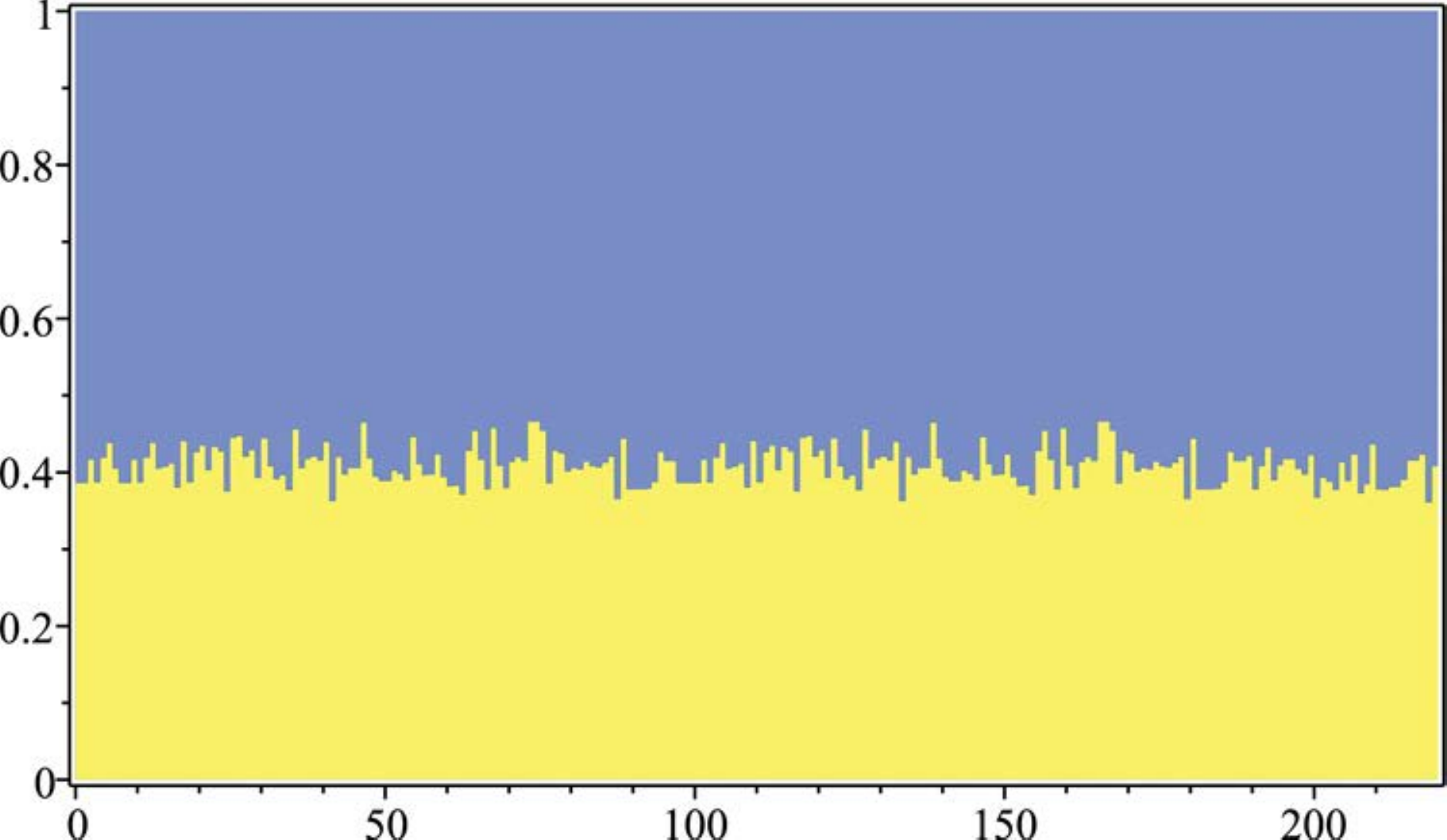}
	
	\caption{Stacked bar diagrams for quark-antiquark and four-quark components of pseudoscalar octet (left column) and scalar octet (right column)
versus simulation number over subsets of $S_\mathrm{I}$, $S_\mathrm{II}$ and $S_\mathrm{III}$ for which $\chi \le \chi^{\rm exp}$.  Quark-antiquark component is plotted in light gray (or yellow in color version) and four-quark component is plotted in  gray (or blue in color version). The three rows from top to bottom respectively correspond to  permutations 127, 137 and 147.}
	\label{F_N8Y8_comps}
\end{figure}

\begin{figure}[!htb]
	\centering
	\includegraphics[width=2.3in]{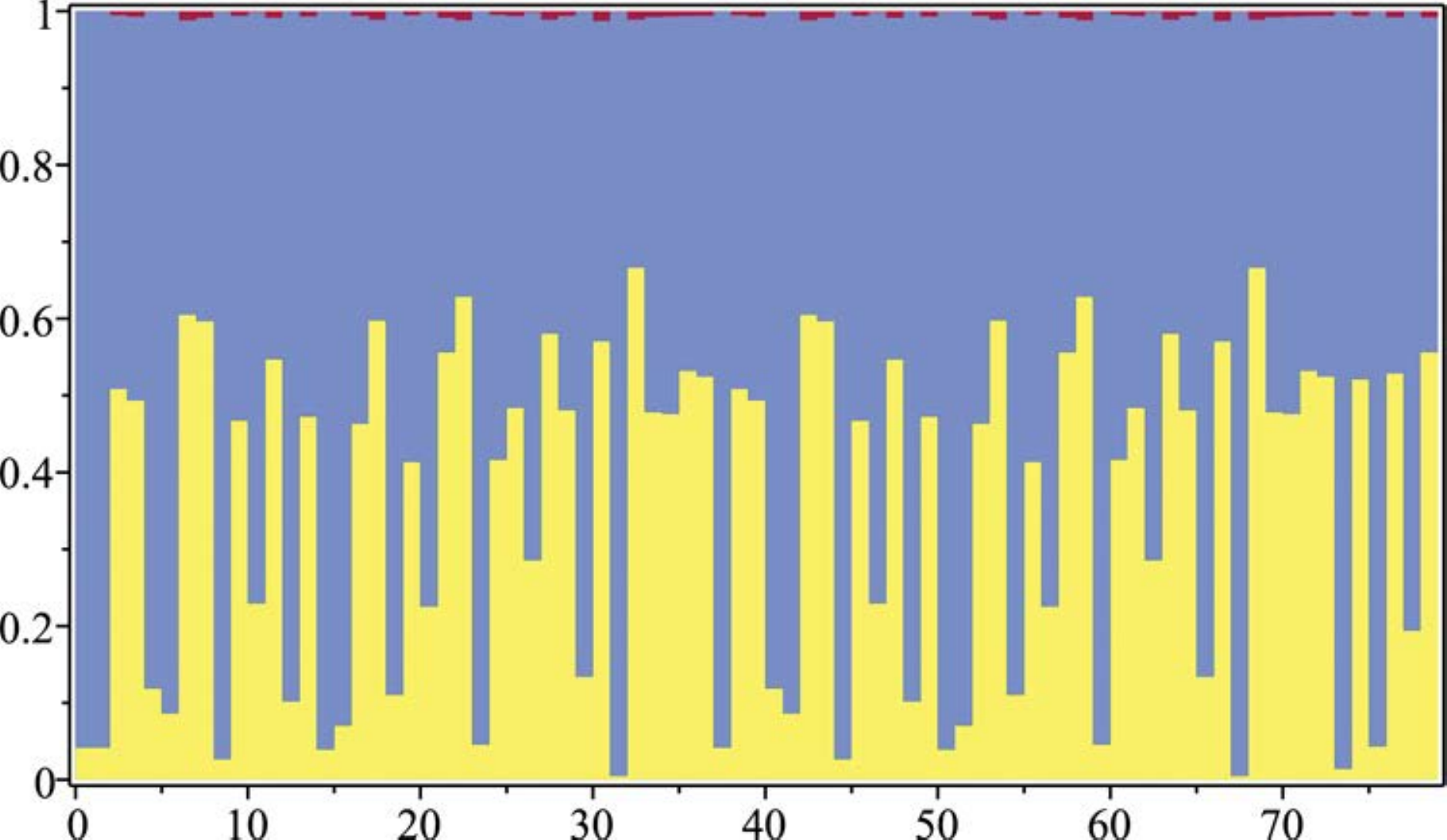}
	\includegraphics[width=2.3in]{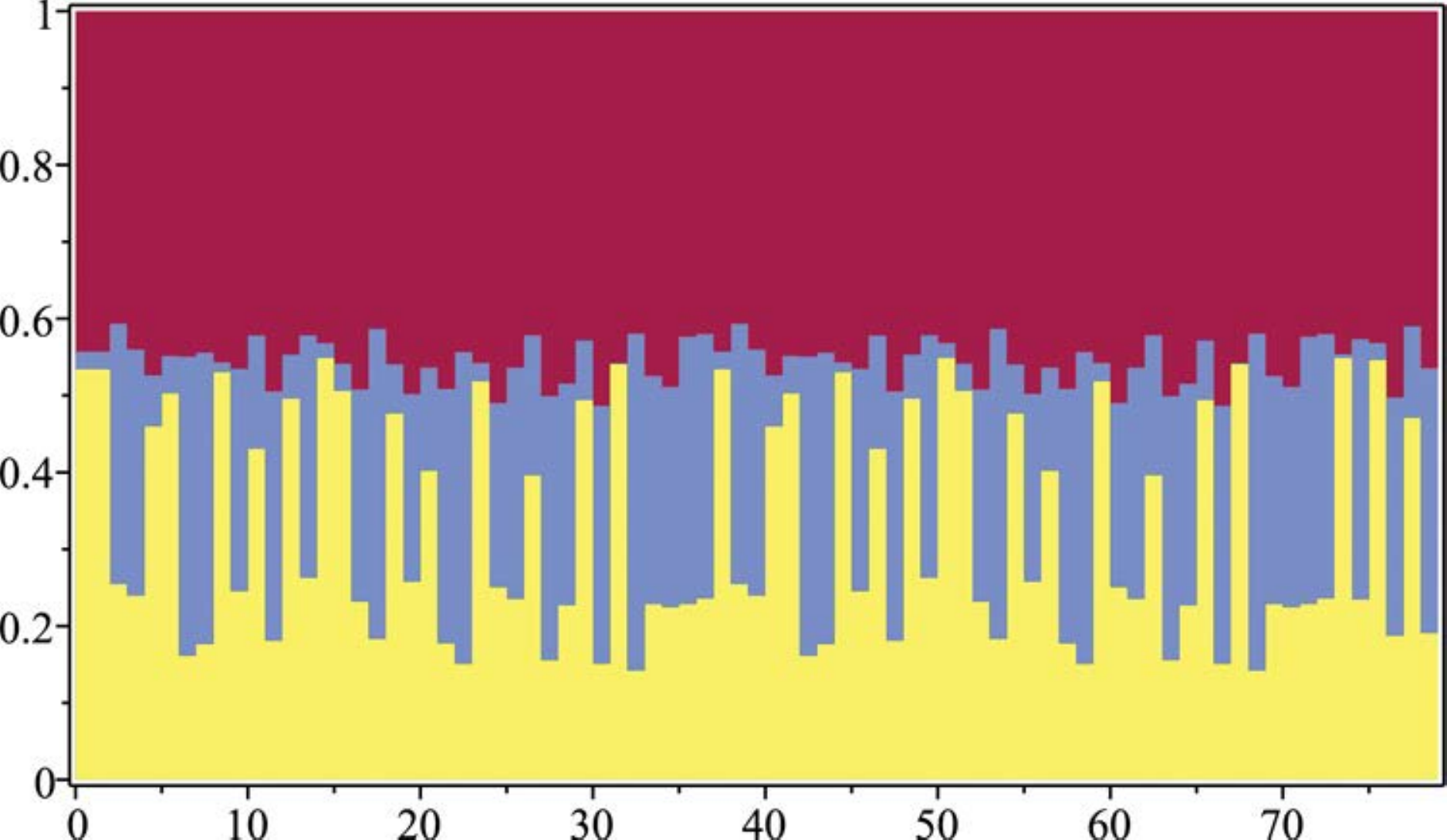}
	\includegraphics[width=2.3in]{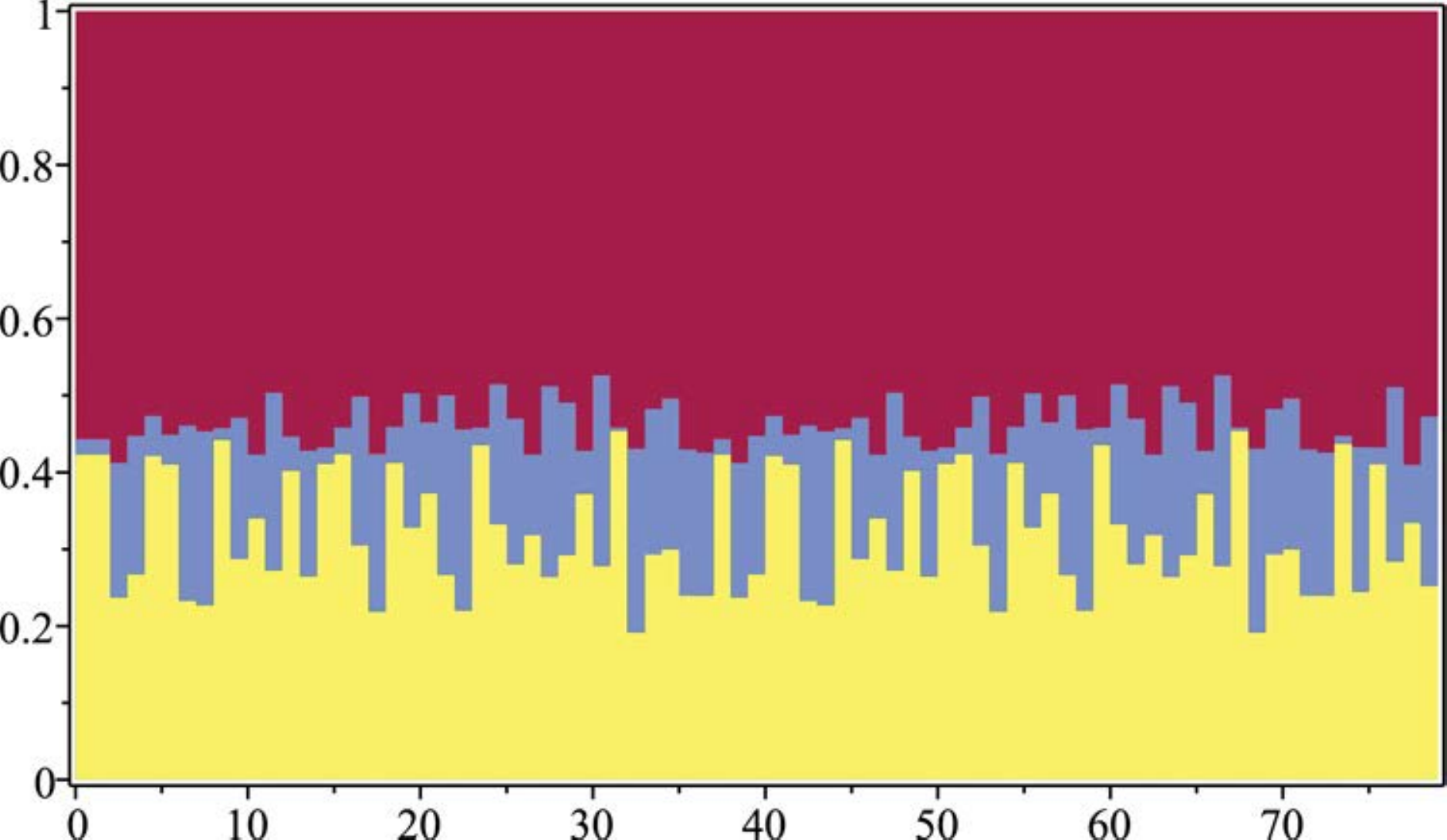}
	
	\includegraphics[width=2.3in]{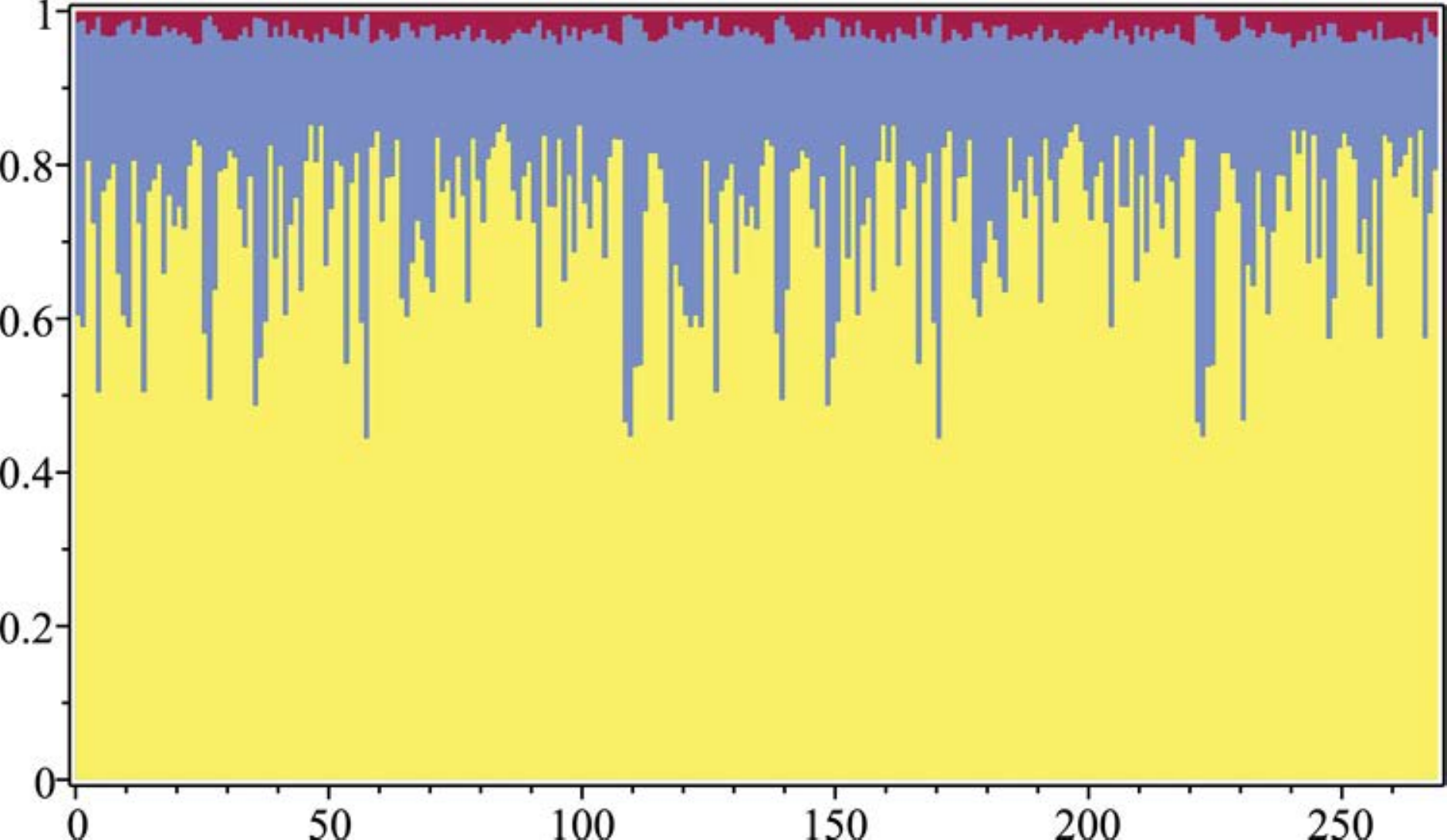}
	\includegraphics[width=2.3in]{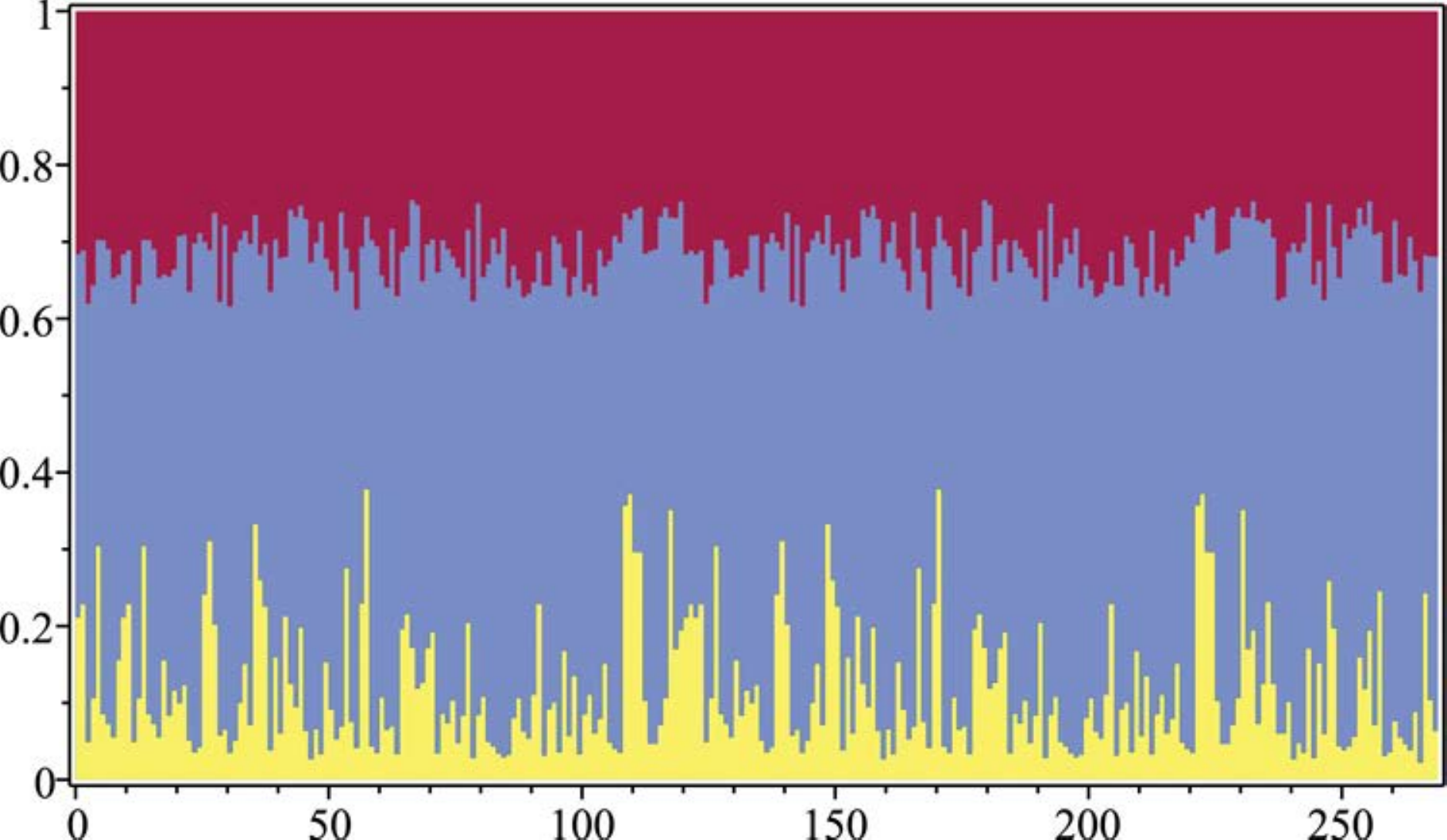}
	\includegraphics[width=2.3in]{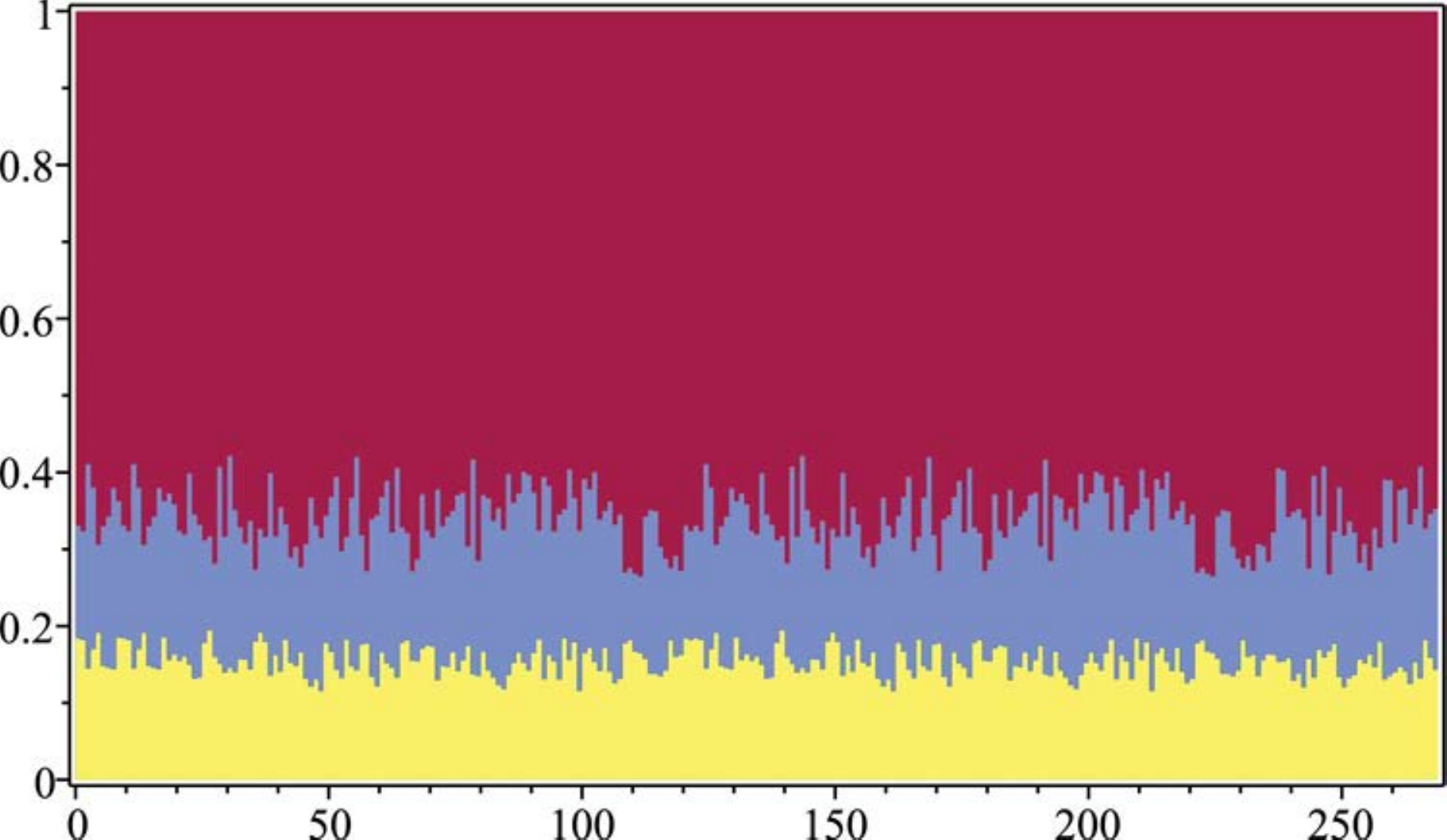}
	
	\includegraphics[width=2.3in]{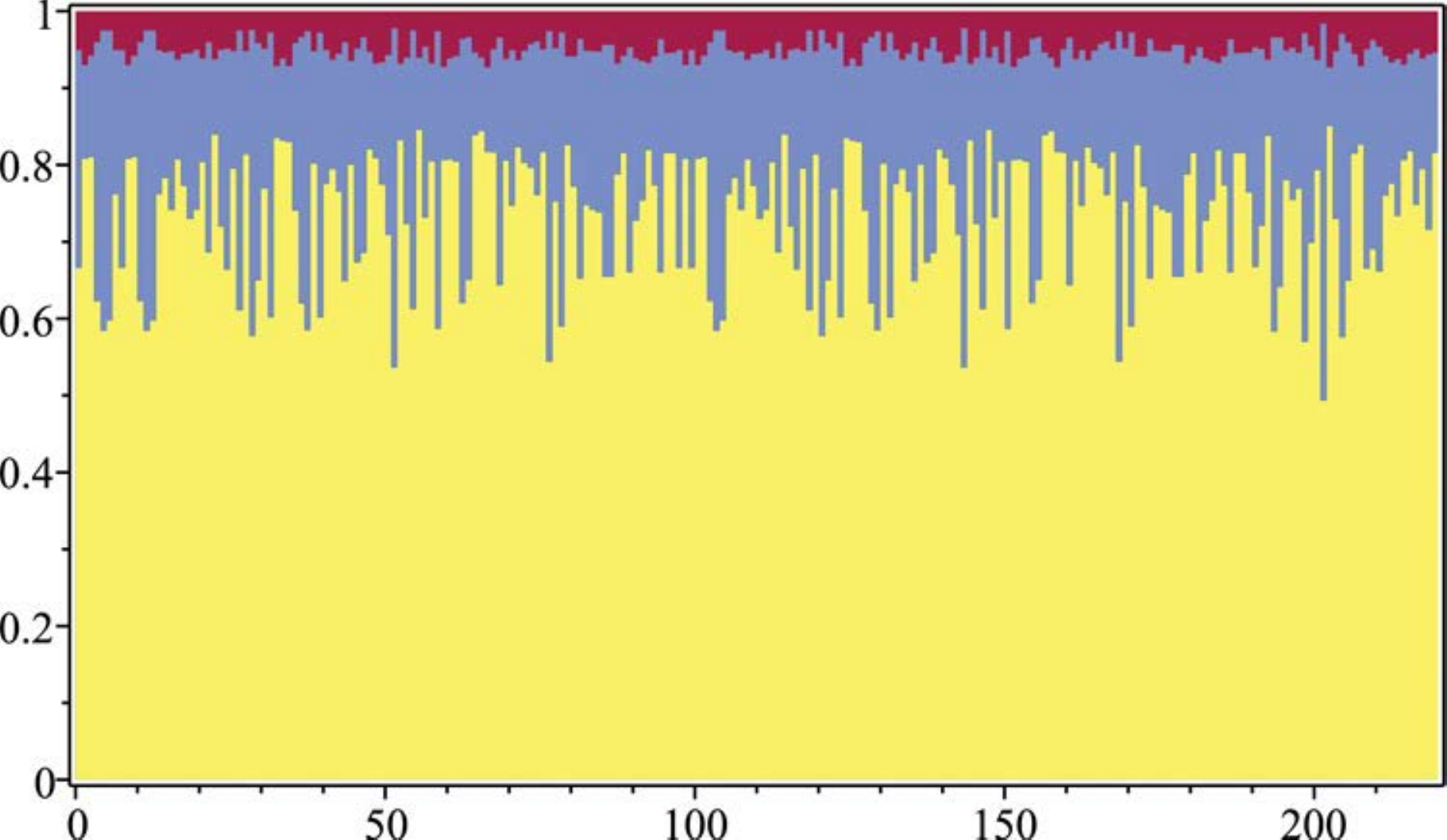}
	\includegraphics[width=2.3in]{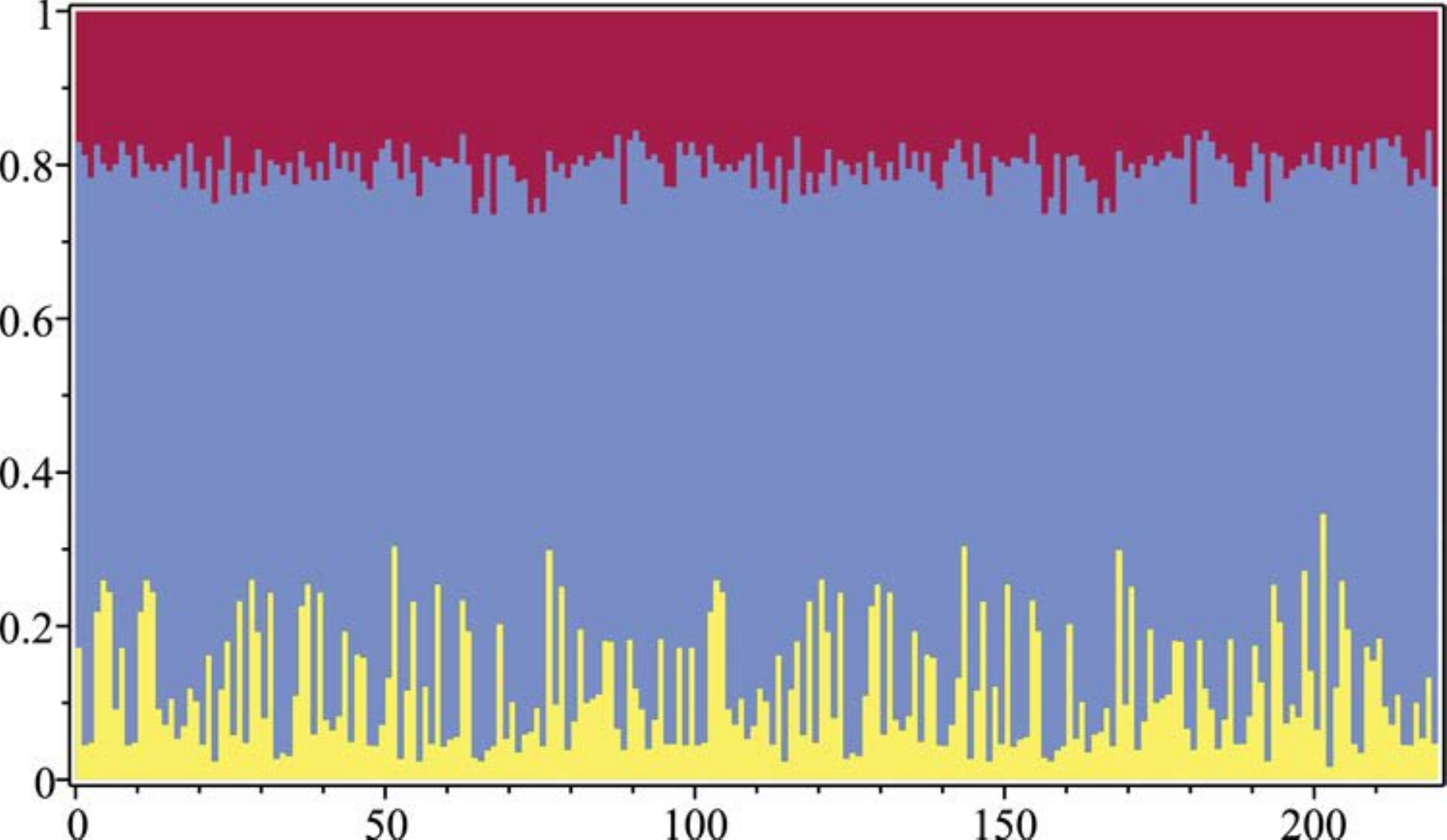}
	\includegraphics[width=2.3in]{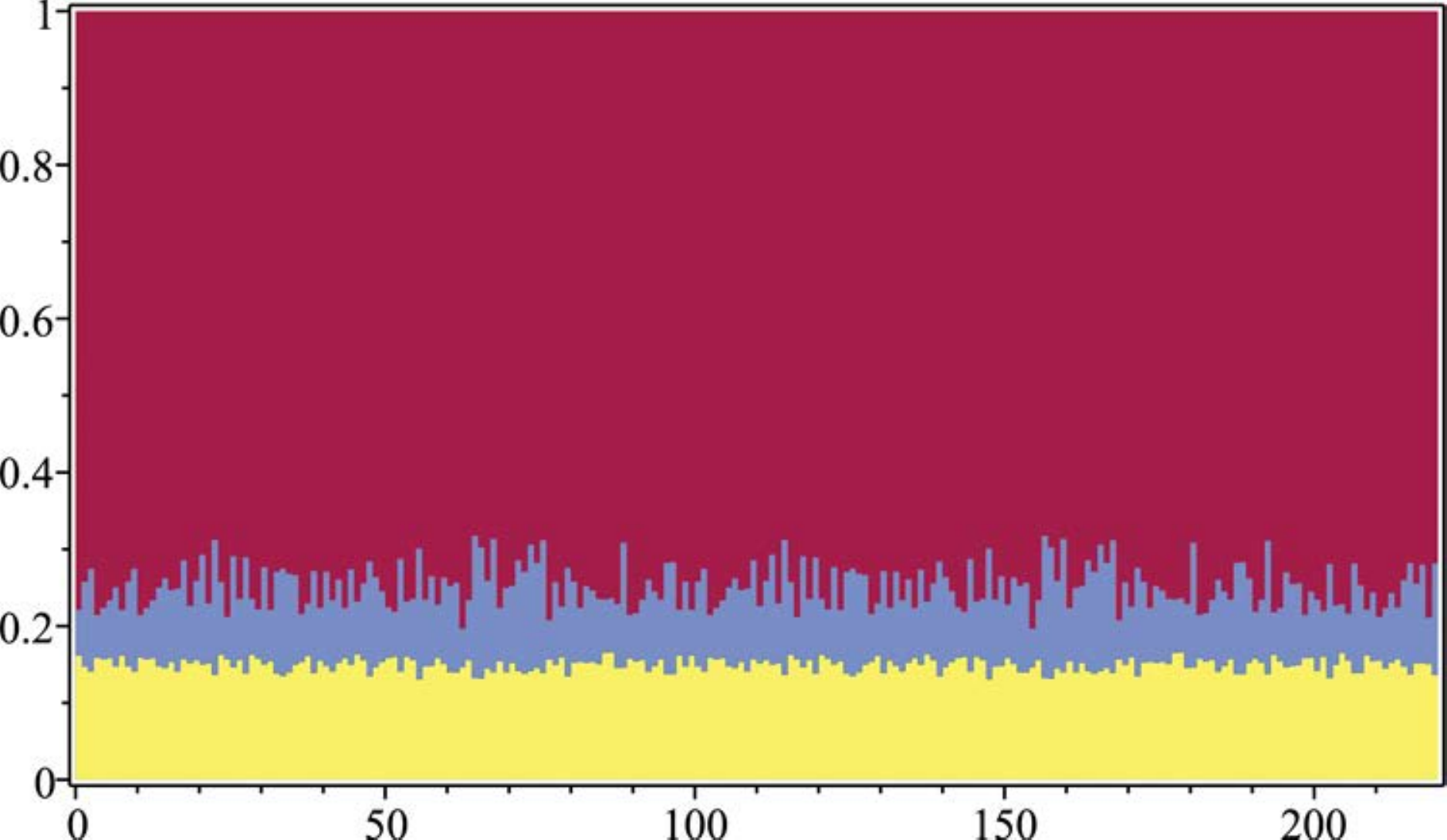}
	
	\caption{Stacked bar diagrams for quark-antiquark and four-quark components of pseudoscalar singlets versus  simulation number
over subsets of $S_\mathrm{I}$, $S_\mathrm{II}$ and $S_\mathrm{III}$ for which $\chi \le \chi^{\rm exp}$.  Quark-antiquark component is in light gray (or yellow in color version), four-quark component is in gray (or blue in color version) and glue component is in dark gray (or red in color version). The three rows from top to bottom respectively correspond to  permutations 127, 137 and 147.}
	\label{F_N0_comps}
\end{figure}

\begin{figure}[!htb]
	\centering
	\includegraphics[width=2.3in]{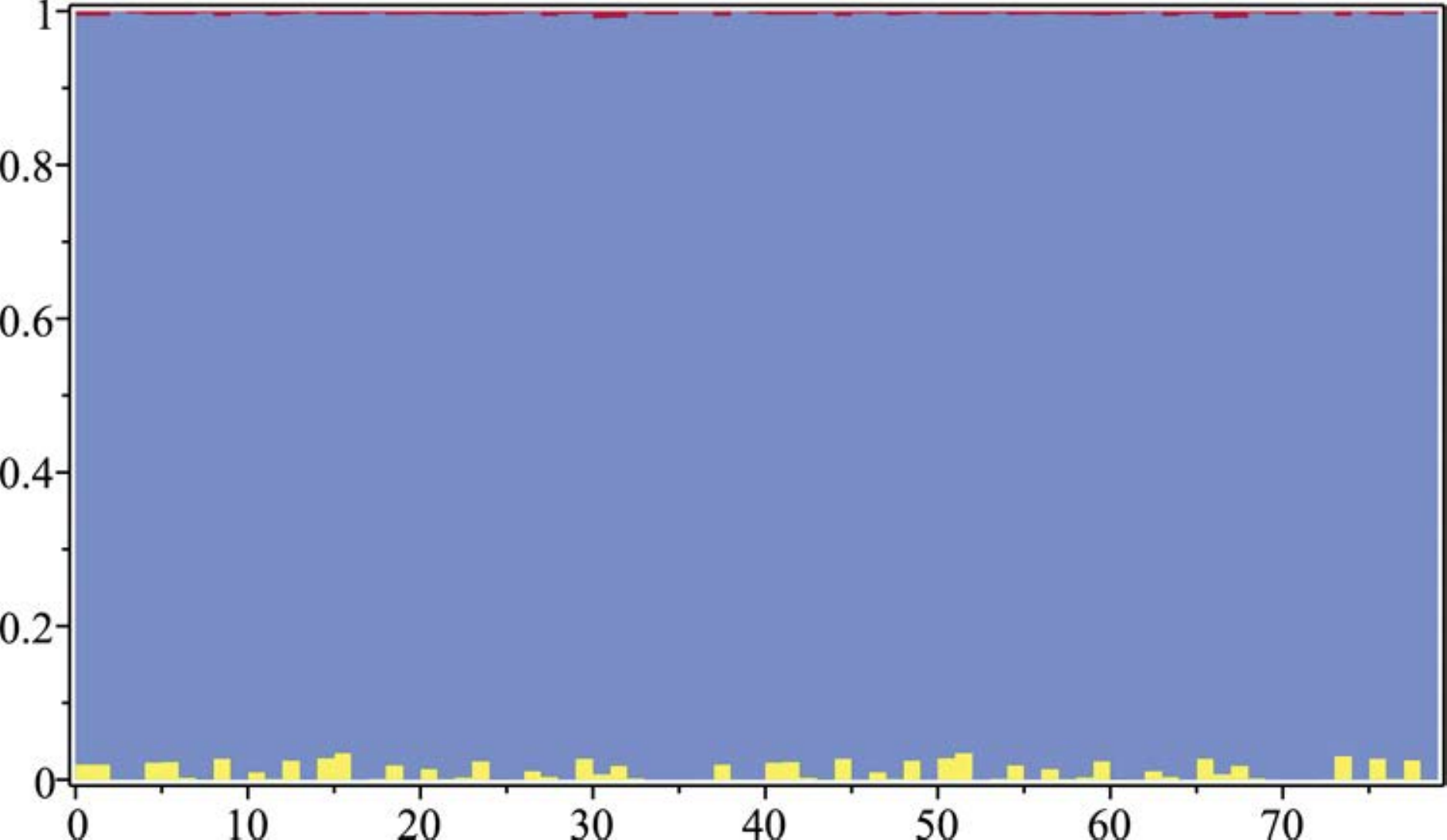}
	\includegraphics[width=2.3in]{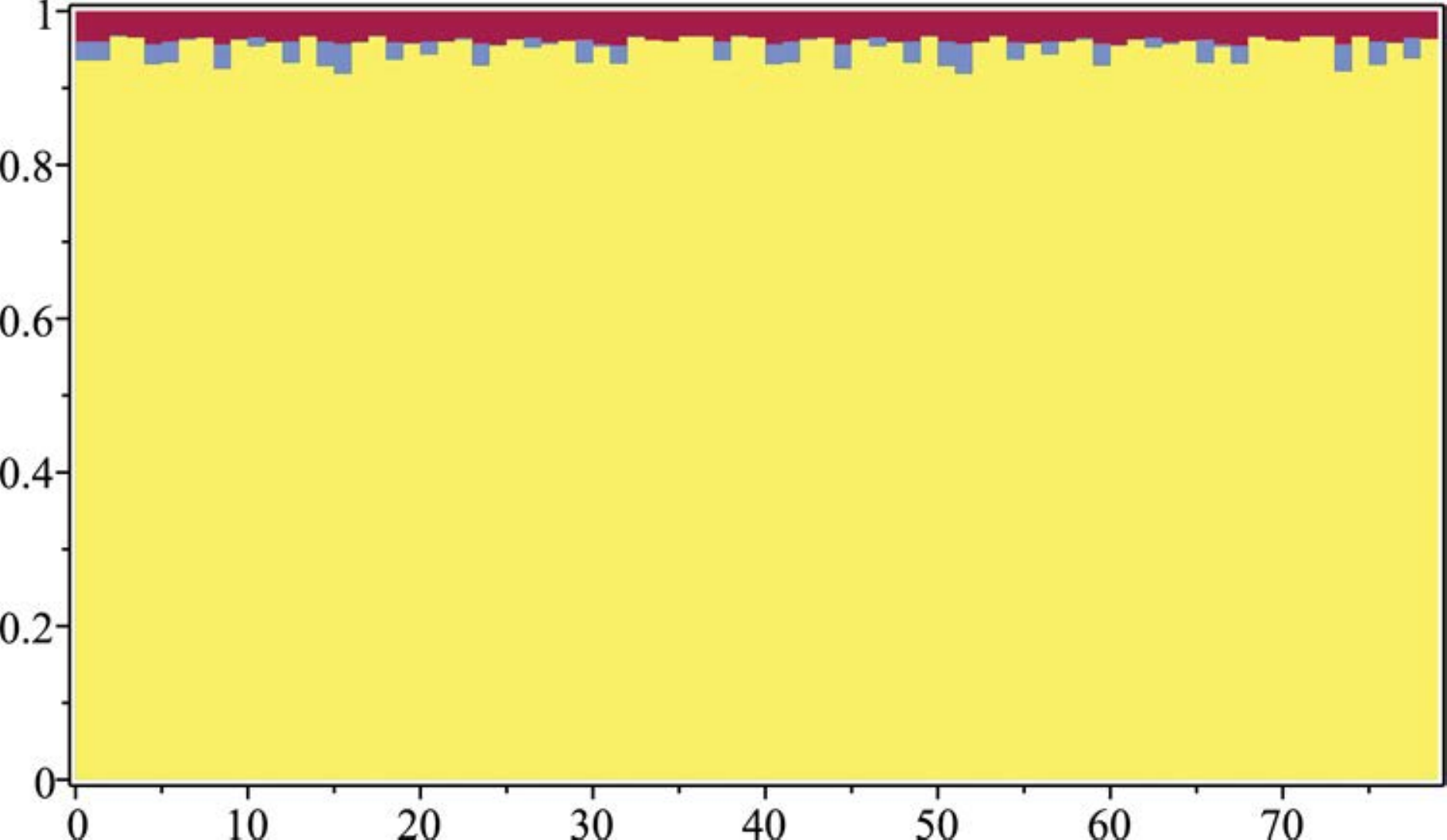}
	\includegraphics[width=2.3in]{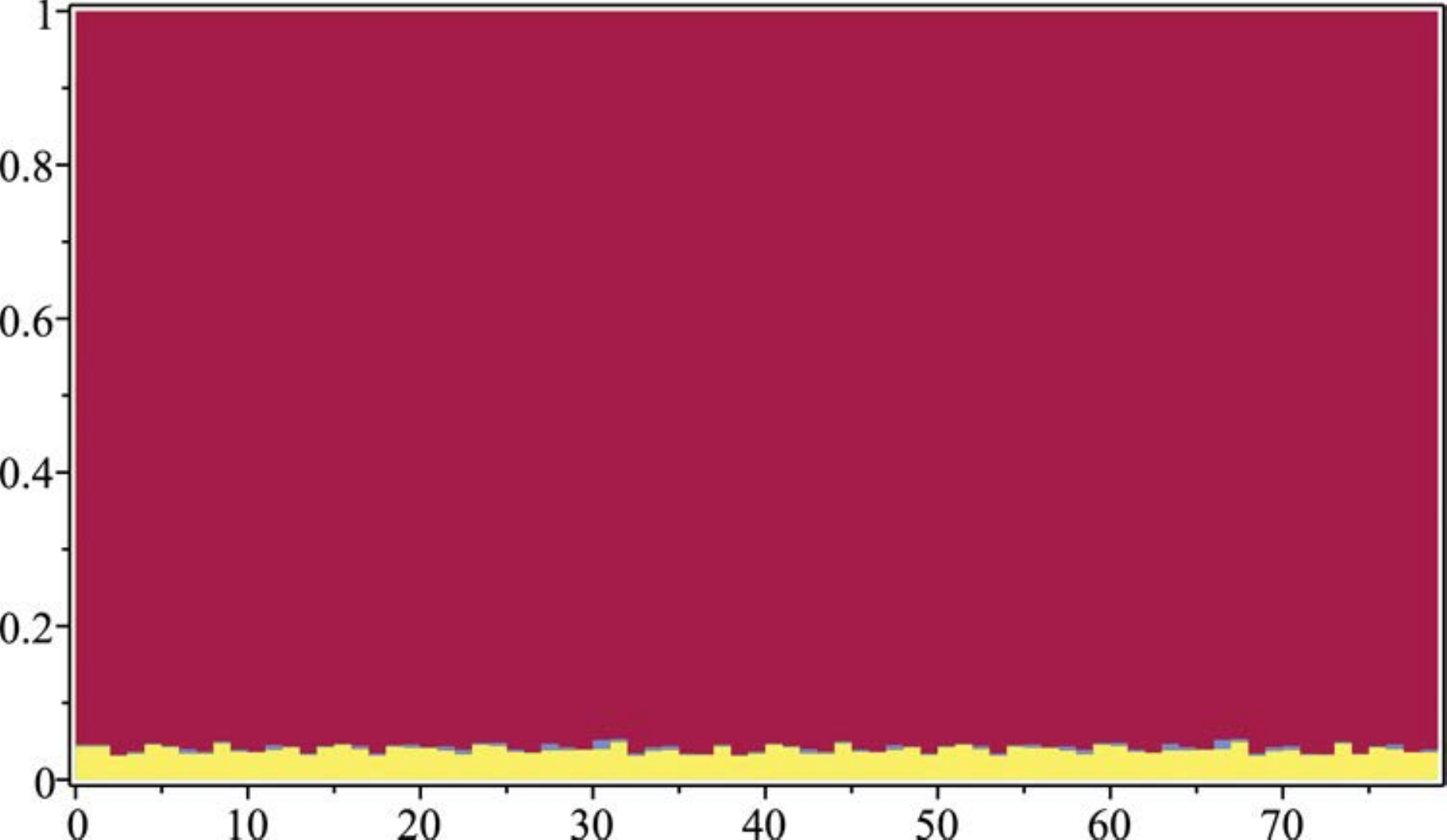}
	
	\includegraphics[width=2.3in]{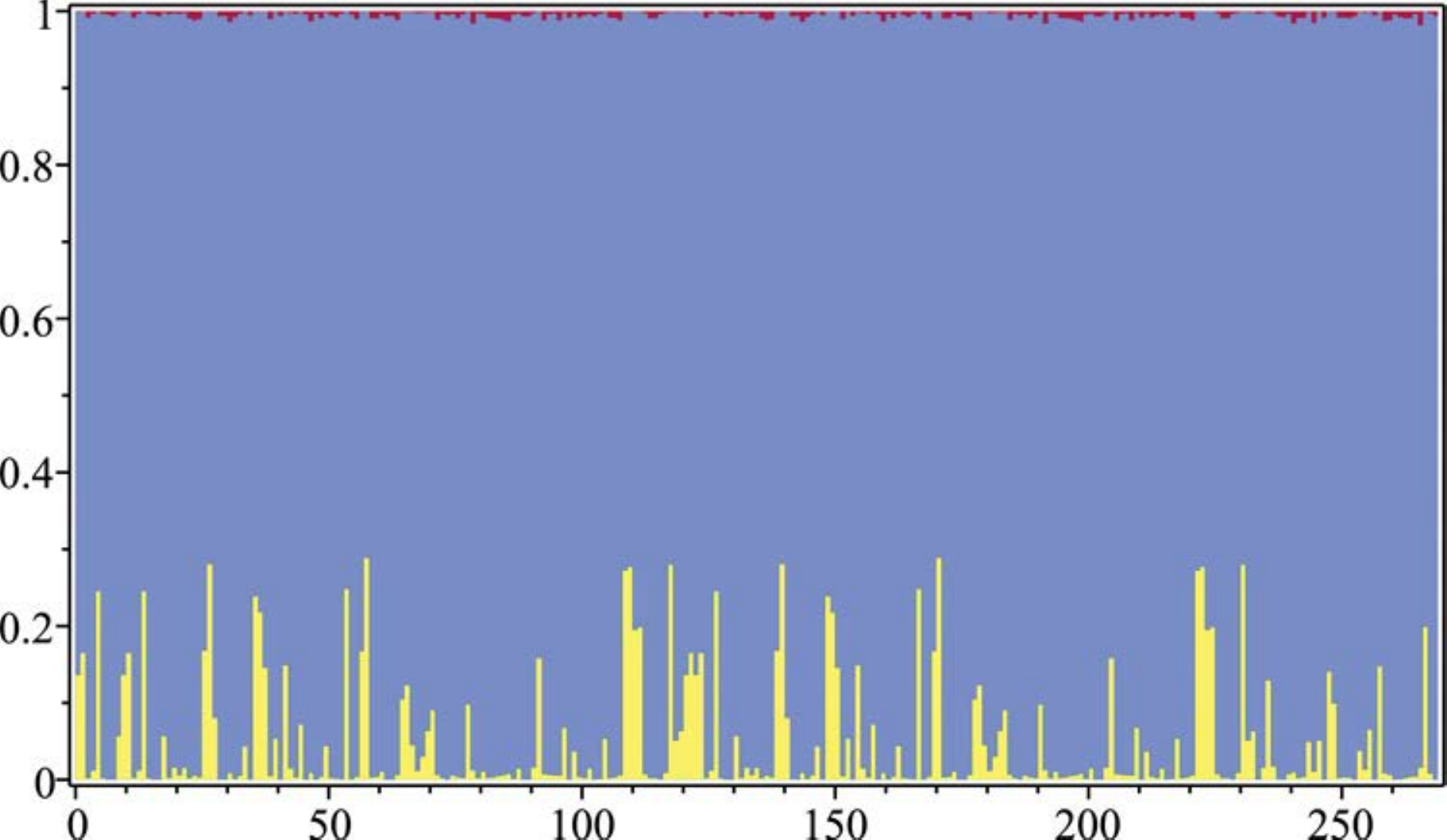}
	\includegraphics[width=2.3in]{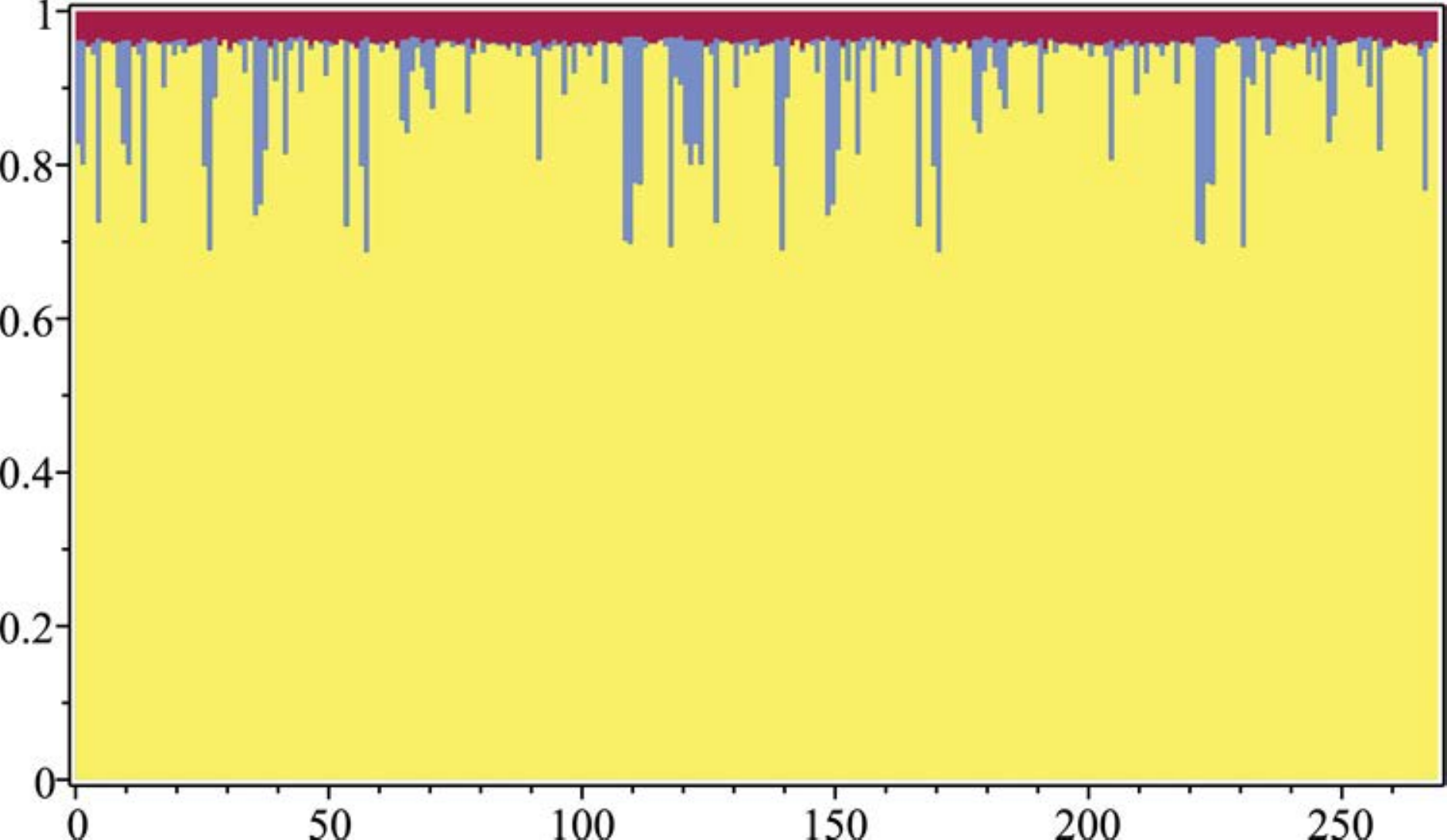}
	\includegraphics[width=2.3in]{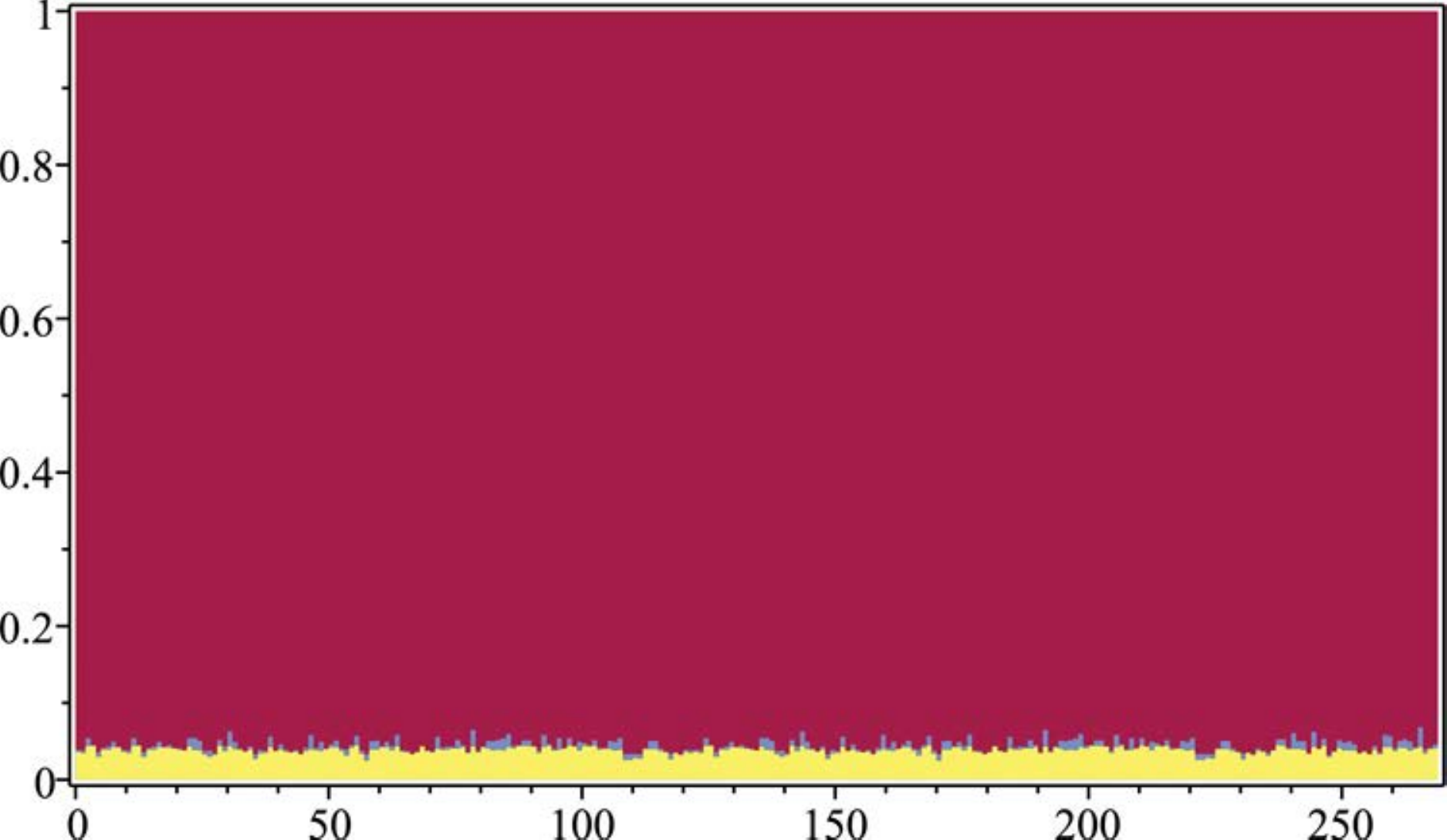}
	
	\includegraphics[width=2.3in]{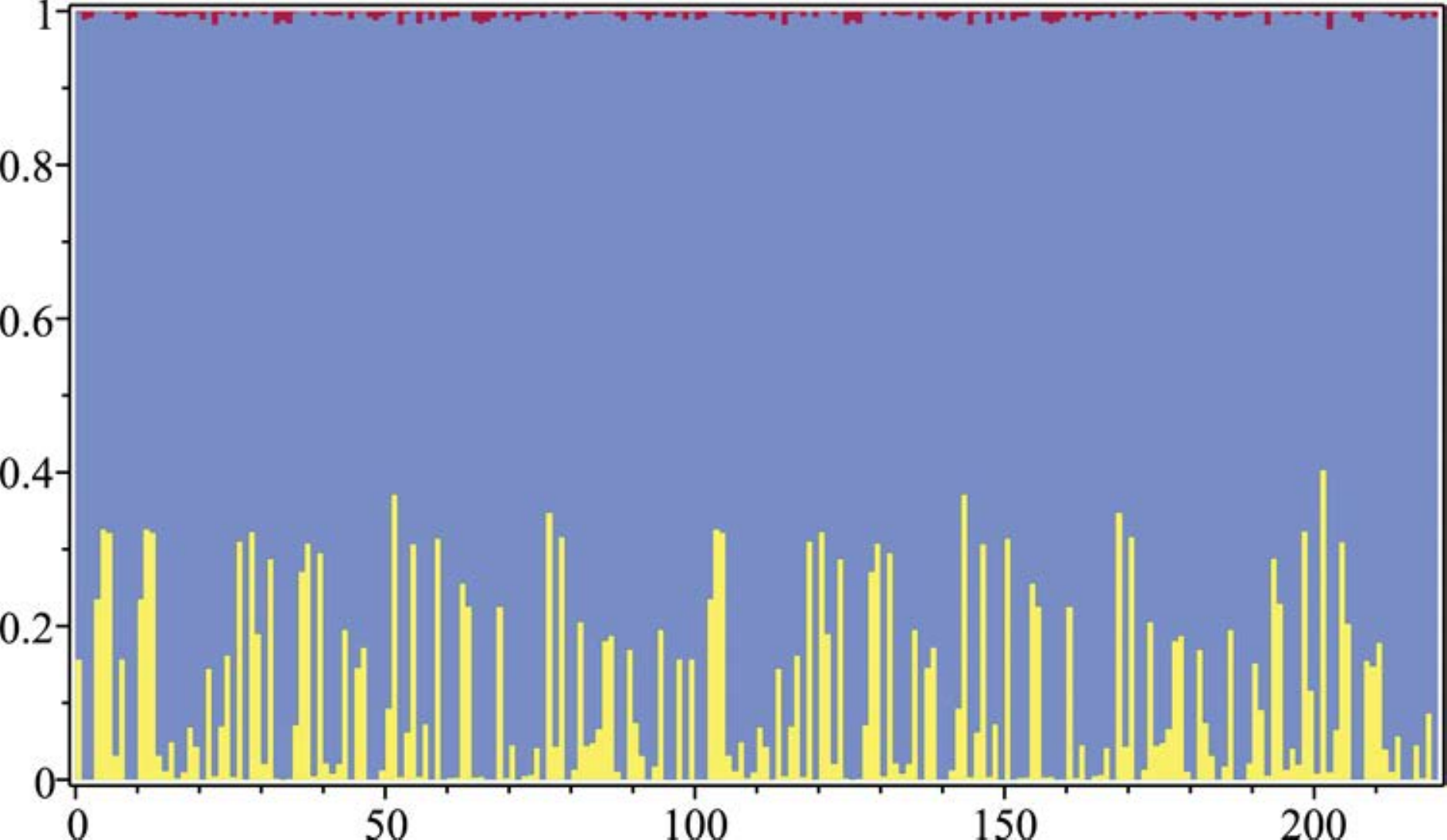}
	\includegraphics[width=2.3in]{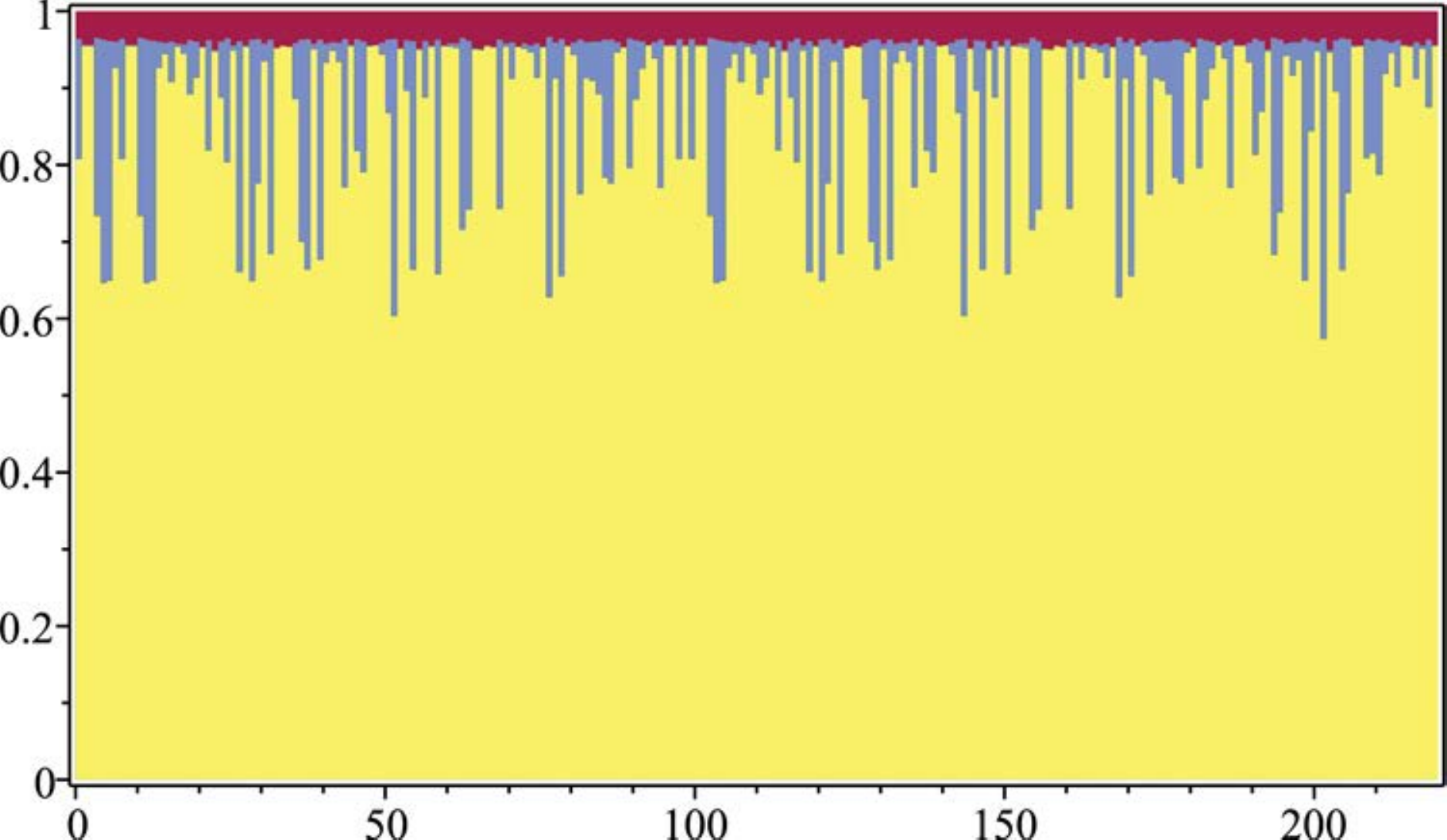}
	\includegraphics[width=2.3in]{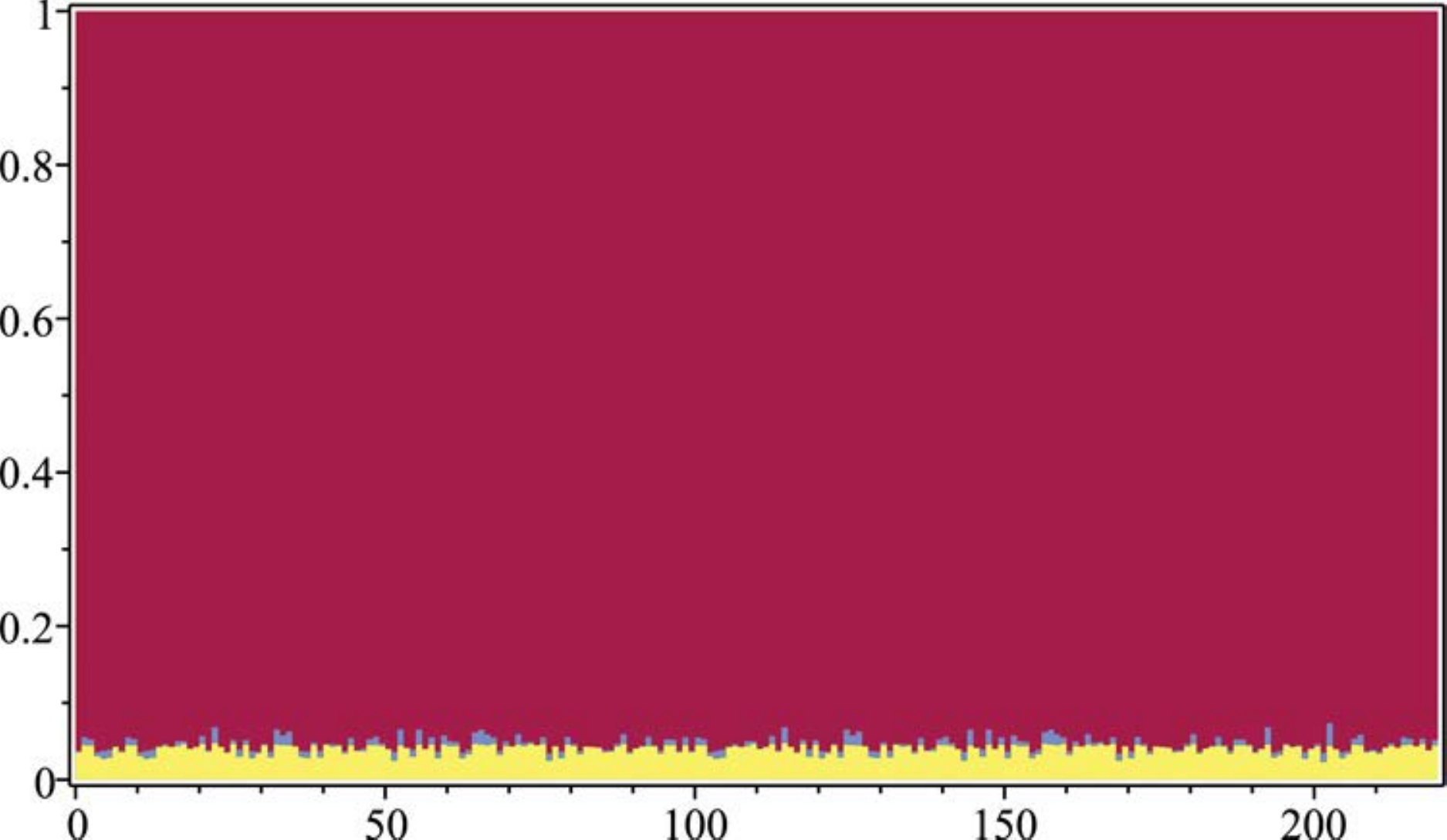}
	
	\caption{
Stacked bar diagrams for quark-antiquark and four-quark components of scalar singlets versus  simulation number
over subsets of $S_\mathrm{I}$, $S_\mathrm{II}$ and $S_\mathrm{III}$ for which $\chi \le \chi^{\rm exp}$.  Quark-antiquark component is in light gray (or yellow in color version), four-quark component is in gray (or blue in color version) and glue component is in dark gray (or red in color version). The three rows from top to bottom respectively correspond to  permutations 127, 137 and 147.}
	\label{F_Y0_comps}
\end{figure}

Our simulations for the substructures of octet states (both pseudoscalars and scalars) are given in Fig. \ref{F_N8Y8_comps} (over the subsets with $\chi\le \chi^{\rm exp}$) showing a fairly stable predictions with a dominant quark-antiquark component for the pseudoscalar octet and a dominant four-quark for the scalar octet.   Similarly, the simulations for pseudoscalar and scalar singlets are overall stable and given in Figs. \ref{F_N0_comps} and \ref{F_Y0_comps}.  For pseudoscalar singlets,  scenario 127 is least stable with sharp competition between quark-antiquark and four-quark components (with latter dominating the former) in substructure of the lighest state.  The same sharp competition can be seen between quark-antiquark, four-quark and glue substructures of the middle and the heaviest pseudoscalar singlets in 127 scenario.  Therefore, while    $\chi_{127}\le \chi^{\rm exp}$, the component predictions are not very stable and we consider 127 scenario disfavored in comparison with 137 and 147 scenarios.   In 137 and 147 cases,  the lightest pseudoscalar singlet is dominantly quark-antiquark (consistent with established phenomenology), middle state contains more four-quark followed by glue and small amount of quark-antiquark components.    The heaviest state is dominantly made of glue with a mass in the range of 2.2 GeV consistent with lattice results for the lightest pseudoscalar glueball mass \cite{Ochs}.    The situation for the scalar singlets is seen in Fig. \ref{F_Y0_comps} in which, in all three scenarios, we see that the substructures of these states (from lightest to heaviest) are dominantly four quark, quark-antiquark and glue.   We see that the heaviest state is almost entirely made of glue and has a mass of about 1.6 GeV which is close to the decoupling limit studied in \cite{Jora25}. For convenience, we have also given the numerical vales for the substructures of all states in Table \ref{table3} (computed at $\chi_{\rm min}$) and in    Table \ref{table4} in which the component averages and standard deviations are given. Table \ref{table5} contains the rotation matrices for all the states involved (also computed at $\chi_{\rm min}$).

\begin{table}[htbp]
	\begin{center}

\caption{Quark-antiquark, four-quark, and glueball components at best $\chi^{\rm exp}$.}

\begin{tabular}{ M{1cm} *{3}{|| *{3}{M{25pt}}  } N}
\hline
\hline
\multirow{2}{*}{States} & \multicolumn{3}{c||}{127} & \multicolumn{3}{c||}{137} & \multicolumn{3}{c}{147} &\\ [2pt]
 & {$q\bar{q}$} & {$qq \bar{q} \bar{q}$}  & {glue} & {$q\bar{q}$} & {$qq \bar{q} \bar{q}$} & {glue} & {$q\bar{q}$} & {$qq \bar{q} \bar{q}$}  & {glue} &\\ [1.2pt]
\hline
\hline
$\psi_{8^+}^{(1)}$ & 
32.3 &  67.7 & --- &  
34.1 &  65.9 & --- &  
37.8 & 62.2 & ---
&\\ [6pt]

$\psi_{8^+}^{(2)}$ &
67.7 &  32.3 & --- & 
65.9 &  34.1 & --- &  
62.2 & 37.8 & ---
\\ [6pt]

\hline

$\psi_{8^-}^{(1)}$ &  
80.8 &  19.2 & --- &  
76.6 &  23.4 & --- &  
74.2 &  25.8 & ---
&\\ [6pt]

$\psi_{8^-}^{(2)}$ &  
19.2 &  80.8 & --- &  
23.4 &  76.6 & --- & 
25.8 & 74.2 & ---
&\\ [6pt]

\hline

$\psi_{0^+}^{(1)}$ &
 0.0 &  96.7 & 3.3 &  
 6.5 &  90.3 &  3.2 &  
 0.0 &  95.5 &  4.5
&\\ [6pt]

$\psi_{0^+}^{(2)}$ & 
99.8 &  0.0 &  0.2 &  
93.5 &  6.5 &  0.0 &  
98.7 &  0.0 & 1.3
&\\ [6pt]

$\psi_{0^+}^{(3)}$ &  
0.2 &  3.4 &  96.4 &  
0.0 & 3.2 &  96.8 &  
1.3 &  4.5 &  94.2
&\\ [6pt]

\hline

$\psi_{0^-}^{(1)}$ &  
59.6 &  17.6 &  22.8 &  
64.2 &  19.4 &  16.4 &  
81.6 &  3.80 &  14.6
&\\ [6pt]

$\psi_{0^-}^{(2)}$ &  
39.4 &  37.9 &  22.7 &  
33.30 &  55.8 &  10.9 &  
11.0 &  77.6 &  11.4
&\\ [6pt]

$\psi_{0^-}^{(3)}$ &  
0.8 &  44.4 &  54.8 &  
2.4 &  24.7 &  72.9 &  
7.2 &  18.5 & 74.3
&\\ [6pt]

\hline
\hline

\end{tabular}
\label{table3}
\end{center}
\end{table}

\begin{table}[htbp]
	\begin{center}

\caption{Central and standard deviation of the percentage of Quark-antiquark, four-quark, and glueball components.}

\begin{tabular}{ M{1cm} *{3}{|| *{3}{M{40pt}}  } N}

\hline
\hline

\multirow{2}{*}{States} & \multicolumn{3}{c||}{127} & \multicolumn{3}{c||}{137} & \multicolumn{3}{c}{147} &\\ [2pt]

 & {$q\bar{q}$} & {$qq \bar{q} \bar{q}$}  & {glue} & {$q\bar{q}$} & {$qq \bar{q} \bar{q}$} & {glue} & {$q\bar{q}$} & {$qq \bar{q} \bar{q}$}  & {glue} &\\ [1.2pt]
\hline
\hline

$\psi_{8^+}^{(1)}$&
38 $\pm$ 5 &  62 $\pm$ 5 & --- &  
41 $\pm$ 5 &  59 $\pm$ 5 & --- &  
41 $\pm$ 2 &  59 $\pm$ 2 & ---
&\\ [6pt]

$\psi_{8^+}^{(2)}$ & 
62 $\pm$ 5 &  38 $\pm$ 5 & --- &  
59 $\pm$ 5 &  41 $\pm$ 5 & --- &  
59 $\pm$ 2 &  41 $\pm$ 2 & ---
&\\ [6pt]

\hline

$\psi_{8^-}^{(1)}$ & 
79 $\pm$ 2 &  21 $\pm$ 2 & --- &  
78 $\pm$ 1 &  22 $\pm$ 1 & --- &  
76 $\pm$ 1 &  24 $\pm$ 1 & ---
&\\ [6pt]

$\psi_{8^-}^{(2)}$ & 
21 $\pm$ 2 &  79 $\pm$ 2 & --- & 
22 $\pm$ 1 &  78 $\pm$ 1 & --- &  
24 $\pm$ 1 &  76 $\pm$ 1 & ---
&\\ [6pt]

\hline

$\psi_{0^+}^{(1)}$ & 
$0.9^{+1.1}_{-0.9}$ &  95 $\pm$ 2 & 4 $\pm$ 1 &  
$5^{+8}_{-5}$ &  91 $\pm$ 7 & 4.0 $\pm$ 0.1 &  
$9^{+12}_{-9}$ &  87 $\pm$ 11 & 4 $\pm$ 1
&\\ [6pt]

$\psi_{0^+}^{(2)}$ & 
99 $\pm$ 1 & 1 $\pm$ 1 &  0.2 $\pm$ 0.2 &  
95 $\pm$ 7 &  $5^{+8}_{-5}$ &  0.5 $\pm$ 0.5 & 
90 $\pm$ 11 &  $9^{+11}_{-9}$ &  0.6 $\pm$ 0.5
&\\ [6pt]

$\psi_{0^+}^{(3)}$ & 
0.3 $\pm$ 0.1 &  4.0 $\pm$ 0.1 & 96 $\pm$ 1 &  
0.4 $\pm$ 0.4 &  4.0 $\pm$ 0.1 & 96 $\pm$ 1 &  
0.5 $\pm$ 0.5 &  4.0 $\pm$ 0.1 & 95 $\pm$ 1
&\\ [6pt]

\hline

$\psi_{0^-}^{(1)}$ &  
36 $\pm$ 22 & 32 $\pm$ 15 &  32 $\pm$ 8 &  
73 $\pm$ 10 & 12 $\pm$ 9 &  15 $\pm$ 2 &  
73 $\pm$ 9 & 12 $\pm$ 8 &  15 $\pm$ 1
&\\ [6pt]

$\psi_{0^-}^{(2)}$ & 
64 $\pm$ 22 &  22 $\pm$ 14 & 14 $\pm$ 8 & 
25 $\pm$ 11 &  57 $\pm$ 8 &  19 $\pm$ 5 & 
22 $\pm$ 10 &  68 $\pm$ 7 &  10 $\pm$ 3
&\\ [6pt]

$\psi_{0^-}^{(3)}$ &
0.4 $\pm$ 0.4 &  46 $\pm$ 3 & 54 $\pm$ 3 &  
3 $\pm$ 1 &  31 $\pm$ 4 & 66 $\pm$ 4 &  
5 $\pm$ 1 &  20 $\pm$ 2 & 75 $\pm$ 3
&\\ [6pt]

\hline
\hline
\end{tabular}
\label{table4}
\end{center}
\end{table}

\begin{table}[htbp]
	\begin{center}

\caption{Rotation matrix for scalar octet (first row), pseudoscalar octet(second row), scalar singlet (third row), and pseudoscalar singlet (last row) at best $\chi \le \chi^{\rm exp}$}
\begin{tabular}{M{80pt} *{3}{||M{120pt}} N}
\hline
\hline
Rotation Matrix & 127 & 137 & 147 &\\ [6pt]
\hline
\hline
$K_{8^+}^{-1}$ & $ \begin {array}{*{2}{D{.}{.}{2.4}}} -0.568 &  0.822 \\ 0.822 &  0.568   \end{array} $ &
$ \begin {array}{*{2}{D{.}{.}{2.4}}} -0.583 &  0.812\\ 0.812 &  0.583 \end{array}  $ &
$ \begin {array}{*{2}{D{.}{.}{2.4}}}  -0.614 &  0.788\\ 0.788 &  0.614 \end {array} $ &\\ 
\hline

$K_{8^-}^{-1}$ & $ \begin {array}{*{2}{D{.}{.}{2.4}}} -0.899 & -0.437 \\ 0.437 & -0.899\end {array}  $ &
$\begin {array}{*{2}{D{.}{.}{2.4}}} -0.875 & -0.483 \\ 0.483 & -0.875\end {array}  $ &
$\begin {array}{*{2}{D{.}{.}{2.4}}} -0.861 & -0.508 \\ 0.508 & -0.861\end {array}  $ &\\ 
\hline

$K_{0^+}^{-1}$ &
$ \begin {array}{*{3}{D{.}{.}{2.4}}}  
0.027 & -0.982 &  0.184\\
-0.998 & -0.018 &  0.051\\ 
0.047 &  0.185 &  0.981
\end {array} $ &

$\begin {array}{*{3}{D{.}{.}{2.4}}}  
-0.255 & -0.949 &  0.181\\
-0.966 &  0.256 & -0.019\\ 
0.028 &  0.180 &  0.983
\end {array}$ &

$\begin {array}{*{3}{D{.}{.}{2.4}}}  
0.011 & -0.9770 &  0.212\\ 
0.992 & -0.014 & -0.117\\ 
0.117&  0.212 &  0.970
\end{array} $ &\\
\hline

$K_{0^-}^{-1}$ &
$\begin {array}{*{3}{D{.}{.}{2.4}}} 
0.772 & -0.420 & -0.476\\
0.628 & 0.615 &  0.475\\
-0.093 &  0.666 & -0.739
\end{array}  $ &

$\begin {array}{*{3}{D{.}{.}{2.4}}} 
-0.801 &  0.440 &  0.404\\ 
0.577 &  0.747 &  0.329\\ 
0.156 & -0.497 &  0.853
\end {array}$ &

$\begin {array}{*{3}{D{.}{.}{2.4}}}
0.903 & -0.195 & -0.380\\ 
0.332 &  0.880 &  0.336\\
-0.269 &  0.431 & -0.861 \end {array} $ &\\

\hline
\hline

\end{tabular}
\label{table5}
\end{center}
\end{table}


\section{Conclusions}

We recently developed the  generalized linear sigma model with glueballs in \cite{Jora25}. We showed that the most general framework contains many terms and therefore is not a  practical starting point,  both from conceptual as well as from computational standpoints. For a practical approach,  a systematic approximation scheme is necessary.   In \cite{Jora25} we considered a special decoupling limit in which the  glueballs decouple from quarkonia and thereby the model becomes exactly solvable and in turn the pure (unmixed) scalar glueball mass can be extracted \cite{Jora26}.  This exactly solvable limit then becomes the foundation for further studies of the evolution of model parameters in a manner that is both intuitive and computationally tractable.

As the next step, in this work we studied the flavor SU(3)  limit of the generalized linear sigma model where the interaction of glueballs with SU(3) singlets are turned on.  Since the interaction of glueballs with quarkonia is expected to be SU(3) symmetric,  at least in first order, we considered this limit to be a necessary step in building the full interaction Lagrangian for quarkonia and glueballs.     We showed that the SU(3) limit, while is an approximation of the underlying strong dynamics,   provides useful insights on some of the main characteristics of the scalar and pseudoscalar states and interactions with glueballs.      Testing the model predictions against experiment did not result in  unique values for the model parameters,  partly due to the large uncertainties of some of the experimental inputs, and partly due to the approximate nature of the model in this limit.  Instead,  we found sets of acceptable parameter values and studied the mass spectra and quark and glue contents of both scalars and pseudoscalars.   With the inputs of the mass of octets, we studied the singlet masses that model predicts.  Our results showed that the lightest  and the heaviest  pseudoscalar singlets (with masses of 0.547 GeV and 2.2 GeV) are simply consistent with $\eta(547)$ and $\eta(2225)$, respectively. 
The middle pseudoscalar singlet could be either $\eta(1295)$ or $\eta(1405)$ under the condition of $\chi \le \chi^{\rm exp}$.    However,  as Fig.  \ref{F_chi_global} showed, scenario 157 in which the middle pseudoscalar singlet is identified with 
$\eta(1475)$ is not too far from  experiment even though $\chi_{157} > \chi_{157}^{\rm exp}$.   This means that our analysis cannot fully rule out  $\eta(1475)$ as one of the singlets which perhaps is due to its proximity to $\eta(1405)$.   In fact, we find that the substructures of the predicted physical states for scenario 157 are close to those found in scenario 147.   
Overall, we find that the states $\eta(1295)$ and $\eta(1405)$ [as well as $\eta(1475)$] have a complicated hybrid substructure with a dominant four-quark component.  This is, at least qualitatively,  in agreement with speculations made about the non-quark-antiquark nature of these states in \cite{07_KZ}, or the possibility that some of these states might be dynamically generated in $\eta f_0(980)$ and $\pi a_0(980)$ channels studied in \cite{ChUA8}.   
We also found that the scalar singlet masses  contain a very light state [consistent with $f_0(500)$ or sigma meson] and a state around 1.2 GeV [consistent with $f_0(1370)$] and the heaviest state around 1.6 GeV [a sign of either $f_0(1500)$ and $f_0(1710)$].

Our numerical simulations gave overall stable results for the substructure of the scalar and pseudoscalar octets and singlets.    The lighter pseudoscalar octet was shown to be mainly of quark-antiquark type (which is consistent with the conventional phenomenology), whereas the lighter scalar octet contained a very large four-quark component (consistent with most investigations on light scalar mesons). Our results also showed that the substructures of pseudoscalar singlets in ascending order of mass   are quark-antiquark (lightest singlet); mainly four-quark and glue (middle singlet); and dominantly glue (heaviest singlet).  Particularly, a clear identification of the heaviest pseudoscalar singlet with $\eta(2225)$ which is dominantly made of glue is closely consistent with lattice QCD results for the mass of pseudoscalar glueball around 2.2 GeV \cite{Ochs}.        The substructures of scalar singlets (in ascending order of masses) are:  dominantly four-quark; dominantly quark-antiquark; and almost entirely glueball consistent with our recent work \cite{Jora25} in the decoupling limit.
	
Several important questions remain, most important of which is to investigate the SU(3) breaking effects on the results presented here which will be presented in our future works.

\appendix

\section*{Acknowledgments}

A.H.F. gratefully acknowledges  the support of College of Arts and Sciences of SUNY Poly in Fall 2018.   M.L. wishes to thank SUNY Poly for a Summer Undergraduate Research Program fellowship.

\section{Simulation Data}

\begin{figure}[!htb]
	\centering
	
	\includegraphics[width=2.5in]{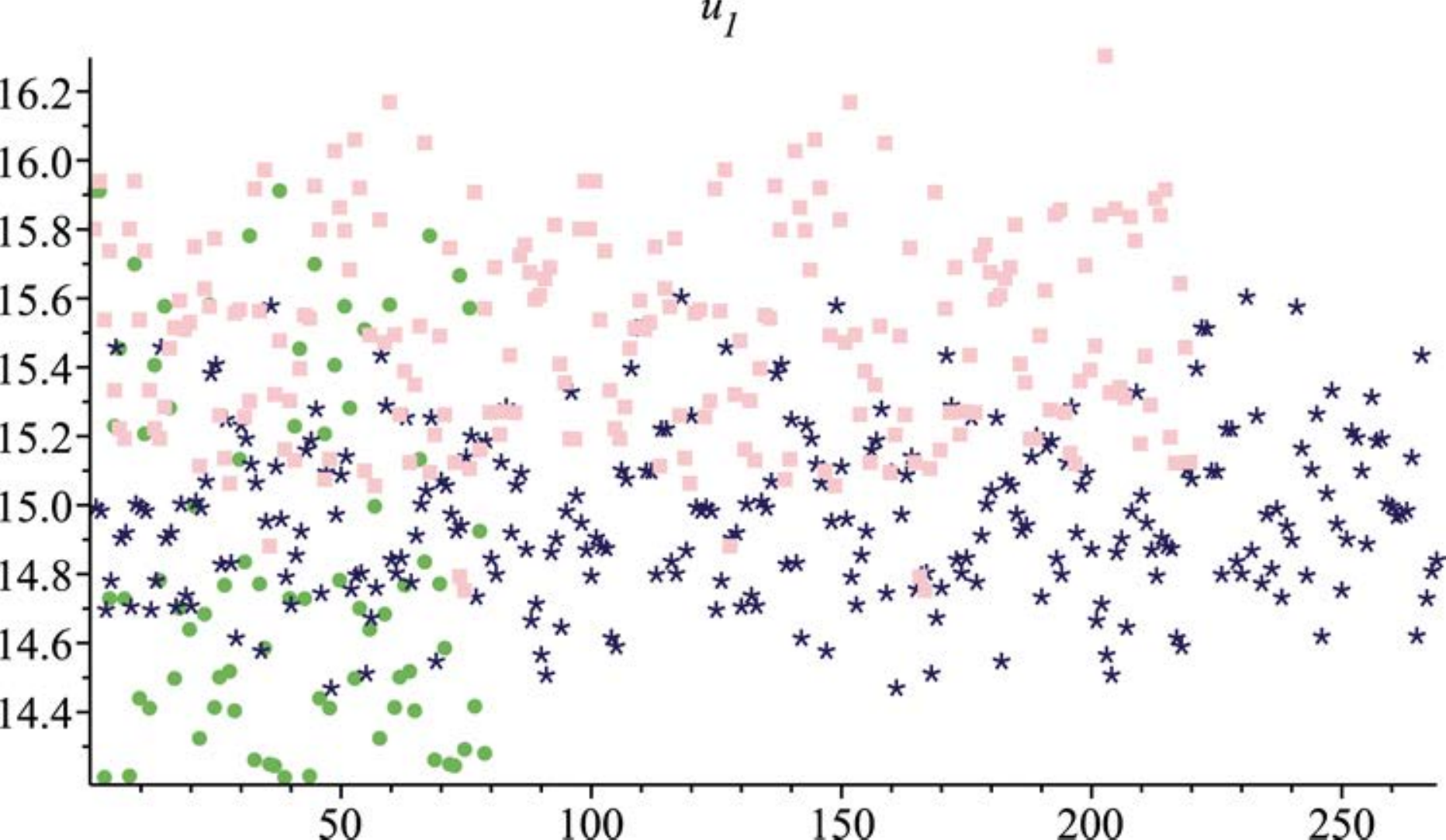}
	\includegraphics[width=2.5in]{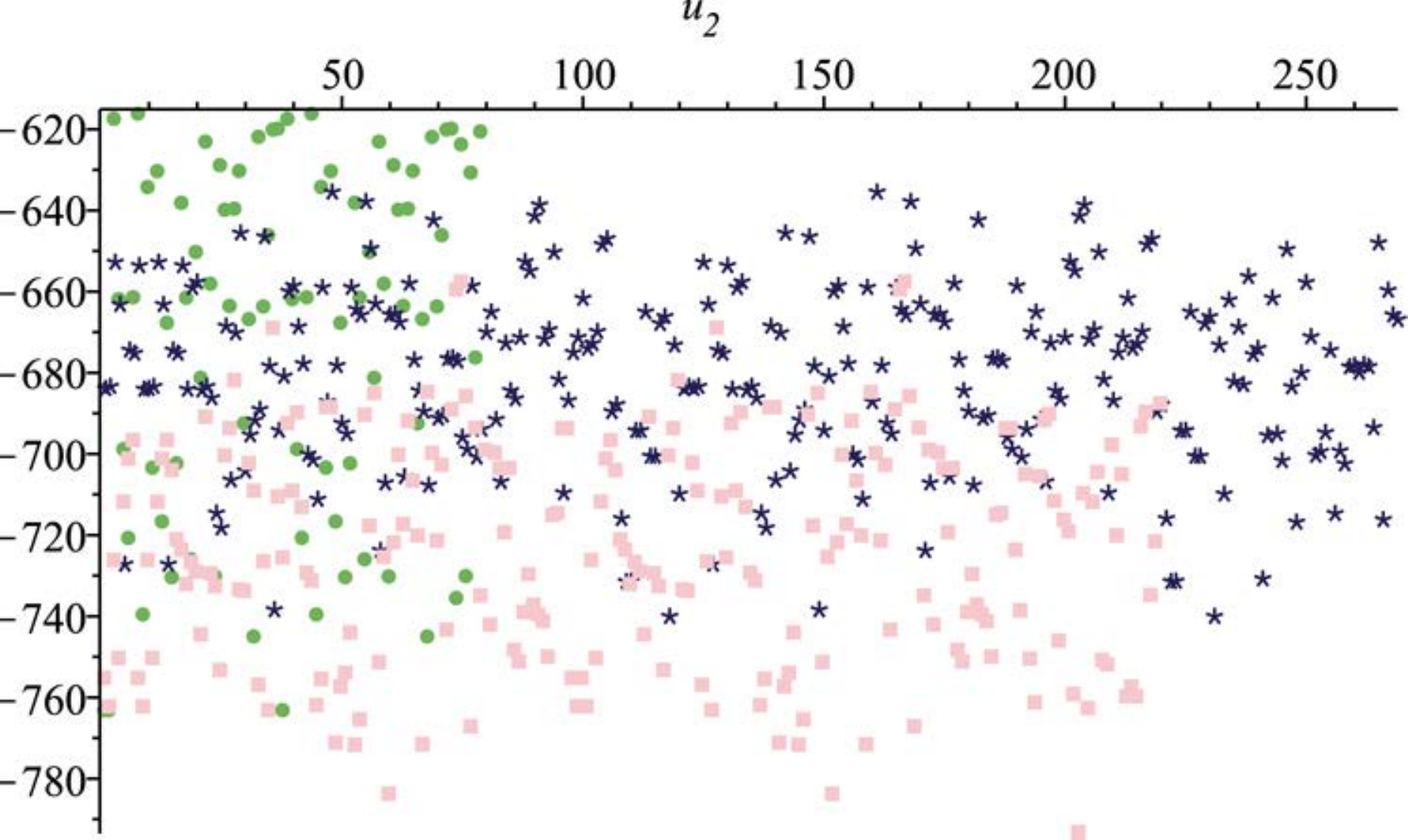}

	\vspace{10pt}	
	
	\includegraphics[width=2.5in]{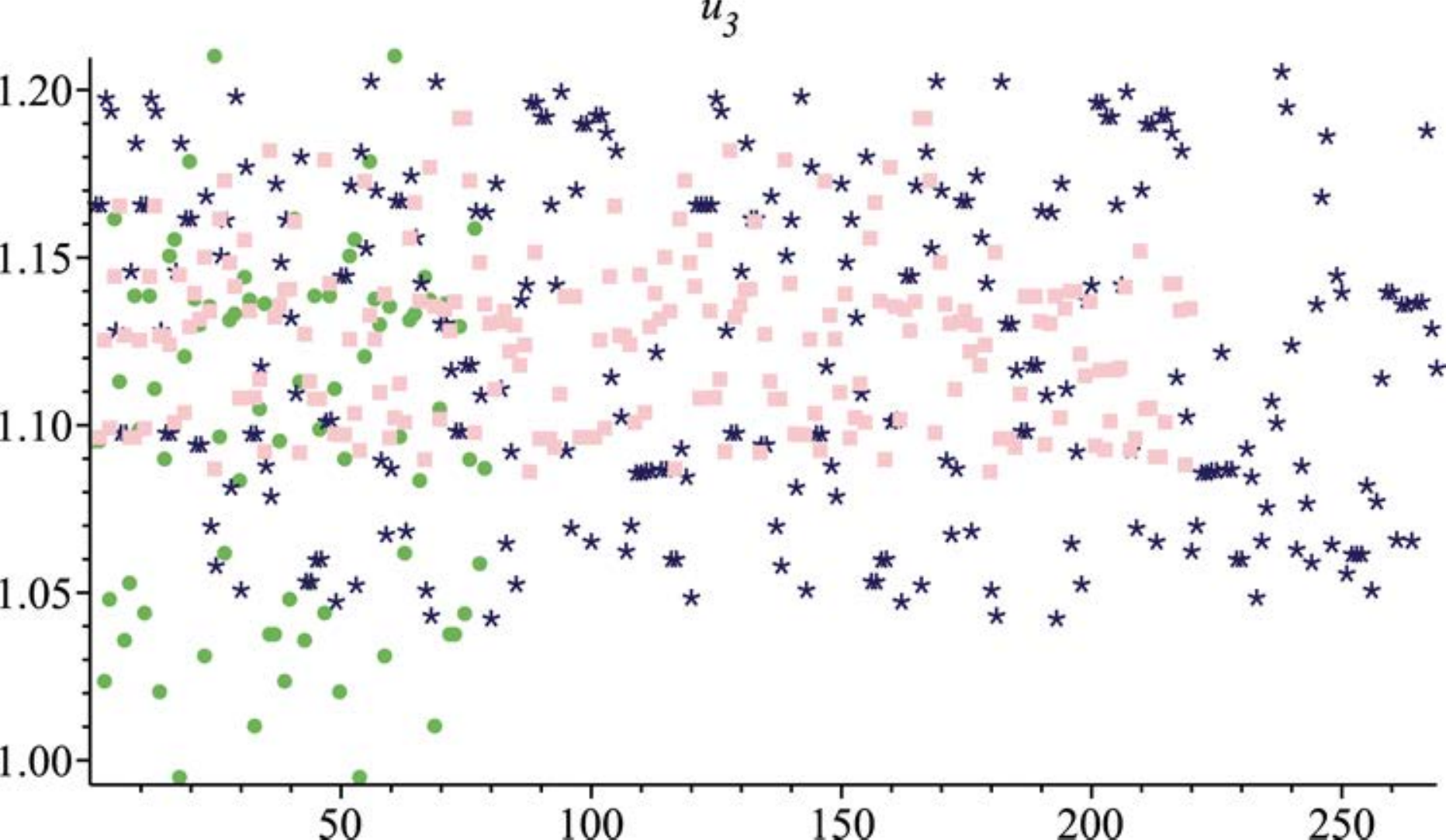}
	\includegraphics[width=2.5in]{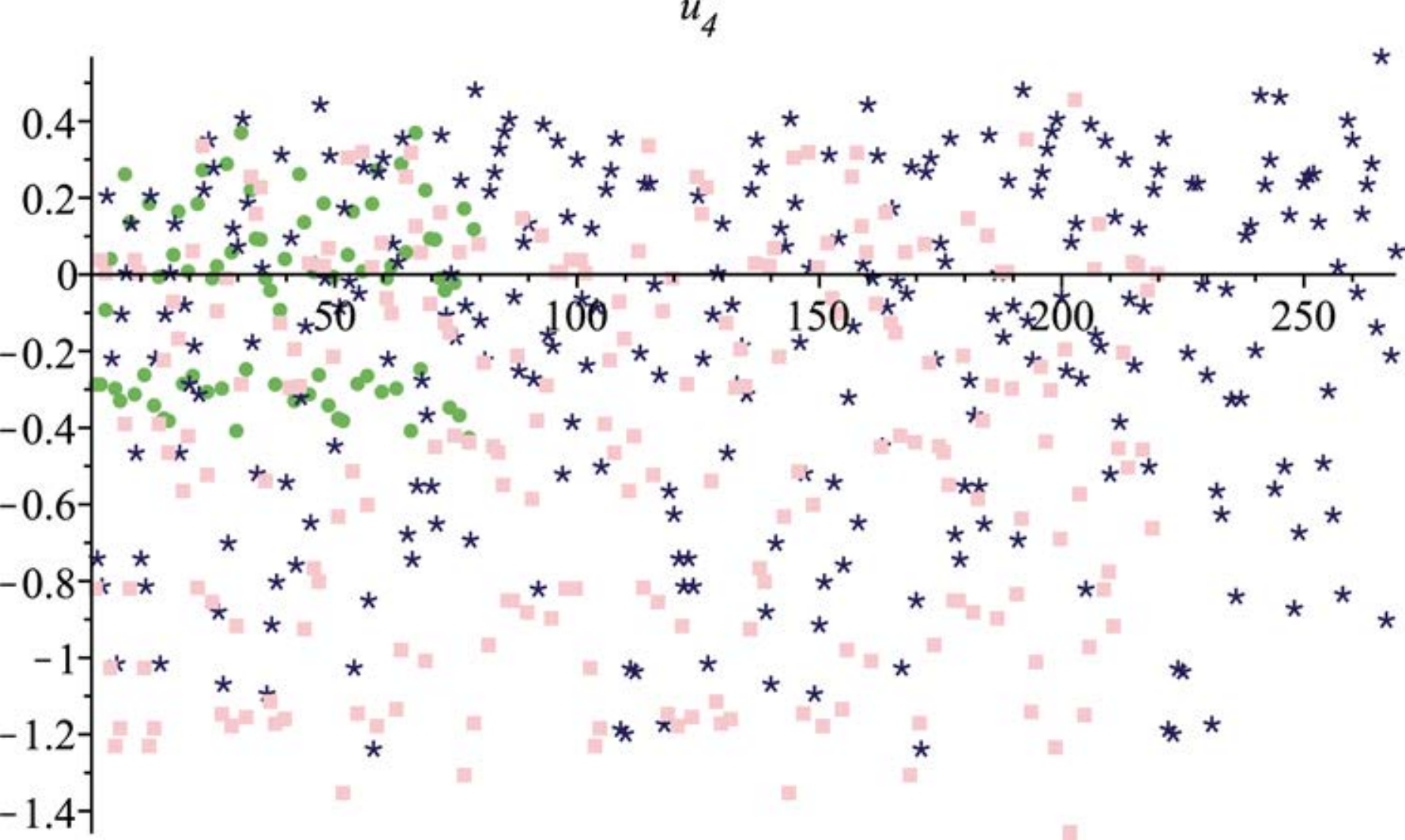}

	\vspace{10pt}	
	
	\includegraphics[width=2.5in]{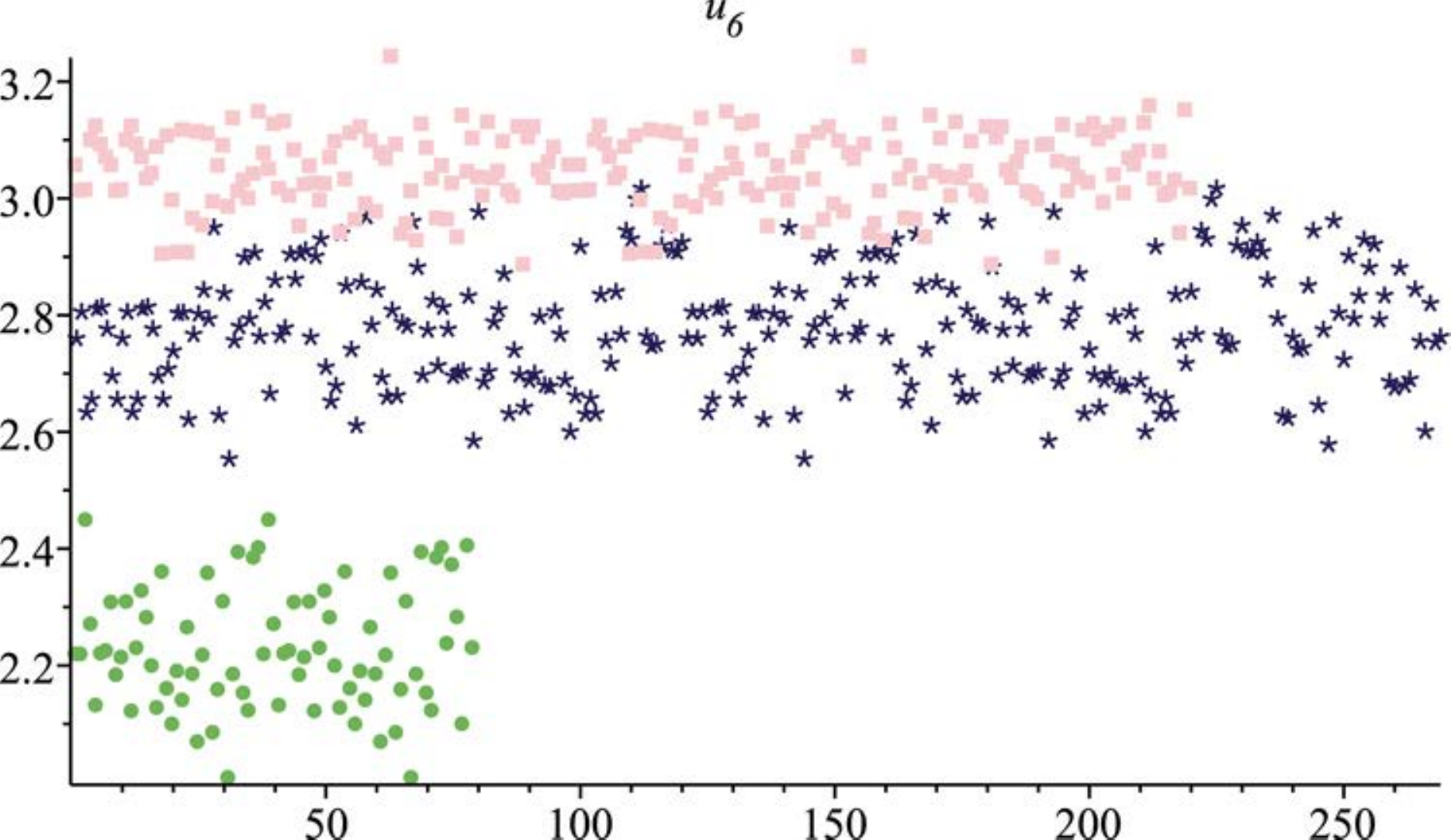}
	\includegraphics[width=2.5in]{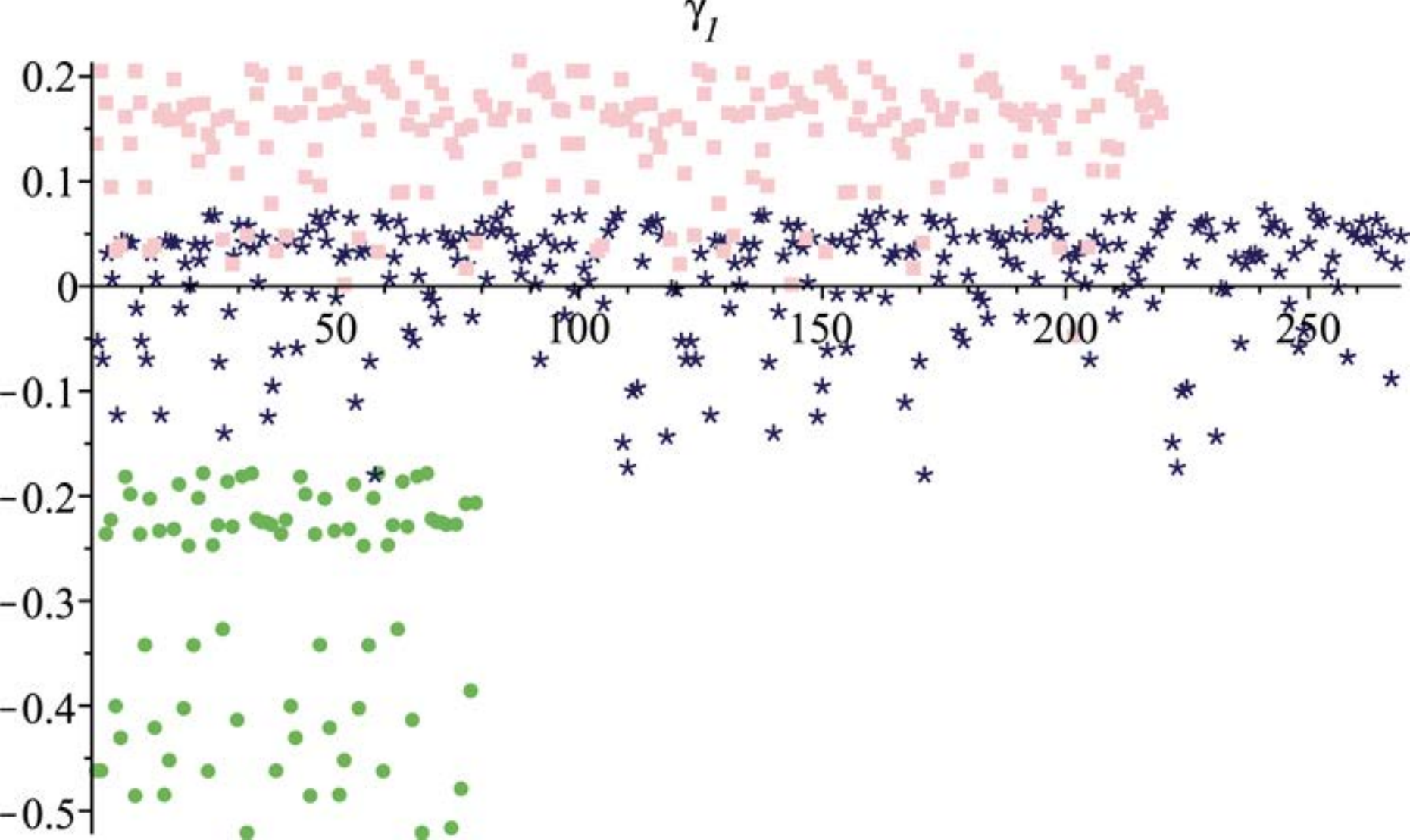}
	
	\caption{Simulation data for model parameters $u_1$, $u_2$, $u_3$, $u_4$, $u_6$ and $\gamma_1$ that result in $\chi< \chi_{\rm exp}$ (the rest of the parameters are given in Fig. \ref{F_parameters_Graph2}). Circles (green), stars (blue) and squares (pink) respectively represent permutations 127, 137 and 147.   In terms of the population of simulation points, it can be seen that permutation 127 is least favored.   The horizontal axis gives the simulation number.}
	\label{F_parameters_Graph1}
\end{figure}

\begin{figure}[!htb]
	\centering

	\includegraphics[width=2.5in]{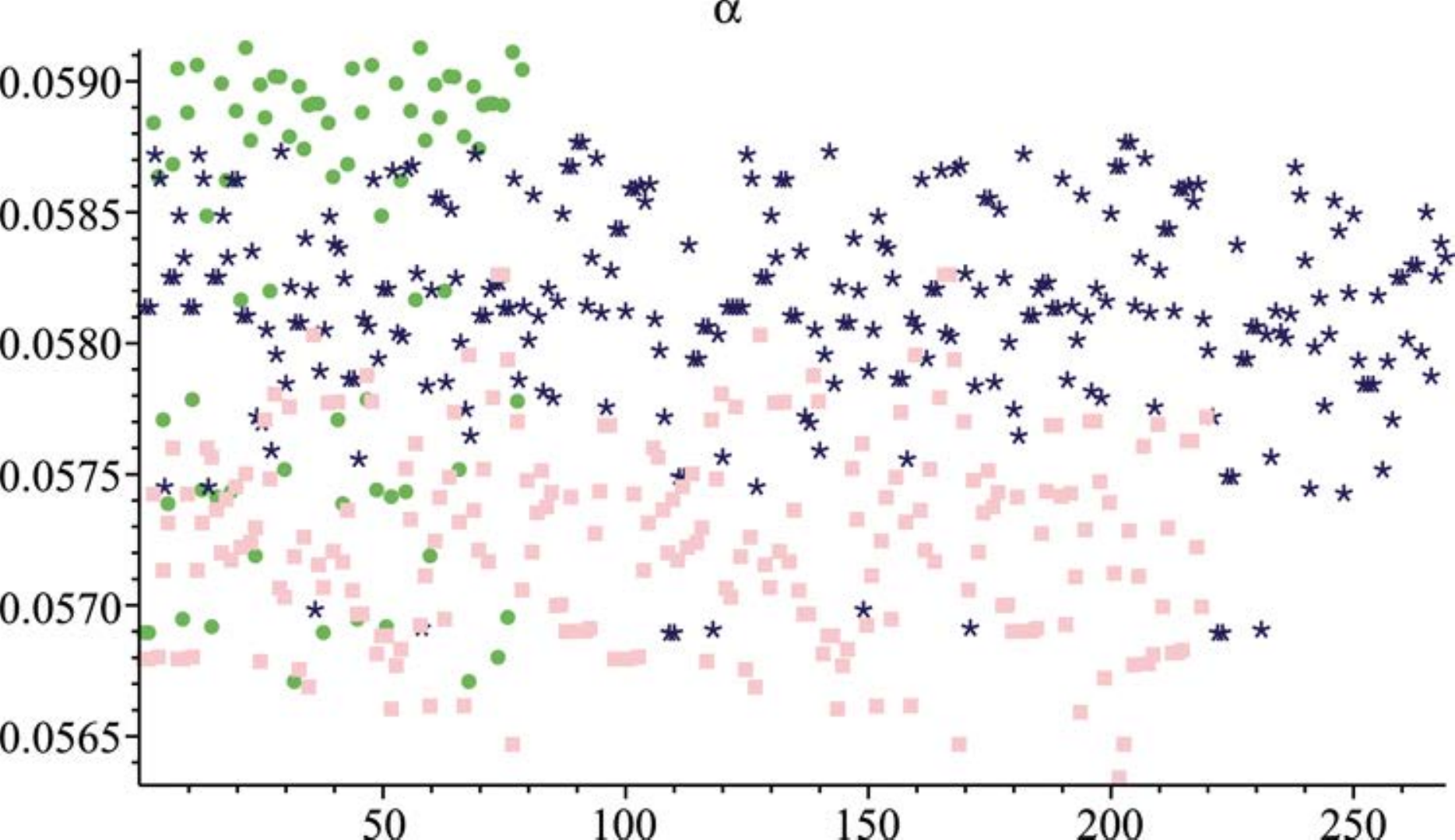}
	\includegraphics[width=2.5in]{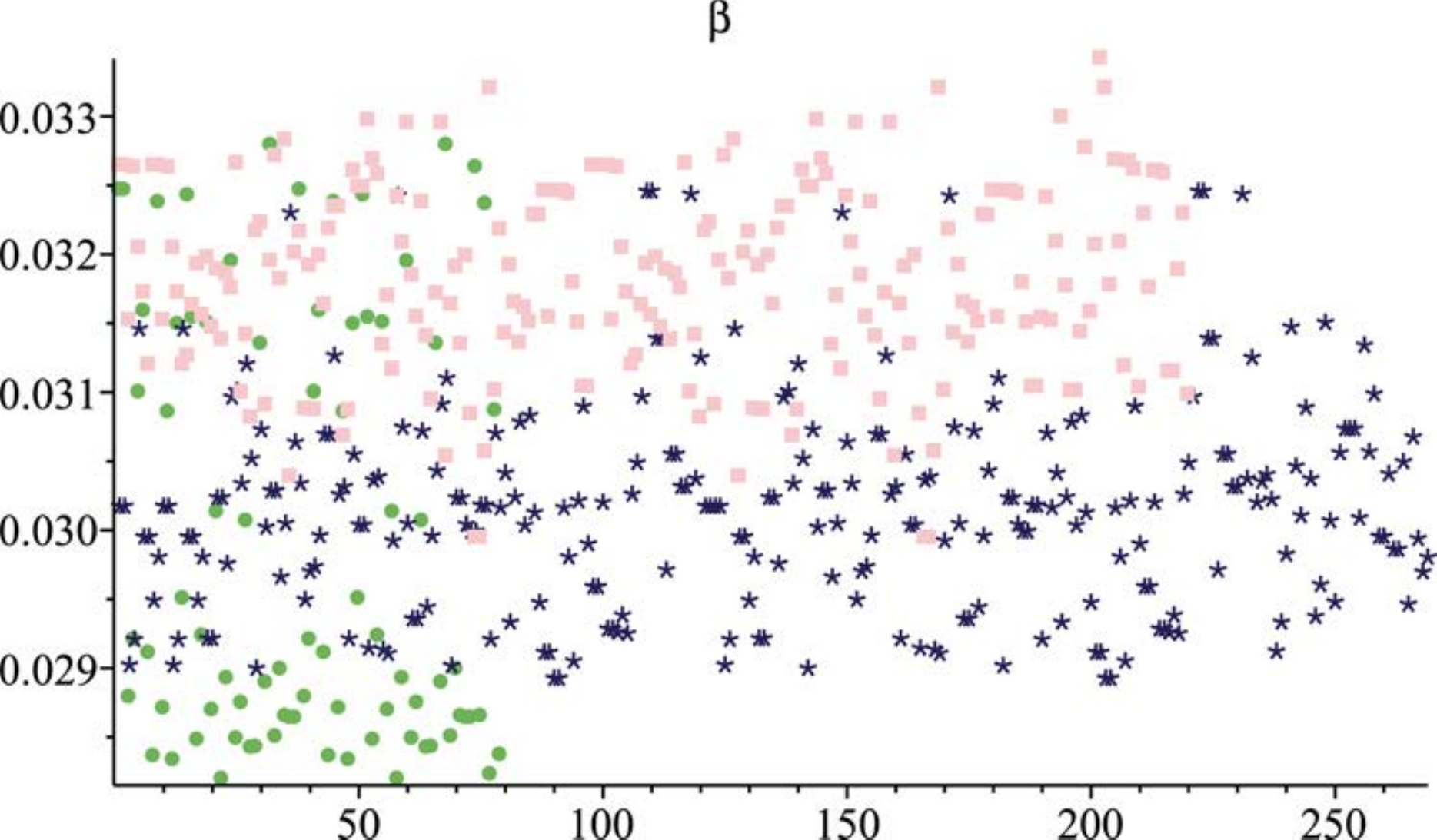}

	\vspace{10pt}		
	
	\includegraphics[width=2.5in]{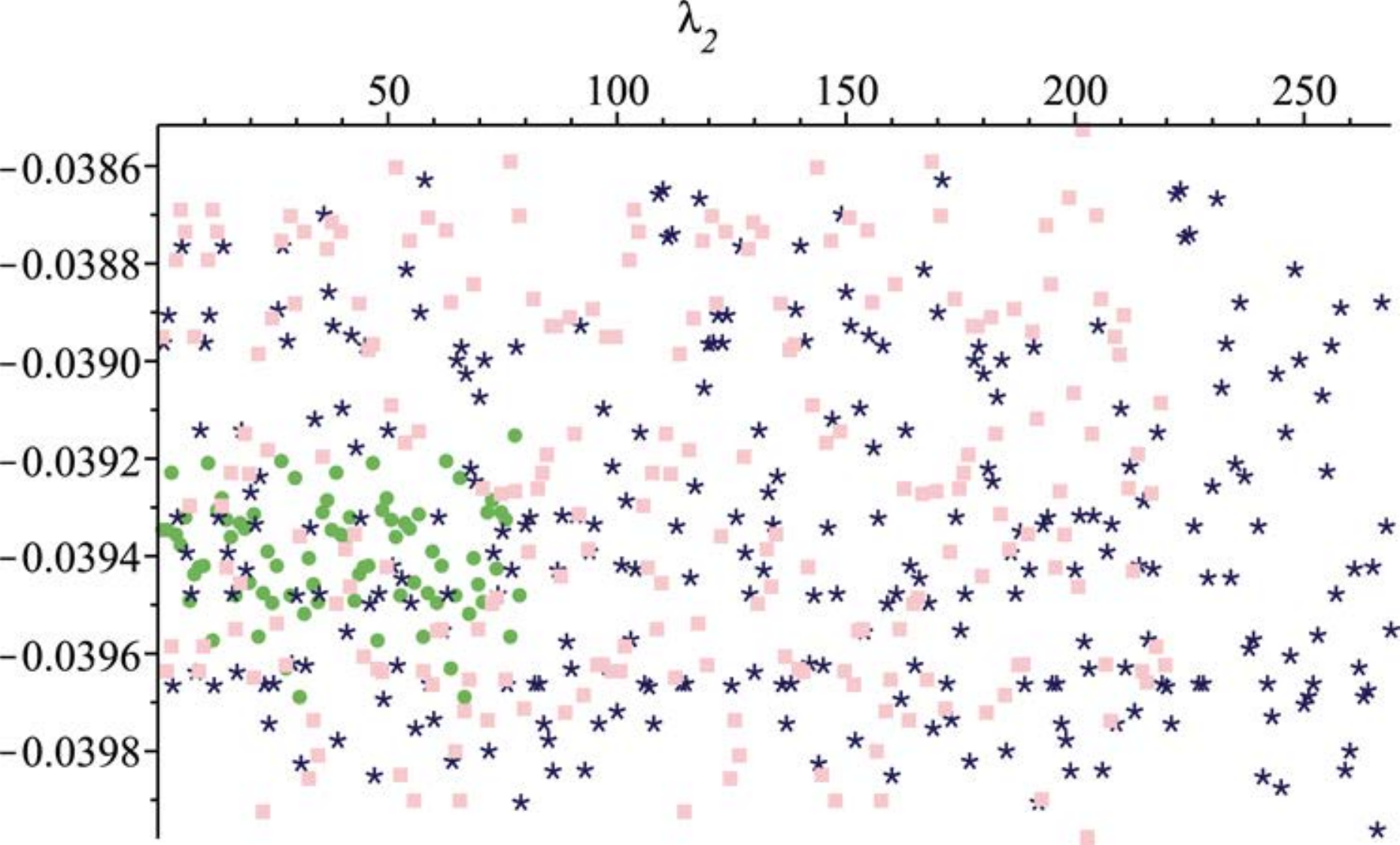}
	\includegraphics[width=2.5in]{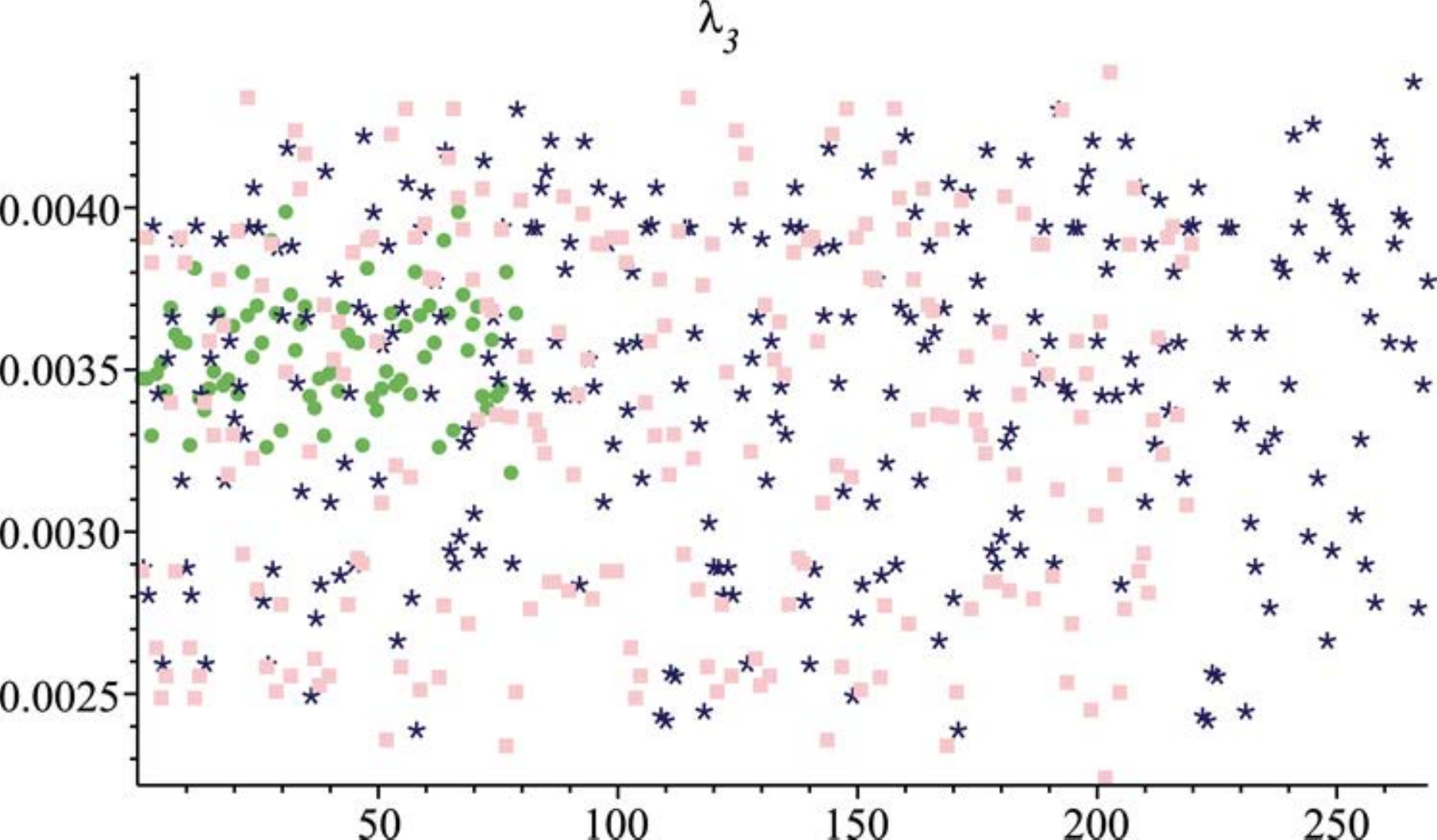}	
	
	\vspace{10pt}	
	
	\includegraphics[width=2.5in]{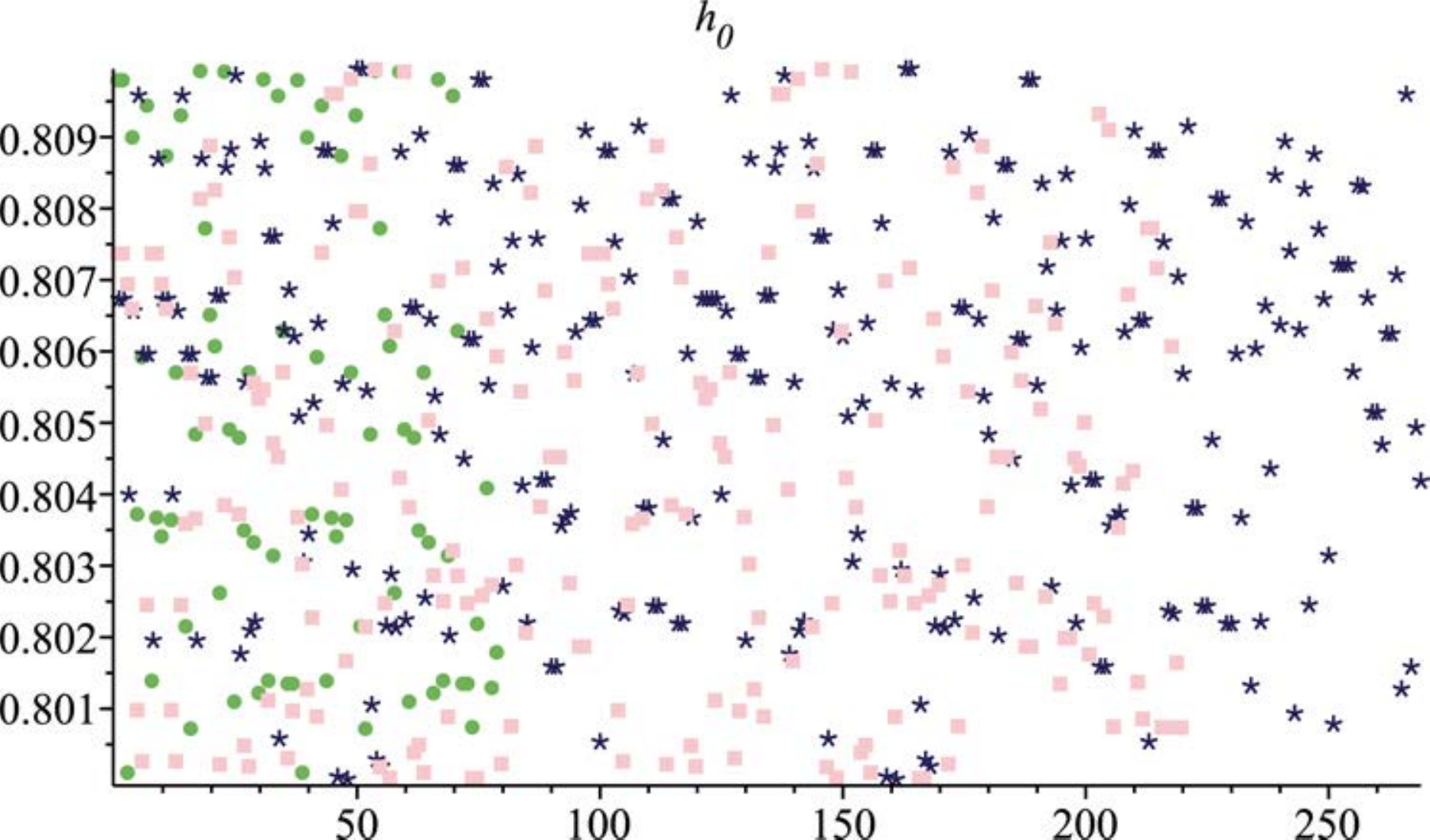}
	\includegraphics[width=2.5in]{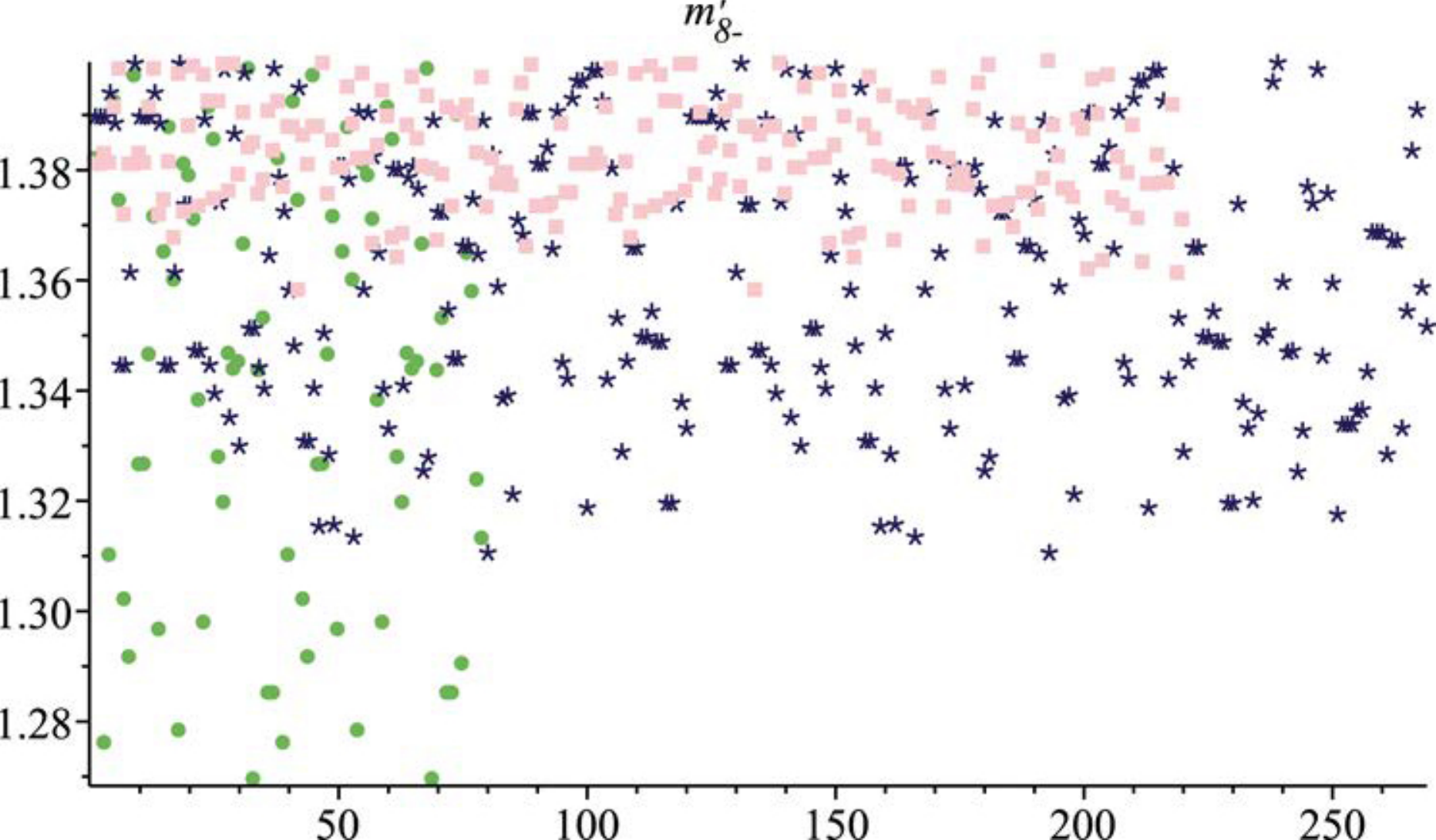}

	\vspace{10pt}
	
	\includegraphics[width=2.5in]{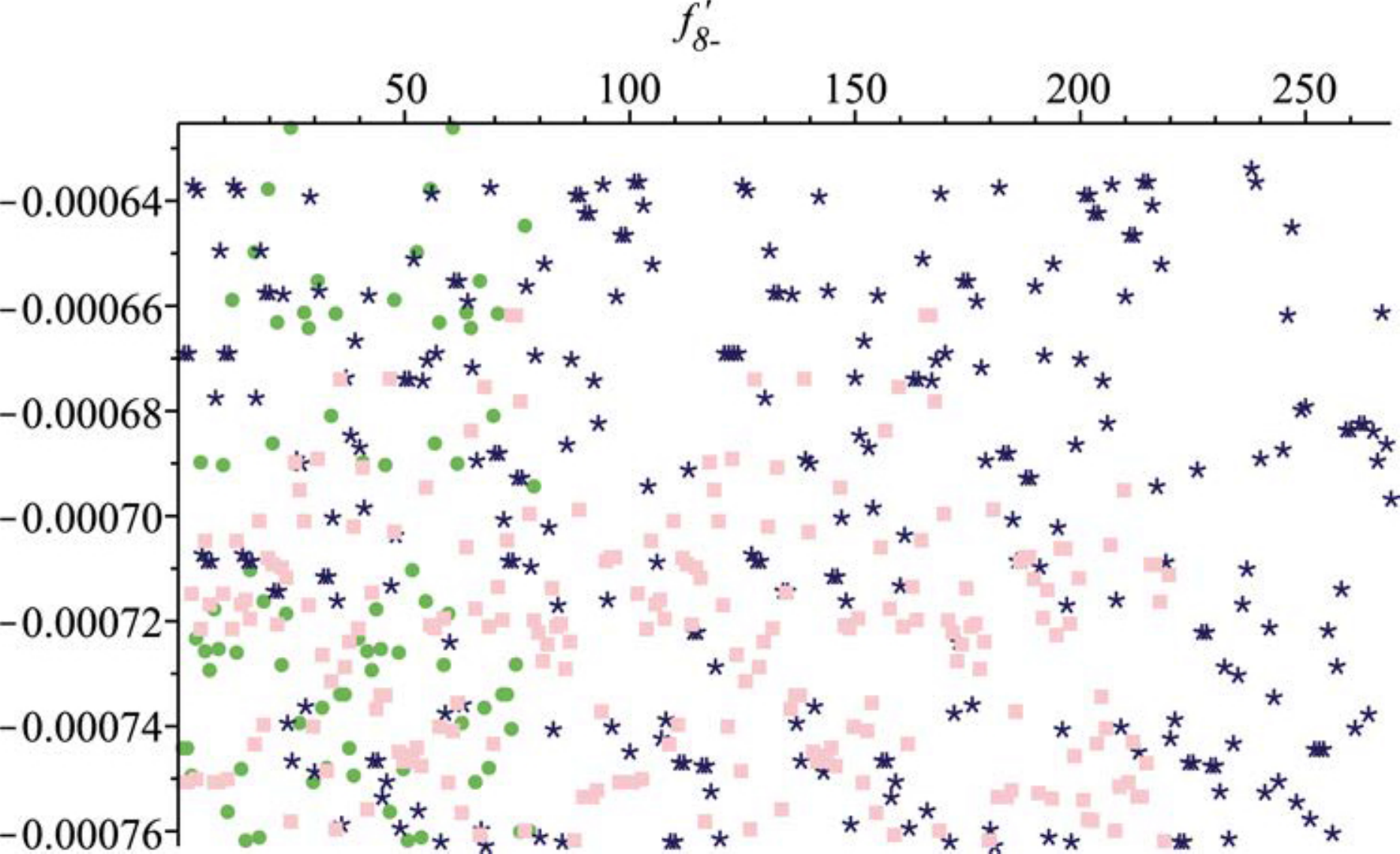}
	\includegraphics[width=2.5in]{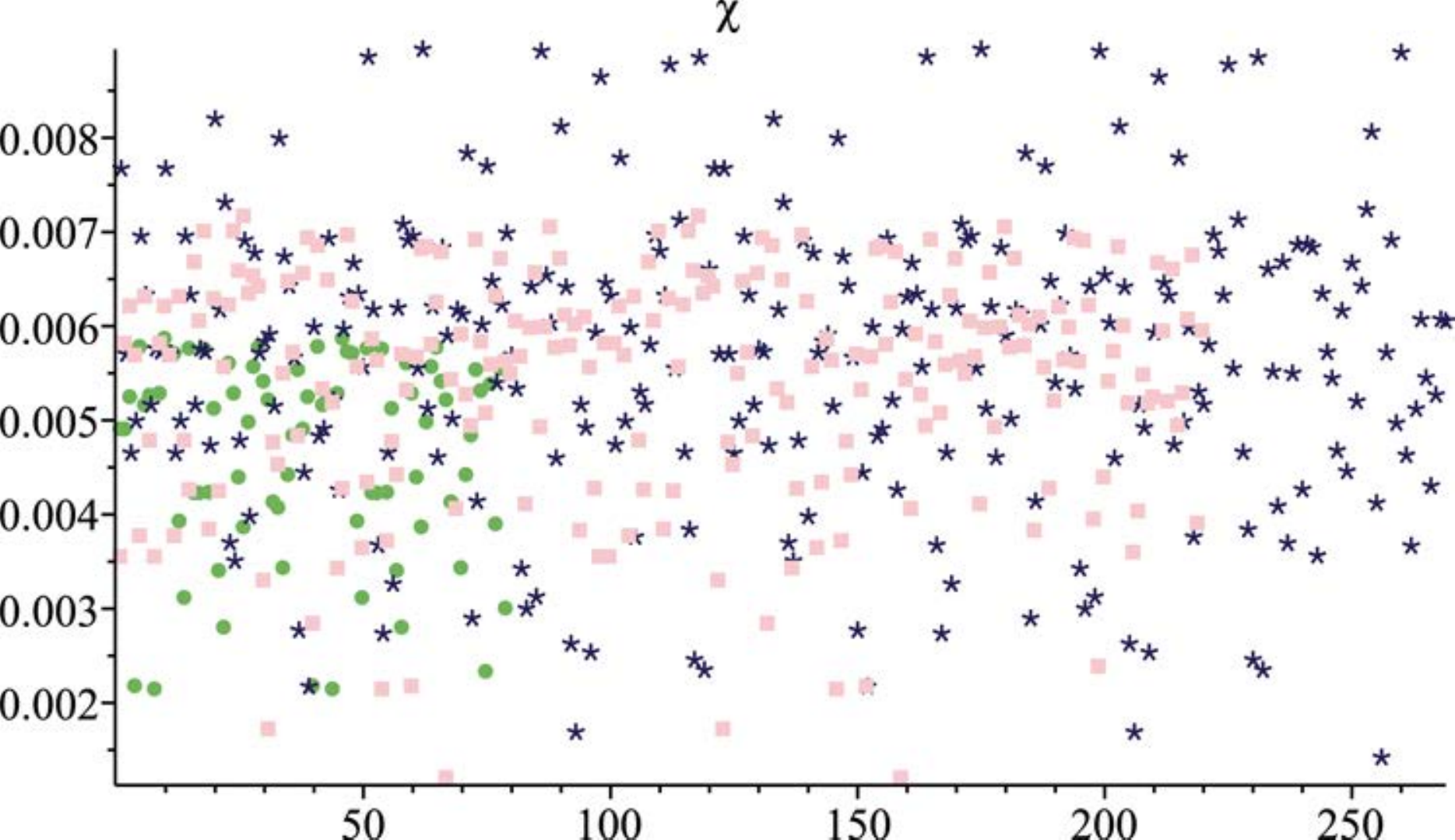}
	
	\caption{Simulation data for model parameters $\alpha$, $\beta$, $\lambda_2$,  $\lambda_3$ and $h_0$ as well as for inputs $m'_{8^-}$ and $f'_{8^-}$, together with the resulting $\chi$ for simulations that give $\chi< \chi_{\rm exp}$ (the rest of the parameters are given in Fig. \ref{F_parameters_Graph1}). Circles (green), stars (blue) and squares (pink) respectively represent permutations 127, 137 and 147.   In terms of the population of simulation points, it can be seen that permutation 127 is least favored.   The horizontal axis gives the simulation number.
 }
	\label{F_parameters_Graph2}
\end{figure}


\begin{thebibliography}{30}

\bibitem{Jora25} A. H. Fariborz and R. Jora, to appear in Phys. Rev. D; arXiv:1807.10927 (2018).
\bibitem{Jora26} A. H. Fariborz and R. Jora, arXiv:1807.10928 (2018).


\bibitem{PDG} M. Tanabashi et al. (Particle Data Group), Phys. Rev. D {\bf 98}, 030001 (2018).


\bibitem{ChPT1} S. Weinberg, Physica A {\bf 96},
327 (1979); J. Gasser and H. Leutwyler, Annalas
Phys. {\bf 158}, 142 (1984); Nucl. Phys. B {\bf
	250}, 465 (1985); H. Leutwyler, Annals of Physics 235, 165-203  (1994);
 I. Caprini, G. Colangelo and
H. Leutwyler, Phys. Rev. Lett. {\bf 96}, 132001 (2006).







\bibitem{ChUA1}
J.A. Oller and E. Oset, Nucl. Phys. A {\bf 620}, 438 (1997); Nucl. Phys. A {\bf 652}, 407 (1999).

\bibitem{ChUA2}
J.A. Oller and E. Oset,  Phys. Rev. D {\bf 60}, 074023 (1999).

\bibitem{ChUA3}
M. Jamin, J.A. Oller and A. Pich, Nucl. Phys. B {\bf 587}, 331 (2000).


\bibitem{ChUA4}
J.A. Oller,  Nucl.  Phys. A {\bf 727}, 353 (2003).

\bibitem{ChUA5}
J.A. Oller, Phys. Rev. D {\bf 71}, 054030 (2005).


\bibitem{ChUA6}
M. Albaladejo, J.A. Oller and L. Roca, Phys. Rev. D {\bf 82}, 094019 (2010).


\bibitem{ChUA7}
Z.H. Guo, J.A. Oller and J. Ruiz de Elvira, Phys. Rev. D {\bf 86}, 054006 (2012).


\bibitem{ChUA8}
M. Albaladejo and J.A. Oller, Phys. Rev. D {\bf 86}, 034003 (2012).











\bibitem{IAM1} A. Dobado and J.R. Pelaez, Phys. Rev. D {\bf 56}, 3057 (1997).

\bibitem{IAM2} J.A. Oller, E. Oset and J. R. Pelaez,  Phys. Rev. Lett. {\bf 80}, 3452 (1998); 
Phys. Rev. D {\bf 59}, 074001 (1999);  Phys. Rev. D {\bf 60}, 099906
(1999);  Phys. Rev. D {\bf 75}, 099903 (2007).

\bibitem{IAM3} J.R. Pelaez, Phys. Rev. Lett. {\bf 92}, 102001 (2004).




























\bibitem{LQCD1}
K.G. Wilson, Phys. Rev. D {\bf 10}, 2445 (1974); K.G. Wilson and J.B. Kogut, Phys. Rept. {\bf 12}, 75  (1974);
J. B. Kogut, D. K. Sinclair and L. Susskind, Nucl. Phys. B {\bf 114}, 199 (1976);
G. P. Lepage and P. B. Mackenzie, Phys. Rev. D {\bf 48}, 2250 (1993).


\bibitem{LQCD2}
M. Wagner, et al., Acta Phys. Polon. Supp. {\bf 6} 847 (2013);
C. Alexandrou, et al,  JHEP {\bf 137}, 1304 (2013);
T. Kunihiro, S. Muroya, A. Nakamura, C. Nonaka, M. Sekiguchi, H. Wada, in proceedings of {\it International IUPAP Conference on Few-Body Problems in Physics (FB 19), Bonn, Germany, 31 Aug - 5 Sep 2009},  EPJ Web Conf.3:03010 (2010);
C. McNeile, in proceedings of {\it 11th Int. Conf. on Meson-Nucleon
	Physics and
	the Structure of the Nucleon}, 10-14 Sept. 2007, J\"ulich, Germany;
C. McNeile and C. Michael (UKQCD Collaboration), Phys. Rev. D {\bf 74},
014508 (2006);
N. Mathur et al, hep-ph/0607110;
A. Hart et al (UKQCD Collaboration),
Phys. Rev. D {\bf 74}, 114504 (2006);
H. Wada (SCALAR Collaboration), Nucl. Phys. Proc. Suppl. {\bf 129}, 432
(2004); T. Kunihiro et al (SCALAR Collaboration), Phys. Rev. D {\bf 70},
034504 (2003);
N. Ishii, H. Suganuma and H.
Matsufuru, Phys. Rev. D {\bf 66}, 014507 (2002);
Xi-Yan Fang, Ping Hui, Qi-Zhou Chen and D. Schutte,
Phys. Rev. D {\bf 65}, 114505 (2002);
M.G. Alford and R.L. Jaffe, Nucl. Phys. B {\bf
	578}, 367 (2000);
C.J.
J. Sexton, A. Vaccarino and D.
Weingarten, Phy. Rev. Lett. {\bf 75}, 4563 (1995);

\bibitem{Bali1}  G. S Bali et al., Phys. Lett. B {\bf 309}, 379 (1993).

\bibitem{Morningstar} C. J. Morningstar and M. J. Peardon, Phys. Rev. D 60, 034509 (1999).

\bibitem{Chen} Y. Chen et al., Phys. Rev. D 73, 014516 (2006).

\bibitem{Ochs} W. Ochs, J. Phys. G 40, 043001 (2013).

\bibitem{Hart1} A. Hart and M. Teper [UKQCD Collaboration], Phys. Rev. D {\bf 65}, 034502 (2002).

\bibitem{Hart2} A. Hart et al. [UKQCD Collaboration], Phys. Rev. D {\bf 82}, 034501 (2010).


\bibitem{Gui:2012gx}
L.~C.~Gui {\it et al.} [CLQCD Collaboration],
Phys.\ Rev.\ Lett.\  {\bf 110}, 021601 (2013).

\bibitem{Kronfeld} A. S. Kronfeld, Ann. Rev. Nucl. Part. Sci. {\bf 62}, 265-284 (2012).


\bibitem{McNeile:2000xx}
C.~McNeile {\it et al.} [UKQCD Collaboration],
Phys.\ Rev.\ D {\bf 63}, 114503 (2001).





\bibitem{SVZ} M. A. Shifman, A. I. Vainshtein and V. I. Zakharov, Nucl. Phys. B {\bf 147}, 385, 448 (1979).

\bibitem{Narison0} S. Narison, Nucl. Part. Phys. Proc. 258-259, 189-194 (2015).

\bibitem{Eli}
R.T. Kleiv, T.G. Steele, A. Zhang and I. Blokland, Phys. Rev. D {\bf 87}, 125018 (2013);
D. Harnett, R.T. Kleiv, K. Moats  and T.G. Steele,  Nucl. Phys. A {\bf 850}, 110 (2011);
J. Zhang, H.Y. Jin, Z.F. Zhang, T.G. Steele and D.H. Lu, Phys. Rev. D {\bf 79}, 114033 (2009);
Fang Shi, T.G. Steele, V. Elias, K.B. Sprague, Ying
Xue and A.H.  Fariborz, Nucl. Phys. A {\bf 671}, 416
(2000);
V. Elias, A.H. Fariborz, Fang Shi and
T.G. Steele, Nucl.  Phys.  A {\bf 633}, 279 (1998).


\bibitem{Narison22} S. Narison and G. Veneziano, Int. J. Mod. Phys. A 4, 2751 (1981).

\bibitem{Narison1} S. Narison, Z. Phys. C {\bf 26}, 209 (1984).

\bibitem{Narison3} S. Narison, Nucl. Phys. Proc. Suppl. 186, 306 (2009).

\bibitem{Narison4} S. Narison, Phys. Lett. B {\bf 673}, 30 (2009).

\bibitem{Narison2} S. Narison, Phys. Lett. B {\bf 706}, 412-422 (2012).


\bibitem{Huang} T. Huang and H. Y. Jin, A. Zhang, Phys. Rev. D {\bf 59}, 034026 (1999).






\bibitem{Schechter1} J. Schechter and Y. Ueda, Phys. Rev. D {\bf 3}, 2874-1893 (1971).
\bibitem{Schechter}J. Schechter, Syracuse University preprint: SU-4217-155, COO-3533-155 (1979).
\bibitem{Schechter2} C. Rosenzweig, J. Schechter and C. G. Trahern, Phys. Rev. D {\bf 21}, 3388 (1980).
\bibitem{Schechter3} C. Rosenzweig, A. Salomone and J. Schechter, Phys. Rev. D {\bf 24}, 2545-2548 (1981).



\bibitem{Jora1} A. H. Fariborz, R. Jora, J. Schechter, Phys. Rev. D {\bf 72}, 034001 (2005), hep-ph/0506170.
\bibitem{Jora2} A. H. Fariborz, R. Jora and J. Schechter, Phys. Rev. D {\bf 77}, 034006 (2008), arXiv: 0707.0843.
\bibitem{Jora3} A.H. Fariborz, R. Jora and J. Schechter,
Phys. Rev. D {\bf 76}, 114001 (2007),
arXiv:0708.3402 [hep-ph].
\bibitem{Jora4} A. H. Fariborz, R. Jora and J. Schechter, Phys. Rev. D {\bf 77}, 094004 (2008), arXiv: 0801.2552.
\bibitem{Jora41} A. H. Fariborz, R. Jora and J. Schechter, Phys. Rev. D {\bf 76}, 014011 (2007).
\bibitem{Jora42} A. H. Fariborz, R. Jora and J. Schechtr, Phys. Rev. D {\bf 77}, 034006 (2008).
\bibitem{Jora43} A. H. Fariborz, R. Jora and J. Schechter, Phys. Rev. D {\bf 76}, 114001 (2007).
\bibitem{Jora5} A. H. Fariborz,  R. Jora, J. Schechter, Phys. Rev. D {\bf 79}, 074014 (2009), arXiv: 0902.2825.
\bibitem{Jora6} A. H. Fariborz, R. Jora, Phys. Rev. D {\bf 95}, 114001 (2017), arXiv:1701.00812.
\bibitem{Jora7} A. H. Fariborz and R. Jora, Phys. Rev. D 96, no. 9, 096021 (2017).
\bibitem{Giacosa11} D. Paranglija and F. Giacosa, Eur. Phys. J. C {\bf 77}, no. 7, 450 (2017).
\bibitem{Giacosa22} F. Giacosa, A. Koenigstein and R. D. Pisarski, Phys. Rev. D {\bf 97}, no. 9, 091901 (2018).




\bibitem{jaffe} R.L.~Jaffe,
Phys.\ Rev.\ D {\bf 15}, 267 (1977).


\bibitem{07_KZ}
E. Klempt and A. Zaitsev, Phys. Rept. 454,1 (2007).

\bibitem{Pelaez:2015qba} 
  J.~R.~Pelaez,
  Phys.\ Rept.\  {\bf 658}, 1 (2016).
  

\bibitem{Weinstein:1990gu} 
  J.~D.~Weinstein and N.~Isgur,
  Phys.\ Rev.\ D {\bf 41}, 2236 (1990).
  doi:10.1103/PhysRevD.41.2236




\bibitem{Ioffe} B. L. Ioffe and K. N. Zyablyuk,Eur. Phys. J. C {\bf 27}, 229 (2003); B. L. IOffe Progr. Part. Nucl. Phys. {\bf 56}, 232 (2006).
\bibitem{Rakow} G. Burgio, F. Di REnzo, G. Marchesini and E. Onofri, Phys. Lett. B {\bf 422}, 219 (1998); R. Horley, P. E. L. Rakow and G. Schierholz, Nucl. Phys. (Proc. Supp.) B {\bf 106}, 870 (2002).
\bibitem{Giacomo1} A. Di Giacomo, Non-perturbative methods, ed. Nariso, World Scientific (1985); M. Campostrini, A. Di Giacomo and G. C. Rossi, Phys. Lett. B {\bf 100}, 481 (1981).
\bibitem{Giacomo2} M. D'EliaA. Di Giacomo and E. Meggiolaro, Phys. Lett. B {\bf 408}, 315 (1997).

\bibitem{Bali} G. Bali, C. Bauer and A. Pineda, Phys. Rev. D {\bf 89}, 054505 (2014); G. Bali, C. Bauer and A. Pineda, arXiv:1403.6477 (2014).
\bibitem{Lee} T. Lee, Nucl. and Particle Phys. Proc.  258-259, 181-184 (2015).




\bibitem{Novikov} V. A. Novikov, M. A. Shifman, A. I. Vaishtein and V. I. Zakharov, Nucl. Phys. B 191, 301 (1981).
















\bibitem{Fritsch} H. Fritsch and P. Minkowski, Nuov. Cim.  30 A, 293 (1975); R. L. Jaffe and K. Johnson, Phys. Lett. B 60, 2011 (1976);
D. Robson, Nucl. Phys. B {\bf 130}, 328 (1977).


\bibitem{Scharre} D. L. Scharre et al., Phys. Lett. B 97, 329 (1980).

\bibitem{Mathieu} V. Mathieu, N. Kochelev and V. Vento, Int. J. mod. Phys. E {\bf 18}, 1 (2009); N. Kochelev and D. P. Min, Phys. Lett. B {\bf 633}, 283 (2006).




\bibitem{Colangelo:2009ra}
P.~Colangelo, F.~Giannuzzi and S.~Nicotri,
Phys.\ Rev.\ D {\bf 80}, 094019 (2009).



\bibitem{Albaladejo:2008qa}
M.~Albaladejo and J.~A.~Oller,
Phys.\ Rev.\ Lett.\  {\bf 101}, 252002 (2008).


\bibitem{Giacosa:2005zt}
F.~Giacosa, T.~Gutsche, V.~E.~Lyubovitskij and A.~Faessler,
Phys.\ Rev.\ D {\bf 72}, 094006 (2005).

\bibitem{Chanowitz:2005du}
M.~Chanowitz,
Phys.\ Rev.\ Lett.\  {\bf 95}, 172001 (2005).


\bibitem{Giacosa:2005qr}
F.~Giacosa, T.~Gutsche, V.~E.~Lyubovitskij and A.~Faessler,
Phys.\ Lett.\ B {\bf 622}, 277 (2005).



\bibitem{Fariborz:2003uj}
A.~H.~Fariborz,
Int.\ J.\ Mod.\ Phys.\ A {\bf 19}, 2095 (2004).



\bibitem{Fariborz:2006xq}
A.~H.~Fariborz,
Phys.\ Rev.\ D {\bf 74}, 054030 (2006).


\bibitem{Minkowski:2004xf}
P.~Minkowski and W.~Ochs,
Eur.\ Phys.\ J.\ C {\bf 39}, 71 (2005).


\bibitem{Close:2001ga}
F.~E.~Close and A.~Kirk,
Eur.\ Phys.\ J.\ C {\bf 21}, 531 (2001).




\bibitem{Lee:1999kv}
W.~J.~Lee and D.~Weingarten,
Phys.\ Rev.\ D {\bf 61}, 014015 (2000).


\bibitem{Sexton:1995kd}
J.~Sexton, A.~Vaccarino and D.~Weingarten,
Phys.\ Rev.\ Lett.\  {\bf 75}, 4563 (1995).


\bibitem{Amsler:1995td}
C.~Amsler and F.~E.~Close,
Phys.\ Rev.\ D {\bf 53}, 295 (1996)


\bibitem{Amsler:1995tu}
C.~Amsler and F.~E.~Close,
Phys.\ Lett.\ B {\bf 353}, 385 (1995).




















\end{thebibliography}
\end{document}